\documentclass[bibyear]{aa} 
\usepackage{txfonts}
\usepackage{graphicx}
\usepackage{amsmath,amssymb,amsfonts,latexsym,mathtools}
\usepackage{natbib}
\usepackage[flushleft]{threeparttable}
\usepackage{booktabs}
\usepackage[subrefformat=parens]{subcaption}
\usepackage{hyperref}
\def\v{\varv}

\newcommand{\red}[1]{\textcolor{black}}%{\bf{#1}}}
\newcommand{\Oka}[1]{\textcolor{black}}%{\bf{#1}}}

\begin{document}

%\title{Monte Carlo simulation for carbon depletion in a protoplanetary disk}
    \title{Effects from different grades of stickiness between icy and silicate particles on carbon depletion in protoplanetary disks}
    \author{T. Okamoto\inst{1,2}\and S. Ida\inst{1}}
    \institute{Earth-Life Science Institute, Institute of Science Tokyo (previously, Tokyo Institute of Technology),
    Meguro-ku, 152-8550 Tokyo, Japan\\
    \email{okamoto.t.aw@m.titech.ac.jp}
    \and
    Laboratoire Lagrange, Centre National de la Recherche Scientifique, Observatoire de la Côte d'Azur, 06304 Nice, France}
    
\abstract
    {The Earth and other rocky bodies in the inner Solar System are significantly depleted in carbon, compared to the Sun and the interstellar medium (ISM) dust. Observations suggest that more than half of the carbon material in the ISM and comets are in a highly refractory form, such as amorphous hydrocarbons 
    and (less refractory) complex organics, 
    which can make up the building blocks of rocky bodies. 
    While amorphous hydrocarbons can be destroyed by
    photolysis and oxidation, previous studies have suggested that the radial transport of solid
particles suppresses carbon depletion. The only
    exception is the case of strictly complex organics as the refractory carbons, which are considerably less refractory than amorphous hydrocarbons.
    }
    {We aim to reveal the conditions for the severe carbon depletion in the inner Solar System, by adding potentially more realistic settings: different levels of stickiness between icy and silicate particles and high-temperature regions in the upper optically thin layer of the disk,
    which were not included in the previous works.
    }
    {We performed a 3D Monte Carlo simulation of 
    radial drift and turbulent diffusion of
    solid particles in a steady accretion disk with the above additional settings as well as ice evaporation and recondensation. 
    We considered the photolysis and oxidation of hydrocarbons in the upper layer as well as the pyrolysis of complex organics to evaluate the radial distribution of carbon fraction in the disk by locally averaging individual particles.
    }
    {The carbon fraction drops off inside the snow line by two orders of magnitude compared to the solar value, under the following conditions: i) when silicate particles
    are much less sticky than icy particles and ii) when there are high-temperature regions in the disk upper layer. The former leads to fast decay of the icy pebble flux, while the silicate particles are still piling up inside the snow line. The latter contributes to the efficient turbulent stirring up of silicate particles to the upper UV-exposed layer. 
     }
    {
    We have identified simulation settings to reproduce a carbon depletion pattern that is consistent with the observed one in the inner Solar System. 
    The conditions are not too restricted and allow for a diverse carbon fraction of rocky bodies. These effects could be responsible for the observed large diversity of metals on photospheres of white dwarfs and may suggest diverse surface environments for rocky planets in habitable zones. 
    }


% \titlerunning{} 
\authorrunning{T. Okamoto \& S. Ida}
\maketitle

\section{Introduction} \label{sec:intro}

\begin{figure}
    \centering
    \includegraphics[keepaspectratio,scale=0.4]{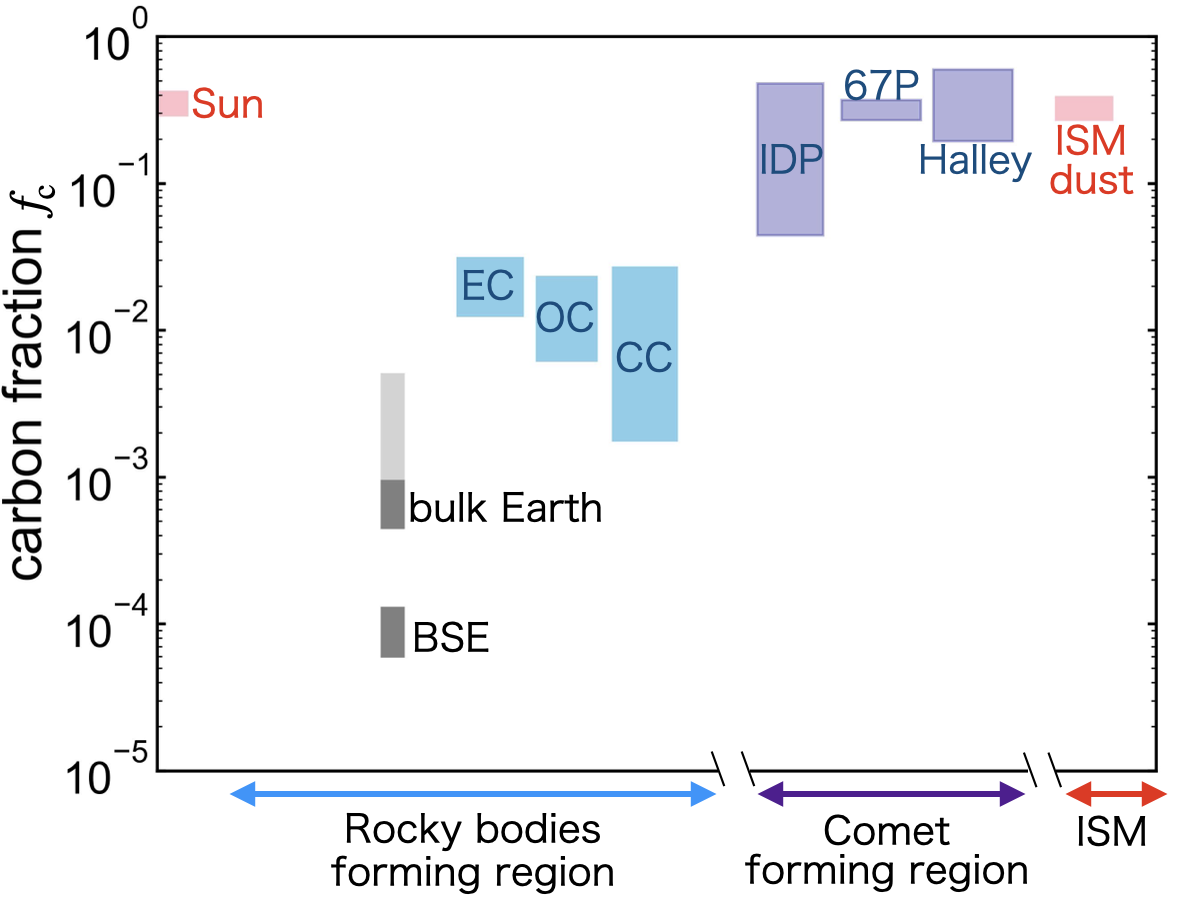}
    \caption{Carbon fraction of Solar System bodies,
    where the carbon fraction, $f_{\rm c}$, is defined by the mass fraction of carbon in the sum of solid carbon and the silicates.
    For ISM dust, comets, and IDPs, this fraction is calculated from the C/Si atomic ratio
    \citep[][and references therein]{Matrajt2005,Bardyn2017,Bergin2015,Binkert2023}.
    %, assuming that all silicon atoms are locked in ${\rm Mg_2SiO_4}$. 
    For comparison, we also plot the ``Sun" values ($f_{\rm c}\simeq 0.25$), which refer to hypothetical dust with solar C/Si.
    The fractions of chondrites are normalized by the matrix volume of each chondrite (see the main text).
    While \cite{Bergin2015} and \cite{Binkert2023} showed $f_{\rm c}$ fraction monotonically decreases as the decrease in the heliocentric radius decreases, the matrix-normalized $f_{\rm c}$ of chondrites may be independent of the distance. 
    The uncertainties of chondrites' $f_{\rm c}$ values come from the variation among each chondrite. The very low $f_{\rm c}$ of BSE and ``bulk Earth" are also discussed in the main text.}
   \label{fig:obs} 
\end{figure}

The bodies in the Solar System are highly depleted in carbon, compared to the Sun. Figure \ref{fig:obs} shows the carbon fraction, $f_{\rm c}$, of the Solar System bodies, where the carbon fraction, $f_{\rm c}$, is defined by the mass fraction of refractory carbon relative to the total refractory components \citep[for details, see][]{Binkert2023}. The carbon fractions for the Sun, interstellar medium (ISM) dust, comets, and interplanetary dust particles (IDPs) are calculated from the C/Si atomic ratio in each body \citep{Binkert2023}. 
The carbon fractions of chondrite meteorites are calculated using the mass fraction of insoluble carbon contents, normalized by their matrix volume \citep[][see below]{Alexander2007}. 
While the $f_{\rm c}$ values of  ISM dust, IDPs, and comets are comparable to the solar value, the
$f_{\rm c}$ values of the bulk silicate Earth (BSE), which is the mantle and crust without the core of the Earth, is lower by 4 orders of magnitudes than the solar value. 

The Earth's mass is not large enough to accrete large amounts of gas. If all carbon carriers are in volatile forms such as CO, CO$_2$, or CH$_4$, it is reasonable that carbon components were not in the Earth's building blocks (pebbles and planetesimals) near the Earth's orbit, resulting in the significant carbon depletion in the Earth. 
However, about half of the carbon carriers in molecular clouds and primordial comets are likely to be in refractory forms.
In the ISM, radio spectrographic observations suggest that 
 % \red{$\sim 60\%$ of the refractory carbonaceous dust (carbon particles) are the highly refractory materials such as graphite or amorphous hydrocarbon with the sublimation temperature $T_{\rm sub} \ga 1000\, \rm K$ \citep{Savage1996}} 
$\sim 60\% $ of the cosmic carbon exists as highly refractory materials including six-membered rings such as amorphous hydrocarbons, polycyclic aromatic hydrocarbons (PAHs), or graphites \citep{Savage1996}.
They would survive until they arrive at the hot regions with $T\gtrsim 1000$ K where they are destroyed by pyrolysis or oxidation by OH \citep[e.g.,][]{Finocchi1997,Gail2017}.
On the other hand, \citet{Fomenkova1997, Fomenkova1999} argued that about 60\% of all solid carbonaceous materials (including less refractory carbon solids) 
in the comet Halley are ``complex organics" such as
%, which is 
a mixture of 
% complex organics  
kerogen and insolvable organic matter (IOM) that
survive until they are pyrolyzed at $T\sim 500$K.
These organics are often called ``refractory organics."
However, because they are much less refractory than amorphous hydrocarbons, graphites, and PAHs, here we refer to them simply as ``complex organics" to avoid confusion.
We summarize the disk mid-plane temperature required for the destruction of each of the refractory carbon materials
 % solid carbonaceous component 
 ($T_{\rm des}$) in Table~\ref{tab:carbon_temp}, which shows that most of the refractory carbon components should be in dust forms around Earth's orbit.
 If Earth's building blocks originated from the same carbon carriers as in molecular clouds, their compositions should be similar to those of ISM dust and comets with the removal of volatile carbon gas that would have been sublimated inside their ice lines. The carbon reduction due to the sublimation of volatile carbon gas should only be on the level of a factor of a few (and certainly not four orders of magnitude). 
 
Assuming that carbon in the Earth's mantle or crust may have been partitioned to its core in the magma ocean stage \citep{Sakuraba2021}, $f_{\rm c}$ values of the bulk Earth could be higher by 1 - 2 orders of magnitudes than that of BSE. However, even in that case, the Earth is still depleted in carbon by more than two orders of magnitude from the solar value. 
%Chondrite meteorites are also highly depleted in carbon. 
Remarkably, even carbonaceous chondrites (CC) are depleted in carbon by more than one order of magnitude from the Sun.
%Therefore, 

The ``bulk" values (rather than matrix-normalized values) of $f_{\rm c}$ for enstatite chondrites (ECs) and ordinary chondrites (OCs) are in between ``bulk Earth" and CC,  monotonically increasing with the orbital radii of the Earth and the inferred parent asteroids \citep[e.g.,][]{Bergin2015,Binkert2023}. 
However, ECs and OCs include substantial chondrules that are carbon-free.
The carbons there might have been lost during the flash heating \citep{Gail2017}.
It would be more appropriate to normalize the carbon fraction by the matrix volumes \citep{Alexander2007}, as in Fig. ~\ref{fig:obs}; 
the normalized $f_{\rm c}$ in ECs, OCs, and CC would be similar independent of the heliocentric radius. 
% If the upper limit is taken for the bulk Earth, $f_{\rm c}$ could be similar in the inner solar system and the $f_{\rm c}$ distribution could be a step-function form with a jump near the snow line.    

 \begin{table}[]
     \centering
 \begin{threeparttable}[]
     \caption{Destruction temperature, $T_{\rm des}$, for each solid carbonaceous component.}
     \begin{tabular}{llc}
     \toprule\toprule
         Solid carbonaceous component & Process & $T_{\rm des}$ [K] \\ \midrule
        Amorphous hydrocarbons & Oxidation & $\gtrsim 1100$\tablefootmark{a}\\
        Graphites & Oxidation & $\gtrsim 1100$\tablefootmark{a}\\
        PAHs & Oxidation & $> 1000$\tablefootmark{b}\\
        Volatile organics & Pyrolysis & $\sim 300 - 400$\tablefootmark{a}\\
        Complex organics & Pyrolysis & $\sim 500$\tablefootmark{c}\\ \bottomrule
     \end{tabular}
     % \begin{tablenotes}
     %     \item[a] \citet{Gail2017} \item[b] \citet{Kress2010} \item[c] \citet{Li2021}
     % \end{tablenotes}
     \tablefoot{“Complex organics" refer to mixed organics such as kerogen and insolvable organic matters (IOM), which are often called ``refractory organics." Because they are moderately refractory in this table, we here use ``complex organics" to avoid confusion.\\
     \tablefoottext{a}{\citet{Gail2017}}\tablefoottext{b}{\citet{Kress2010}}\tablefoottext{c}{\citet{Li2021}}}
     \label{tab:carbon_temp}
 \end{threeparttable}
 \end{table}
 % Since terrestrial planets cannot accrete the gas, they can take in the carbonaceous volatile molecules that evaporate near the earth's orbit, such as carbon monoxide. However, refractory carbon particles, such as amorphous hydrocarbon and complex organics, should be taken in pebbles and planetesimals as dust near the earth's orbit. Thus, if the components of rocky bodies were the same as the ISM or comets, the carbon fraction of the earth should be lower only by about two factors of magnitude than the sun. It is a big mystery that carbon fractions of the earth and other rocky bodies are lower by two orders of magnitude than the sun.

These observational data strongly suggest that carbons in the refractory forms should have been destroyed by some mechanism in the inner Solar System. 
The question of what specific mechanism is at work remains a mystery. 

Carbon depletion may have also occurred in many exoplanetary systems.
About 25 to 50\% of white dwarfs (WDs) are polluted by heavy elements (metals) such as Fe, Si, O, and C in their photospheres \citep[e.g.,][]{Zuckerman2003,Zuckerman2010,Koester2014,Hollands2017}; however, these metals should quickly turn to sediment to the WDs' interiors in the WD's strong gravitational fields \citep{Paquette1986,Koester2010}.
This suggests that rocky planets or asteroids fall onto the host WDs through accretion disks \citep[e.g.,][and references therein]{Okuya2023},
implying that the observed metal abundance should reflect the bulk compositions of the rocky bodies and it shows that C/Si is distributed from the solar value down to the BSE value \citep{Zuckerman_Young2017}. 
The carbon depletion problem is not only an issue in the Solar System but also in exoplanetary systems.

Oxidation destroys amorphous hydrocarbons, PAHs, and graphites. 
%Oxidation is considered to occur
When an oxygen atom collides with dust, a carbon atom is stripped to form a gaseous CO molecule. \citet[]{Lee2010} showed that enough amounts of hot oxygen atoms to oxidize graphite are produced by far-ultraviolet (FUV) irradiation in an upper layer over the disk photo-surface.
The oxidation timescale of graphite is also applied for amorphous hydrocarbons and PAHs.
% \ida{What is ''amorphous hydrocarbons”? Is it different from amorphous hydrocarbon? Oxidation does not occur for graphite? What about complex organics?}. 
Amorphous hydrocarbons may also be directly destroyed by FUV photons \citep{Anderson2017}; small hydrocarbons are released from the surface of an amorphous hydrocarbon particle that contains $10^6$ or more C atoms \citep{Alata2014,Alata2015}.
In this paper, we refer to this destruction as ``photolysis.'' 

The complex organics are thermally destroyed at $T_{\rm des} \sim 500 \, \rm K$ (Table~\ref{tab:carbon_temp}),
%\citep{Li2021}
which we call ``pyrolysis." 
Although they are converted to gas molecules, half of pyrolysis products may be hydrocarbons with $T_{\rm des} \gtrsim 1000\, \rm K$ \citep[e.g.,][]{Chyba1990,Gail2017}. The experiments of \citet{Nakano2003} showed that remnants of pyrolysis can exist even in regions where the temperature exceeds 700 K, although the mass fraction of the remnants is unclear.
\citet{Bergin2015} and \citet{Li2021} assumed only complex organics as a source of refractory carbons in ISM to propose that migration of their sublimation line (``soot line") solves the carbon depletion problem in the inner Solar System.
However
if the remnants of the pyrolysis are hydrocarbons, the question of how hydrocarbons are destroyed must be discussed, even if the ISM refractory carbons do not include the highly refractory materials: hydrocarbons, PAHs, or graphites.

\citet{Klarmann2018} and \citet{Binkert2023} studied the carbon depletion in the inner Solar System, through Eulerian diffusion and advection simulations with the two-components fluid approximation of disk gas and dust.
% These previous studies did not calculate the dust dynamics in a disk and some studies indicated that dust transportation hinders carbon depletion in the inner solar system. 
\citet[]{Klarmann2018} calculated the destruction of hydrocarbon particles due to oxidization and FUV photolysis, taking into account the dust drift and diffusion due to disk gas accretion and turbulence.
They concluded that it is difficult to explain the observed carbon depletion because the dust radial drift timescale is much shorter than the timescale required to destroy most of the refractory carbon particles there.

\citet[]{Binkert2023} considered both the photolysis of amorphous hydrocarbons and pyrolysis of complex organics. They suggested that the carbon depletion could be explained if the FU Ori-type outburst sublimates all of the complex organics and if the dust surface density and the optical depth for FUV are low enough to promote photolysis. However, it is not clear that the FU Ori-type outburst events occurred in the proto-Solar System and enough amount of the irradiated dust particles stay to be building blocks of the inner Solar System bodies.
Furthermore, they assumed most of the refractory carbons are complex organics.
% were converted only to gaseous molecules by pyrolysis. 
If their conversion to amorphous hydrocarbons is taken into account, their conclusion may change. 
% \citep[e.g.,][]{Chyba1990,Gail2017}.
%Moreover, the low accretion rate for dust caused carbon destruction in the comets-forming region. Therefore, it is difficult to explain the carbon abundance for all bodies in the solar system.

These previous studies assumed that dust is composed only of silicates and carbon materials. However, beyond the snow line, H$_2$O components are added. % Although they assumed the fragmentation velocity for dust is that of only silicate dust, 
Conventional results from experiments and numerical simulations suggested that ice may be stickier than silicate \citep[e.g.,][]{Blum2000,Zsom2011,Wada2011}, and icy particles grow to larger sizes than silicate particles.
Here, we assume that carbonaceous solid materials move with silicate particles.
%\red{The larger} icy particles drift faster than the silicate particles, resulting in rapid decay of the icy pebble flux and the pile-up of silicate particles inside the snow line. 
% the icy pebbles grow up enough to drift inward rapidly and 
If the icy pebbles envelop many small silicate particles, the silicate particles are released by sublimation of the icy mantle. Because of their small size, they are coupled to gas and piled up like traffic jam near the snow line, which may result in rocky planetesimal formation \citep[e.g.,][]{Saito2011,Ida&Guillot2016,Ida2021} and an enhancement of the crystalline to amorphous ratio of silicates even beyond the snow line \citep{Okamoto2022}.

These previous studies did not consider the effect of the vertical temperature profile on the particle motions, because they estimated the vertical diffusion timescale by using the mid-plane temperature of the disk. However, in the optically thin upper layer, the local gas temperature is higher than in the mid-plane. It enhances turbulent motions there to lift particles more. At the same time, the lower gas density weakens the gas drag coupling, which suppresses particle stirring.
Because oxidation and photolysis are regulated by how much the particles are stirred up to the upper layer,
it is important to include the above effects. Our Lagrangian particle tracking method can more easily incorporate these effects than the Eulerian method adopted by the previous studies.

In this paper, we investigate the effect of the different sizes between icy and silicate particles and that of the vertical temperature profile on the carbon depletion problem.
We adopt a global 3D Monte Carlo simulation \citep{Ciesla2010,Okamoto2022} to calculate the radial drift and turbulent diffusion motions of solid particles in a disk and the chemical reactions \citep{Ishizaki2023} such as photolysis in the upper FUV-exposed layer by tracking motions of individual particles. %We assumed the initial carbon components as all amorphous hydrocarbon, all complex organics, and amorphous hydrocarbon : 
Because our model also calculates the particle size evolution due to coagulation and fragmentation,
the pebble flux decay is consistently calculated.

% In this paper, we aim to investigate the effect of the different fragmentation velocities between icy and silicate particles on the carbon fraction of the inner Solar system bodies.%On the other hand, we consider the difference in stickiness and perform a 3D Monte Carlo simulation to evaluate the carbon distribution in a steady accretion disk. 
% We perform a global 3D Monte Carlo simulation to calculate the motion of solid particles in a disk and estimate the carbon destruction effect. We calculate the FUV-exposed time for each particle by tracking the vertical particle motion. %We assumed the initial carbon components as all amorphous hydrocarbon, all complex organics, and amorphous hydrocarbon : 

In Sect. \ref{sec:method}, we describe our Monte Carlo methods for particle motions 
%advection 
%and diffusion of dust particles in the steady accretion gas disk 
and refractory carbon destruction models. 
%by photolysis (Section~\ref{subsec:photo}) and oxidation (Section~\ref{subsec:oxi}) that we adopt. 
In Sect.~\ref{sec:resu}, we show the simulation results. The carbon fraction drops off inside the snow line by two or more orders of magnitude from the solar value in a fiducial case.
We also investigate the dependence on simulation parameters.
% due to the fast decay of icy pebble flux and silicate particles piling up inside the snow line. We also show the vertical temperature of the disk can contribute to the lower carbon fraction.
In Sect.~\ref{sec:dis}, we %discuss other carbon destruction effects and 
compare the simulation results to the observational data. 
Section~\ref{sec:con} presents our conclusions.
%, we summarize the is paper.

\section{Method} \label{sec:method}
\subsection{Disk model}
\label{subsec:disk}
We assumed 
a self-similar protoplanetary gas disk \citep{Hartmann98} with characteristic radius $r_{\rm disk}=100\,\rm au$. 
At $r \ll r_{\rm disk}$, the disk gas accretion rate ($\dot{M}_{\rm g}$) is independent of $r$ (steady accretion) and $\dot{M}_{\rm g}$ decays as $\propto (1 + t/t_{\rm disk})^{-3/2}$, where $t_{\rm disk} \simeq 10^7 (\alpha/10^{-3})^{-1}(r_{\rm disk}/100\,\rm au) \rm yrs$. %\ida{$\leftarrow$ OK?}. 
We assumed the conventional gas accretion model due to turbulent diffusion and
do not consider disk-wind driven accretion, for simplicity.  
The gas surface density is given by
\begin{equation}
    \begin{split}
        \Sigma_{\rm g}&=\frac{\dot{M}_{\rm g}\mu m_{\rm H}\Omega_{\rm K}}{3\pi \alpha k_{\rm B}T_{\rm mid}}\\ &\simeq 1200 \left(\frac{\dot{M}_{\rm g}}{10^{-8}M_{\odot}/{\rm yr}}\right)\left(\frac{\alpha}{10^{-3}}\right)^{-1}\left(\frac{T_{\rm mid}}{300\, {\rm K}}\right)^{-1}\left(\frac{r}{1\,{\rm au}}\right)^{-3/2}{\rm g/cm^2},  
    \end{split}
    \label{eq:Sigma_T}
\end{equation}
where $\alpha$ is a parameter of turbulent strength and $\Omega_{\rm K}$ is Keplerian frequency. 
% In our calculation, \Oka{we decay the gas and dust disk at the disk radius $r_{\rm disk}$ as a self-similar time decaying disk,} and we set $r_{\rm disk}=100$ au. We show the results in case gas and dust disk decay by exponential low at $r_{\rm disk}$ in the appendix \ref{app:exp}.

The hydrostatic %vertical structure of the gas disk is calculated assuming hydrostatic equilibrium and the the 
local gas density is given by
\begin{equation}
    \rho_{\rm g}=\frac{\Sigma_{\rm g}}{\sqrt{2\pi}H_{\rm g}}\exp\left(-\frac{z^2}{2H_{\rm g}^2}\right),
    \label{eq:H_T}
\end{equation}
where $H_{\rm g}$ is the disk gas scale height. The disk aspect ratio is
% of the gas disk $h_{\rm g}$ is given by
\begin{equation}
    h_{\rm g}=\frac{H_{\rm g}}{r}=0.034\left(\frac{T}{300\,{\rm K}}\right)^{1/2}\left(\frac{r}{1\,{\rm au}}\right)^{1/2}.
\end{equation}

\begin{figure}
    \begin{tabular}{c}
        \begin{minipage}[t]{\hsize}
            \centering
            \includegraphics[keepaspectratio, scale=0.55]{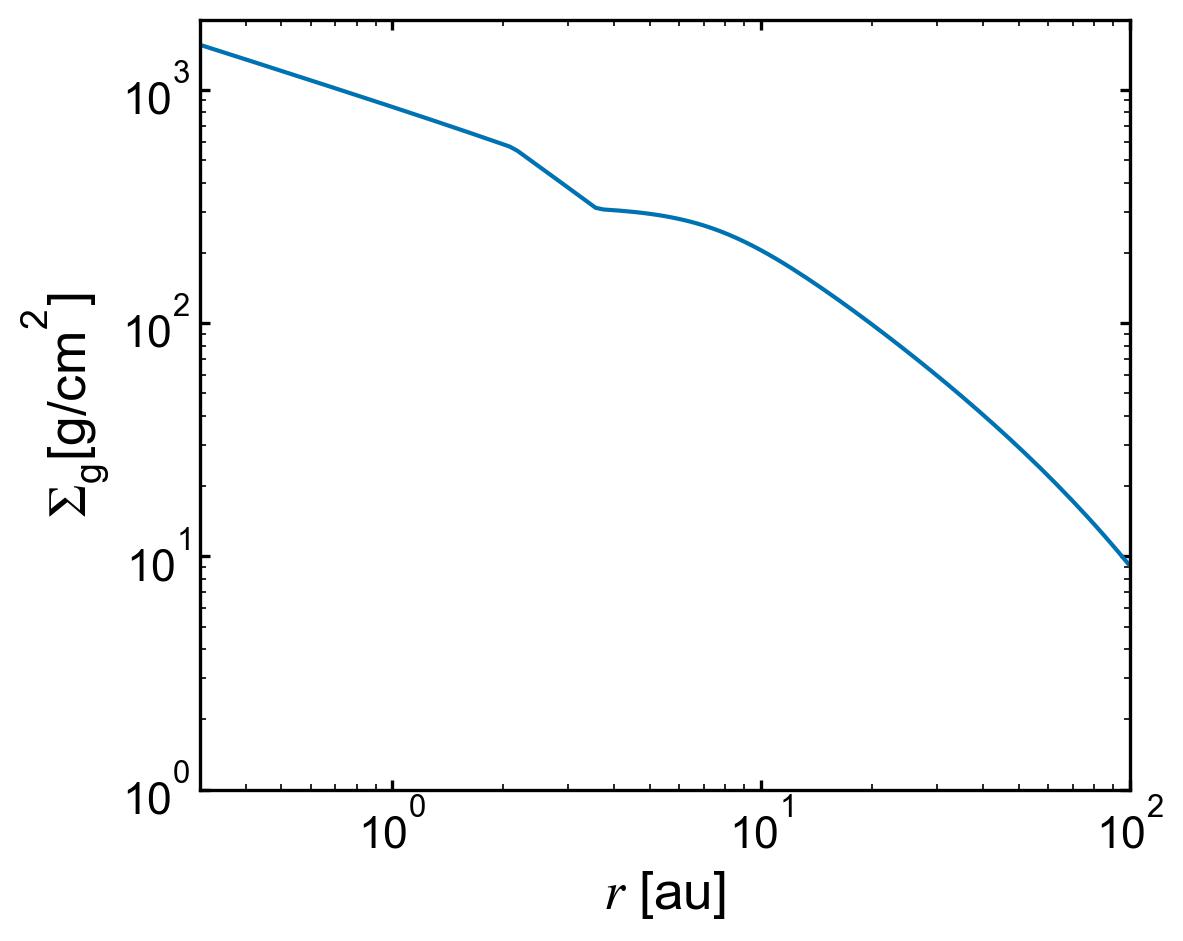}
        \end{minipage}\\
        \begin{minipage}[t]{\hsize}
            \centering
            \includegraphics[keepaspectratio, scale=0.55]{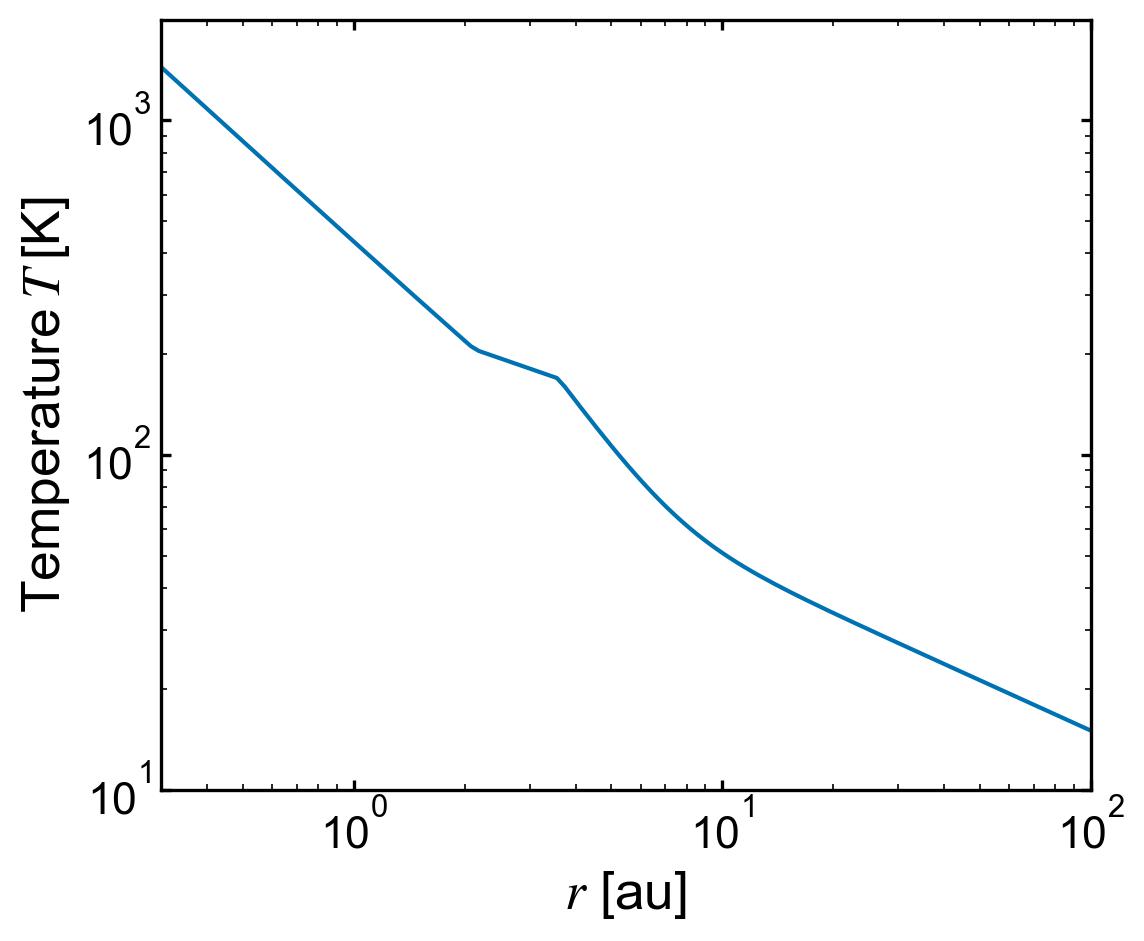}
        \end{minipage}
    \end{tabular}
    \caption{Gas surface density, $\Sigma_{\rm g}$, (upper panel) and the mid-plane temperature (lower panel) in the fiducial run.}
    %\ida{Add $\Sigma_{\rm g}$ and $T$ in the $y$ axis; g/cm$^2$ must be RITTAI; $r$ in the horizontal axis must be italic.}
    \label{fig:setting}
\end{figure}

We set the gas disk temperature distribution as follows. 
We calculate the mid-plane temperature $T_{\rm mid}$ by $T_{\rm mid}^4=T_{\rm vis}^4+T_{\rm irr}^4$,
where $T_{\rm vis}$ and $T_{\rm irr}$ are the temperature determined by viscous heating and irradiation, respectively.
%considering irradiation and viscous heating. We calculate 
The viscous heating temperature with the opacity given by \citet[]{Bell1994}: 
$T_{\rm vis}=\min(T_{\rm vis,ice},\max(T_{\rm vis,eva},T_{\rm vis,sil}))$, where $T_{\rm vis,ice}$ and $T_{\rm vis,sil}$ are the temperatures with ice and silicate dust opacity, respectively, and  $T_{\rm vis,eva}$ is that with the opacity for a transition from ice to silicate due to the evaporation of ice, which are given by 
%\ida{You need a reference to this temperature model, if any.}
\begin{align}
    T_{\rm vis,ice} & =1141\left(\frac{\dot{M}_{\rm g}}{10^{-8}\,M_{\odot}/{\rm yr}}\right)^{2/3}\left(\frac{\alpha}{10^{-3}}\right)^{-1/3}\left(\frac{r}{1\,{\rm au}}\right)^{-3/2}{\rm K}.
    \label{eq:T_vis,ice}\\
    T_{\rm vis, eva} & =270\left(\frac{\dot{M}_{\rm g}}{10^{-8}\,M_{\odot}/{\rm yr}}\right)^{1/6}\left(\frac{\alpha}{10^{-3}}\right)^{-1/12}\left(\frac{r}{1\,{\rm au}}\right)^{-3/8}{\rm K}, \\
    T_{\rm vis,sil} & =433\left(\frac{\dot{M}_{\rm g}}{10^{-8}\,M_{\odot}/{\rm yr}}\right)^{4/9}\left(\frac{\alpha}{10^{-3}}\right)^{-2/9}\left(\frac{r}{1\,{\rm au}}\right)^{-1}{\rm K}.
    \label{eq:Tvissil}
\end{align}
On the other hand, the temperature by irradiation is set to \citep{Oka2011,Ida2016}
\begin{equation}
\label{eq:Tirr}
    T_{\rm irr}=150\left(\frac{r}{1\,{\rm au}}\right)^{-3/7}\, \rm K.
\end{equation}
The transition between $T_{\rm vis}$ and $T_{\rm irr}$ occurs at
\begin{equation}
\label{eq:r_vis-irr}
    r_{\rm vis-irr}=6.6 \left(\frac{\dot{M}_{\rm g}}{10^{-8}\,M_{\odot}/{\rm yr}}\right)^{28/45}\left(\frac{\alpha}{10^{-3}}\right)^{-14/45}\,{\rm au}.
\end{equation}
Figure~\ref{fig:setting} shows the gas surface density and the mid-plane temperature distributions in the fiducial run.

\begin{figure}
    \centering
    \includegraphics[keepaspectratio,scale=0.4]{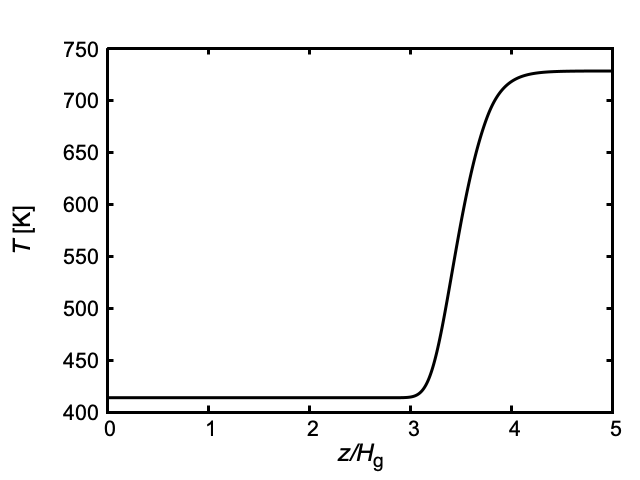}
    \caption{Vertical temperature profile at 1 au given by Eq.~(\ref{eq:T_z}).}
    \label{fig:T_z}
\end{figure}

The vertical temperature distribution is set as follows. 
In the upper optically thin FUV-exposed layer (at $|z| \ga (3$-$4)\, H_{\rm g}$),
the temperature is generally higher than near the mid-plane, which is given by \citep{Klarmann2018} 
\begin{equation}
    T_{\rm FUV}=750\left(\frac{r}{1\,{\rm au}}\right)^{-3/5}\,{\rm K}.
\end{equation}
We connect $T_{\rm FUV}$ with $T_{\rm mid}$ in the mid-plane as
\begin{equation}
    \label{eq:T_z}
    T_z^4=T_{\rm mid}^4\left(1-\exp(-\tau_{{\rm FUV},z})\right)+T_{\rm FUV}^4\exp(-\tau_{{\rm FUV},z}),
\end{equation}
where $\tau_{{\rm FUV},z}$ is the vertical optical depth for FUV given by Eq.~(\ref{eq:tau_z}). 
Figure~\ref{fig:T_z} shows $T_z$ at 1 au.

\subsection{Monte Carlo simulation}
\label{subsec:MC}

%\subsubsection{Overview}
%\label{subsubsec:MC_overview}
\subsubsection{Representative refractory carbons}
\label{subsubsec:rep_carbon}
We considered the two types of refractory carbons: amorphous hydrocarbon as the highly refractory materials and complex organics as modest refractory materials, although, in many simulation sets, we considered only amorphous hydrocarbons by the reason in Sect. \ref{sec:intro}. 
Because graphites and PAHs would have similar destruction temperatures (Table~\ref{tab:carbon_temp}) and, in an observational sense, it is not clear what fraction of interstellar carbon exists in individual forms of refractory materials, the highly refractory carbons are represented by the amorphous hydrocarbons (in this paper). 
% We will test the initial conditions that all the refractory carbons are in the form of with $T_{\rm des}\ga 1100\,\rm K$.
%, because in the point of a refractory degree as described in Tab.~\ref{tab:carbon_temp}, they may be represented by the amorphous hydrocarbons. 
We assumed that amorphous hydrocarbon particles are destructed by the photolysis and/or oxidation in the FUV-exposed layer. 
%The timescales of photolysis and oxidation are discussed in Section~\ref{subsec:photo} and~\ref{subsec:oxi}.

\subsubsection{Photolysis and oxidation}
\label{subsubsec:photo_oxy}

% \ida{I changed the order of 2.2.2 and 2.2.3}

When the carbonaceous particles are lifted to the upper layer by turbulence, the particles are assumed to be ``photodegraded" or ``oxidized." In this paper, ``photolysis" is defined as the reaction that small hydrocarbon volatile molecules such as methane are released from the surface of the amorphous hydrocarbon particles by FUV, as suggested by \cite{Alata2015}, and ``oxidation" means the reaction that single C atoms are removed from the carbon particle surface by the collisions with oxygen atoms as gaseous carbon monoxide molecules. Since the oxygen atoms should be abundant only in the upper layer of the disk \citep[e.g.,][]{Lee2010}, we calculated the oxidation rate only when the particle exists in the FUV-exposed layer (see Sect.~\ref{subsec:opa}). Although oxidation by OH could occur \citep[e.g.,][]{Bauer1997, Finocchi1997}, the abundance of OH is uncertain near the mid-plane and the oxygen atoms are more abundant than OH in the FUV-exposed layer \citep[][]{Lee2010}. 
Accordingly, we did not consider oxidation by OH.
The rates of photolysis and oxidation are described in Sects.~\ref{subsec:photo} and \ref{subsec:oxi}.

When the Stokes number (St), which is defined by the stopping time due to gas drag scaled by the inverse of local Kepler frequency, is smaller than the viscosity parameter $\alpha$ for the released and piled-up silicate particles at the snow line (and when ${\rm St} > \alpha$ for icy pebbles), the silicate particles are stirred up more highly than the icy pebbles.
In this case, the piled-up silicate particles
block FUV radiation from the central star to the region beyond the snow line, which is also known as the ``shadow area'' \citep[e.g.,][]{Ueda2019,Ohno2021}. Therefore, we assumed that amorphous hydrocarbons are not destroyed outside the snow line.
% Although the piled-up particles also block stellar radiation of longer wavelengths and the temperature in the shadow area becomes lower, this structure could not change our results significantly, because icy pebbles 
%also grow up and 
% drift fast near the midplane with small scale height \ida{Is this what you wanted to write here?}.  
%inward fast in the shadow area. 

% We summarize the assumptions of these carbon destruction reactions in Table~\ref{tab:reac}. 
%  \ida{I don't think this table is helpful. Because ``Carbon - C,'' and ``losing a carbon atom" are not clear, this table is rather confusing.}

\begin{figure*}
    \centering
    \includegraphics[keepaspectratio,scale=0.25]{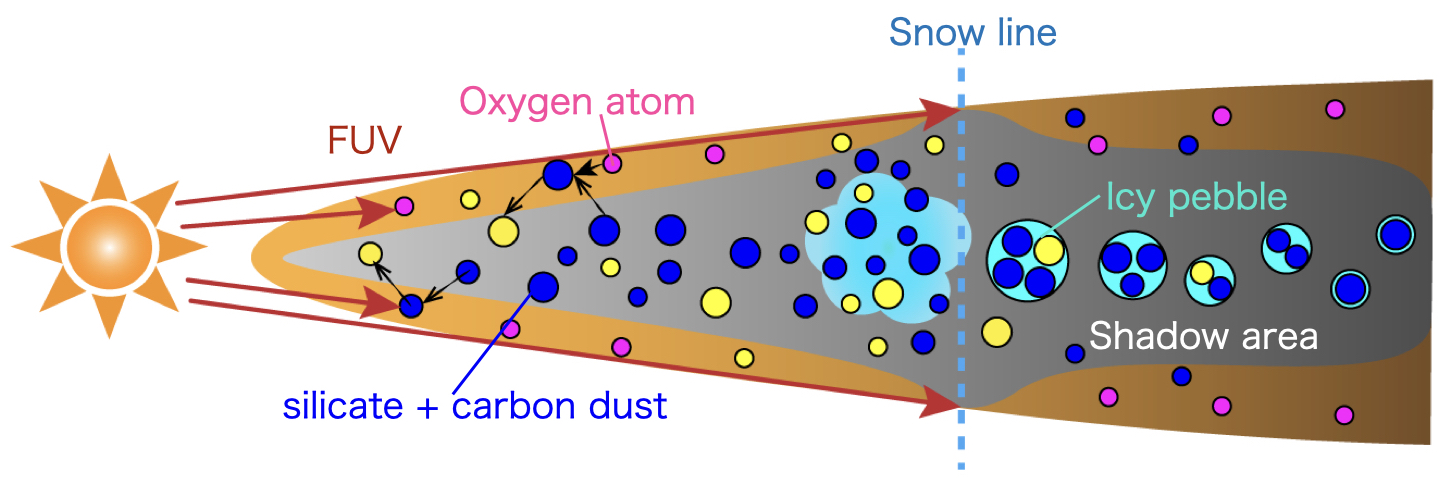}
    \caption{Overview of our models for gas disk and particle evolution in 
    different disk regions.
    The blue and yellow particles represent silicates with and without refractory carbon. The light blue particles represent icy pebbles. The pink particles show the oxygen atoms that are formed by the destruction of oxygen molecules because of the FUV radiation which is described as red arrows. 
The light blue dashed line is the snow line. 
Icy pebbles enclosing many small silicates drift to the snow line (blue dashed line) from the outer region. Orange upper and lower layers are FUV-exposed regions, where oxygen atoms exist.}
\label{fig:model}
\end{figure*}
\renewcommand{\arraystretch}{1.3}
% \begin{table}[]
% \centering
%     \caption{Assumptions of the carbon destruction reactions in this paper.}
%     \begin{tabular}{l||c|c|c}\hline\hline
%          & Oxidation & Photolysis & Pyrolysis\\ \hline \hline
%         Reactants & \multicolumn{2}{c|}{Amorphous hydrocarbon and complex organics}&Only complex organics\\ \hline
%         Products & Carbon - C + CO & Carbon - C + hydrocarbon gas &  Amorphous hydrocarbon + CO + ${\rm CO_2}$ + ${\rm CH_4}$ \\ \hline
%         Region & FUV-exposed layer ($\tau_r\le 1$) & FUV-exposed layer & Hot disk region \\ \hline\hline
%     \end{tabular}
% \label{tab:reac} 
%     \\ \vspace{1mm} Note: ``Carbon - C'' means the amorphous hydrocarbon particles and complex organics losing a carbon atom.
% \end{table}
% {
% % \tabcolsep = 1pt
% \begin{table}[]
% \centering
%     \caption{Assumptions of the carbon destruction reactions in this paper.}
%     \begin{tabular}{l||c|c}\hline\hline
%          & Oxidation & Photolysis\\ \hline \hline
% %        Released gas & CO & hydrocarbon gas \\ \hline
%         Region & FUV-exposed layer ($\tau_r\le 1$) & FUV-exposed layer \\ \hline\hline
%     \end{tabular}
% \label{tab:reac} 
%     % \\ \vspace{1mm} Note: ``Carbon - C'' \ida{What shortened form is ``Carbon - C''?} means the amorphous hydrocarbon particles and complex organics losing a carbon atom \ida{I don't understand what ``losing a carbon atom" mean (maybe, English issue).}. 
%     % \ida{I don't think this table is helpful. Because ``Carbon - C,'' and ``losing a carbon atom," 
%     % This table is rather confusing.}
% \end{table}
% }

\subsubsection{Fragmentation limit velocity}
\label{subsubsec:v_frag}

In the fiducial case, we adopted the conventional
%We set the 
fragmentation limit velocity for ice particles as $\v_{\rm frag,ice} = 10{\rm m/s}$ \citep[e.g.,][]{Wada2013,Gundlach2015} and  for silicate particles as $\v_{\rm frag,sil}= 1{\rm m/s}$ \citep[e.g.,][]{Wada2013}.
We note that this conventional notion is now challenged by new experiments and recent observations of the protoplanetary disks (see below). 
%in the fiducial case. 
In this fiducial case, silicate particles cannot grow enough and stay small. We assumed that amorphous hydrocarbons are included in the silicate particles. 
Outside the snow line, water vapor condenses on the small silicate particles and these ice-enveloped particles would stick together according to the larger fragmentation velocity for ice to form relatively large aggregates, known as ''icy pebbles." 
When the particles pass inward the snow line, the icy mantle evaporates, and the small silicate particles inside the icy pebbles are released. 
As we will show later, the released silicate particles are piled up because they are coupled more strongly to disk gas than the icy pebbles are. 
The sizes of the silicate particles and the icy pebbles are calculated automatically by the model considering the fragmentation limit (Sect.~\ref{subsec:size}). 

Realistic fragmentation velocities of the icy and silicate particles remain uncertain. As mentioned above, although the previous studies showed that icy particles can stick easier than silicates, recent experiments showed that the stickiness of icy particles is similar to that of silicates \citep[e.g.,][]{Musiolik2019,Schrapler2022}. 
Furthermore, recent observations could also show that the size of mass-dominant icy particles is $\la 1 \, \rm mm$ in outer regions of protoplanetary disks, which corresponds to the fragmentation velocity of icy particles of $\la 1 \, \rm m/s$
\citep{Kataoka2015,Ueda2024}.
However, even if the fragmentation velocities of silicate and icy particles are similar to each other, the icy particles could become larger by recondensation of water vapor near the snow line \citep{Ros2019}, which is not incorporated in this study. In this case, the results can be similar to the fiducial case. 
A more detailed investigation is left for future work.
% \citep{Okamoto2022}.}

We set the fragmentation limit velocities for amorphous hydrocarbons to be the same as silicate particles to highlight the effect of the fragmentation velocity difference, for simplicity. We note that the stickiness of carbonaceous materials is also unclear. Although some previous studies suggested complex organics on the particle's surface make the silicate particle stickier \citep[e.g.,][]{Kudo2002,Homma2019}, a recent experiment showed that their stickiness is similar to dry silica dust \citep{Bischoff2020}. %Some silicate particles are lifted to the FUV-exposed layer and amorphous hydrocarbon inside them is destructed by photolysis or oxidation.

\renewcommand{\arraystretch}{1}
\subsubsection{Orbital evolution of particles}

In order to follow the orbital evolution of silicate particles described in
Sect. \ref{subsubsec:v_frag}, 
we employed the 3D Monte Carlo simulation method for 
radial drift
%advection 
and diffusion of silicate particles developed by \cite{Ciesla2010,Ciesla2011}.
%\cite{Ciesla2010,Ciesla2011} showed that the particles' trajectories are described by the following 3D Monte Carlo simulations when t
The surface density evolution of the icy and silicate particles is determined by the concentration equation: 
\begin{equation}    \label{conti}
\dfrac{\partial \Sigma_{\textrm{d}}}{\partial t}=\dfrac{1}{r}\dfrac{\partial }{\partial r}\left[r\Sigma_{\textrm{d}}\left(\v_{r}-\frac{D}{\Sigma_{\textrm{d}}/\Sigma_{\textrm{g}}}\frac{\partial}{\partial r}\left(\frac{\Sigma_{\textrm{d}}}{\Sigma_{\textrm{g}}}\right)\right)  \right],
\end{equation} 
where $\Sigma_{\textrm{d}}$ is the surface density of the icy and silicate particles, $D$ is their diffusivity, and $\v_r$ is their radial drift velocity.
The vertical advection-diffusion equation is given by
\begin{equation}
    \frac{\partial\rho_{\rm d}}{\partial t}=\frac{\partial}{\partial z}\left(\rho_{\rm g}D\frac{\partial}{\partial z}\left(\frac{\rho_{\rm d}}{\rho_{\rm g}}\right)\right)-\frac{\partial}{\partial z}(\rho_{\rm d}\v_{z}),
\end{equation}
where $\rho_{\rm d}$ is the local spatial density
%densties 
of solid particles.
In a steady accretion disk, the changes in the positions in Cartesian coordinates ($x, y, z$) of each particle after the timestep $\delta t$ are given by
\begin{align}
    \delta x & = \v_{r}\times\frac{x}{r} \delta t +\mathcal{R}_x\sqrt{6D(x')\delta t}, \label{eq:diffx}\\
    \delta y & = \v_{r}\times\frac{y}{r} \delta t +\mathcal{R}_y\sqrt{6D(y')\delta t}, \label{eq:diffy}\\
    \delta z & = \left[\v_{z}+\frac{1}{\rho_{\textrm{g}}}\frac{\partial(D\rho_{\textrm{g}})}{\partial z}\right] \delta t +\mathcal{R}_z\sqrt{6D(z')\delta t}, \label{eq:diffz}
\end{align}
where 
$\delta t$ is given by the inverse of the local Keplerian frequency,
$\delta t = \Omega_{\rm K}^{-1}$, and 
$x', y'$, and $z'$ are given by
\begin{align}
    x'=x+\frac{1}{2}\frac{\partial D}{\partial x}\delta t \: ; \: \:
    y'=y+\frac{1}{2}\frac{\partial D}{\partial y}\delta t \: ; \: \:
    z'=z+\frac{1}{2}\frac{\partial D}{\partial z}\delta t.
    \label{eq:dD}
\end{align} 
The first and second terms on the right-hand sides represent
the advection and diffusion.
In the diffusion terms, \(\mathcal{R}_x\), \(\mathcal{R}_y\), and \(\mathcal{R}_z\) are independent random numbers in a range of [-1,1]
with a root mean square of $1/\sqrt{3}$. We set the particles' diffusivity considering collective effect as \citep{Hyodo2021,Okuya2023}
\begin{equation}
    D=\frac{\Lambda}{1+\Lambda^2{\rm St^2}}\nu
,\end{equation}
where $\nu$ is the gas turbulent viscosity given by
\begin{align}
\nu = \alpha h_{\rm g}^2 r^2 \Omega_{\rm K}, 
\end{align}
and ${\rm St}$ is the Stokes number given by
\begin{equation}
\label{eq:St}
    {\rm St}=\left\{
    \begin{array}{ll}
        \displaystyle{ \frac{\rho_{\rm bulk}\,s}{\rho_{\rm g}\v_{\rm th}}\Omega_{\rm K}} & \displaystyle{ \left[s\le \frac{9}{4}\lambda_{\rm mfp}:\;\textrm{Epstein}\right],}\\ 
        & \\
        \displaystyle{ \frac{\rho_{\rm bulk}\, s}{\rho_{\rm g}\v_{\rm th}}\frac{4s}{9\lambda_{\rm mfp}}\Omega_{\rm K}} & \displaystyle{\left[s\ge \frac{9}{4}\lambda_{\rm mfp}: \; \textrm{Stokes}\right],}
    \end{array}
    \right.
\end{equation}
where \(\v_{\rm th}=\sqrt{8/\pi}\,c_s\), $c_s$ is the sound speed of gas, \(\lambda_{\rm mfp}\) is the mean free path of gas, $\Lambda=(1+\rho_{\rm d}/\rho_{\rm g})^{-1}$, and $\rho_{\rm bulk}$ is the bulk density of the particle, which is assumed to be $3.0 \, \rm g/cm^{3}$ for the silicate particles and $1.5 \, \rm g/cm^{3}$ for icy pebbles.

Because we have adopted a disk temperature model where $T$ is raised at $|z| \gtrsim {\rm a \: few}\, H_{\rm g}$ (Fig.~\ref{fig:model}), we chose to adopt a vertical diffusivity that is
calculated by local temperature.
% corresponding to the local disk temperature given by Eq.~(\ref{eq:T_z}). 
On the other hand, because most of the particles should be within the gas scale height,
%settled near the mid-plane, 
we set the radial diffusivity calculated  by the mid-plane temperature 
without 
%and does not have the 
a vertical dependency.
% \ida{I don't understand $\rightarrow$} and approximate that $x'\sim x$.
In this case, the radial dependence of diffusivity can be neglected
in Eq.~(\ref{eq:dD}). 
In contrast, $D$ can change significantly for $z\rightarrow z'$ due to the steep increase or decrease of the temperature, as shown in Fig.~\ref{fig:T_z}. Thus, we can assume $D(x')\sim D(x)$ and $D(y')\sim D(y)$.
% \Oka{In this case, $(\partial D/\partial r) \delta t$ is negligible as $\alpha\ll 1$
% \ida{If this is the reason, why you can't do $z'\sim z$ as well?
% I understand $|\partial D/\partial r| \sim h |\partial D/\partial z|$.
% Even considering this, why you can't do $z'\sim z$?}, we approximate that $x'\sim x$ and $y'\sim y$.
% }

The radial drift velocity due to gas drag is given by
\citep[][]{Ida&Guillot2016,2017Schoonenberg}
\begin{equation}   
\label{v_r0}    
\v_{r} \simeq -\frac{\Lambda}{1+\Lambda^2\,{\rm St}^2}
\left(2\Lambda\,{\rm St} \,\eta \, \v_{\textrm{K}}+u_{\nu}\right),
\end{equation}
%$\v_{\textrm{drag},r}=- 2 \tau_{\textrm{s,sil}}\eta \v_{\textrm{K}}+u_{\nu}$,
where $\v_{\rm K}$ is the local Keplerian velocity, \(u_{\nu}\) is the disk gas (inward) accretion velocity given by $u_{\nu} = 3\nu/2r=(3\alpha h^2_{\textrm{g}}/2)\v_{\textrm{K}}$,
and \(\eta\) is the degree of deviation of the gas rotation angular velocity \(\Omega\) from the Keplerian one, given by
\begin{equation}
    \eta\equiv\frac{\Omega_{\textrm{K}}-\Omega}{\Omega_{\textrm{K}}}=-
    \frac{1}{2}\frac{d\ln P}{d\ln r} h_{\textrm{g}}^2 =
    C_{\eta}h_{\textrm{g}}^2.
    \label{eq:mu}
\end{equation}
For simplicity, we take \(C_{\eta}=11/8\) for $T\propto r^{-1/2}$. 
The vertical advection velocity \(\v_{z}\) is given by \citep{Ciesla2010}:
\begin{equation}
    \v_{z}=-\frac{{\rm St}}{1+{\rm St}}\Omega \, z. \label{eq:v_dragz}
\end{equation}
Therefore, Eqs.~(\ref{eq:diffx}) to (\ref{eq:diffz}) can be rewritten as
\begin{align}
    \delta x &=- \frac{\Lambda}{1+\Lambda^2{\rm St}^2}\left[2 \Lambda C_{\eta} {\rm St}+\frac{3}{2}\alpha\right]h^2_{\textrm{g}}x+\mathcal{R}_x\sqrt{\frac{6\alpha\Lambda}{1+\Lambda^2{\rm St}^2}}\,h_{\textrm{g}}r, \\
    \delta y &=- \frac{\Lambda}{1+\Lambda^2{\rm St}^2}\left[2 \Lambda C_{\eta} {\rm St}+\frac{3}{2}\alpha\right]h^2_{\textrm{g}}y+\mathcal{R}_y\sqrt{\frac{6\alpha\Lambda}{1+\Lambda^2{\rm St}^2}}\,h_{\textrm{g}}r, \\
    \delta z &=-\left(\frac{{\rm St}}{1+{\rm St}}+\alpha\right)z +\mathcal{R}_z\sqrt{\frac{6\alpha\Lambda}{1+\Lambda^2{\rm St}^2}}\, h_{\textrm{g}}(z')r.
\end{align}

In our simulation, at every timestep,
we calculated the instantaneous surface and local densities of solids (icy and silicate particles) from the super-particle distribution to evaluate size growth and collective effect. The solid surface density at $r$ is given 
with the number of super-particles, \(\Delta N_{r}\), in the radial width 
of $\Delta r$ around $r$ given by
\begin{equation}
    \Sigma_{\textrm{d}}= \frac{m\,\Delta N_{r}}{2\pi r\Delta r}=\frac{m\,\Delta N_{r}}{2\pi r^2\times 2.3(\Delta\log_{10}r)}.
\label{eq:sigcal}
\end{equation}
where \(m\) is the super-particle mass. The solid local density at $r$ and $z$ is calculated by the number of super-particles \(\Delta N_{r,z}\) in the radial and vertical width of $\Delta r$ and $\Delta z$, given by
\begin{equation}
    \rho_{\textrm{d}}= \frac{m\Delta N_{r,z}}{2\pi r\Delta r\times 2\Delta z}=\frac{m\Delta N_{r,z}}{4\pi r^2\times 2.3(\Delta\log_{10}r)\Delta z}.
\end{equation}

When the initial state of the surface density of particles is given by \(\Sigma_{\rm d}=Z_0\Sigma_{\rm g}\), the initial number of particles in each cell \(\Delta N_{r,0}\) is given by
\begin{align}
    \label{eq:N_par}
    \Delta N_{r,0}&=\frac{Z_0\Sigma_{\rm g}\times 4.6\pi r^2(\Delta\log_{10}r)}{m}\\ &\simeq 81\left(\frac{Z_0}{0.01}\right)\left(\frac{\Delta\log_{10}r}{0.05}\right)\left(\frac{\Sigma_{\rm g}}{1000\,{\rm g/cm^2}}\right)\left(\frac{m}{10^{-8}\,M_{\odot}}\right)^{-1}\left(\frac{r}{1\,{\rm au}}\right)^2.
\end{align}

We summarize the simulation parameters in Table~\ref{tab:fidu}.
% \ida{Please also add the variations of the parameters you studied in this paper to the table.}
{
\tabcolsep = 1pt
\begin{table}[]
    \centering
    \caption{Parameters we use in this paper.}
    \begin{tabular}{lcc}\hline \hline
       Quantity  & Symbol & Value \\ \hline
       Stellar FUV luminosity [$L_{\odot}$] & $L_{\rm FUV}$ & 0.01\\
       Gas accretion rate [$M_{\odot}/$yr] & $\dot{M_{\rm g}}$ & $10^{-9},\,{\bf 10^{-8}}$ \\
       Turbulence strength & $\alpha$ & $5\times 10^{-4},\,{\bf 10^{-3}}$ \\
       Disk radius [au] & $r_{\rm disk}$ & 100 \\
       Initial solid-to-gas ratio & $Z_0$ & {\bf 0.01}, 0.05\\
       Fragmentation velocity for ice [m/s]& $\v_{\rm frag,ice}$ & 1, {\bf 10}\\
        Fragmentation velocity for silicate [m/s]& 
       $\v_{\rm frag,sil}$ & 0.1, {\bf 1}, 10 \\ \hline
    \end{tabular}
    \tablefoot{The bold letters show the parameters of the fiducial run.}
    \label{tab:fidu}
\end{table}
}
\subsubsection{Size evolution of particles}\label{subsec:size}

We used a super-particle approximation. One super-particle represents a huge number of small dust particles. We assigned a single dust size to each super-particle that has the same total mass.
We calculated the particle size growth by the ``two population model'' developed by \citet[]{Birnstiel2012}. The size of the large super-particle is given by
\begin{equation}
    s(t+\delta t)=\min\left[s_{\rm max},\,s(t)\exp(Z\Omega_{\rm K}\delta t )\right],
    \label{eq:size}
\end{equation}
where
\begin{align}
s_{\rm max} = \min(s_{\rm frag},\,s_{\rm drift}),
\label{eq:s_max}
\end{align}
which is the local maximum size at the mid-plane corresponding to the Stokes numbers (Eq.~(\ref{eq:St}))
and $s_{\rm frag}$ and $s_{\rm drift}$ are the limits
by fragmentation and radial drift,
%. The Stokes number limited by fragmentation ${\rm St}_{\rm frag}$ is 
given by
\begin{align}
    {\rm St}_{\rm frag} & =0.37\,\frac{\v_{\rm frag}^2}{3\, \alpha \, c_s^2}
    \simeq 2.3 \times 10^{-2}\left( \frac{\v_{\rm frag}}{10 \, \rm m/s} \right)^2 \left( \frac{\alpha}{10^{-3}}\right)^{-1} \left( \frac{T}{150 \, \rm K} \right)^{-1}
    \label{eq:frag} ,\\
    {\rm St}_{\rm drift} & = 0.40\,\frac{\Sigma_{\rm d}}{\Sigma_{\rm g}C_{\eta}}\left(\frac{r\Omega_{\rm K}}{c_s}\right)^2
    \simeq 0.49 \left( \frac{\Sigma_{\rm d}/\Sigma_{\rm g}}{10^{-3}} \right)
    \left( \frac{r}{1 \, \rm au} \right)^{-1} \left( \frac{T}{150 \, \rm K} \right)^{-1}.
    \label{eq:drift}
\end{align}
%We derive $s_{\rm frag}$ and $s_{\rm drift}$ from ${\rm St}_{\rm frag}$ and ${\rm St}_{\rm drift}$ with Eq.(\ref{eq:St}).
%where $s_{\rm frag}$ is the maximum size \red{limited by} fragmentation, given by
%\begin{equation}
%    s_{\rm frag}=\frac{2f_{\rm f}\Sigma_{\rm g}}{3\pi\rho_{\rm bulk}\alpha} \left(\frac{\v_{\rm frag}}{c_s}\right)^2,
 %   \label{eq:frag}
%\end{equation}
% where $\rho_{\rm bulk}$ is the bulk density of dust and $c_{\rm s}$, and $s_{\rm drift}$ is the maximum size of drift limit given by
% \begin{equation}
%     s_{\rm drift}=\frac{f_{\rm d}\Sigma_{\rm d}}{\pi\rho_{\rm bulk} C_{\eta}}\left(\frac{r\Omega_{\rm K}}{c_s}\right)^2.
% \end{equation}
We set $\rho_{\rm bulk}=3.0\,{\rm g/cm^3}$ inside the snow line and $\rho_{\rm bulk}=1.5\,{\rm g/cm^3}$ outside the snow line. 

When $s$ of a super-particle exceeds $s_{\rm frag}$, 
we assume that a collision occurs with $\v > \v_{\rm frag}$
and randomly chose a new representative particle size by the weight of $dN(s) \propto s^{-3.5}ds$, where $N(s)$ is a cumulative size distribution function, with the maximum and minimum sizes of $s_{\rm frag}$ and 0.1 $\rm \mu$m, conserving the mass of the super-particle. The minimum size is fixed, while $s_{\rm frag}$ is updated at every timestep.

% We assume that if the particle size becomes larger than $s_{\rm frag}$, particles are fragmented in the power-law of $dN(s) \propto s^{-3.5}ds$ assuming that they collide with a higher relative velocity than fragmentation velocity. Conserving the total mass of individual super-particles, we re-assign a representative particle size by a random number generation with a weighting corresponding to $dN(s) \propto s^{-3.5}ds$ and the maximum and minimum sizes of $s_{\rm frag}$ and 0.1 $\rm \mu$m. \Oka{We assume the size of small particles is 0.1 $\rm \mu$m constantly.}

% We assume two types of super-particles: \red{consisting of normal} large ($> 0.1\,\rm \mu m$) particles and \red{special} small ($\sim 0.1\,\rm \mu m$) particles.

\subsection{Evaporation and recondensation of ice}

We consider the evaporation and recondensation of icy particles, following \citet[]{Ciesla2006}. 
The decrease rate of the particle mass ($m_{\rm p}$) is given by a balance between recondensation and evaporation as \citep{Lichtenegger1991}\begin{align}
\frac{dm_{\rm p}}{dt} & 
% = \left(\frac{dm_{\rm p}}{dt}\right)_{\rm evp} - \left(\frac{dm_{\rm p}}{dt}\right)_{\rm cnd} \nonumber\\ & 
 = 4\pi s^2 \varv_{\rm th} (\rho_{\rm eq} - \rho_{\rm vap}) = 
2 s^2 \, \varv_{\rm th}^{-1} %\frac{\rho_{\rm vap}}{P_{\rm vap}} 
( P_{\rm eq} - P_{\rm vap}), 
\label{eq:subl0}
\end{align}
where $\rho_{\rm vap} = \Sigma_{\rm vap}/\sqrt{2\pi}\, H_{\rm g}$ is the vapor density,
the averaged normal component of the velocity passing through the particle surface $\varv_{\rm th}$ is given by
\begin{align}
\varv_{\rm th} = \frac{c_s}{\sqrt{2\pi}} = 
\frac{1}{\sqrt{2\pi}} 
\left( \frac{k_{\rm B}T}{\mu_{\rm H_2O}}\right)^{1/2},
\end{align}
with $\mu_{\rm H_2O}$ being the molecular weight of water and
we used 
\begin{align}
  P_{\rm vap} = \rho_{\rm vap} \, c_s^2.     
  % = \frac{\Sigma_{\rm vap}}{\sqrt{2\pi}H_{\rm g}}c_s^2
\end{align}
The equilibrium pressure $P_{\rm eq}$ is given by the Clausius-Clapeyron equation \citep{Lichtenegger1991}:
\begin{equation}
    P_{\rm eq}= 1.14\times 10^{13}\cdot \exp\left(-\frac{6062\,{\rm K}}{T}\right)\,{\rm g\,cm^{-1}\,s^{-2}}.
\end{equation}

For each super-particle, $P_{\rm eq}$ and $P_{\rm vap}$ at the location of the particle are calculated at each timestep.
When $P_{\rm eq} > P_{\rm vap}$, the evaporation probability during $\delta t$ is evaluated as  
\begin{align}
    p_{\rm evp} & = \min \left( \frac{\dot{m}_{\rm p}\,\delta t}{m_{\rm p}},\,1 \right) =\min\left(\frac{2}{\v_{\rm th}} \, \frac{3}{4\pi\, \rho_{\rm bulk}\,s}\, (P_{\rm eq}-P_{\rm vap}) \, \delta t,\,1\right).
%    & =\min\left(\sqrt{\frac{8\pi\mu_{\rm H_2O}}{k_{\rm B}T}}\cdot\frac{3}{4\pi \, \rho_{\rm bulk}s}\cdot P_{\rm eq}\delta t,\,1\right).
    \label{eq:peva}
\end{align}
% \ida{Why $P_{\rm eq}$ but not $(P_{\rm eq} - P_{\rm vap})$? See $p_{\rm cond}$ below.}
% where $\rho_{\rm bulk}$ is the bulk density of the particle.
For a generated random number $\mathcal{R}$ in [0,1],
if $\mathcal{R} \le p_{\rm evp}$, we regard that the icy super-particle evaporates.
% The condensation probability is described by \ida{Write how to derive this equation; not in details, insert an intuitive equation like Eq.~(\ref{eq:peva})}

When $P_{\rm eq} < P_{\rm vap}$, recondensation should occur immediately. The change in the water vapor surface density, $\Delta \Sigma_{\rm vap}$, should be corresponding to $P_{\rm vap}-P_{\rm eq}$. The recondensation probability is evaluated as
\begin{equation}
    % p_{\rm cnd}=\min\left(2H_{\rm g}\cdot\frac{\mu_{\rm g}}{k_{\rm B}T}(P_{\rm vap}-P_{\rm eq}),\,1\right).
    p_{\rm cnd}=\min\left(\frac{\Delta \Sigma_{\rm vap}}{\Sigma_{\rm vap}},1\right)=\min\left(\frac{P_{\rm vap}-P_{\rm eq}}{P_{\rm vap}},1\right).
\end{equation}
Similarly,
%to evaporation, we generate a random number for condensation $\mathcal{R}_{\rm cond}$ in a range of [0,1] and 
the icy super-particle condenses if $\mathcal{R} \le p_{\rm cnd}$ for a newly generated $\mathcal{R}$.

We assume water vapor re-condenses on surfaces of both ice and silicate dust particles. \citet{Ros2019} suggested water vapor tends to re-condense on the icy pebbles' surface rather than the silicate dust surface. 
Even if we restrict recondensation to silicate particle surfaces, our results hardly change.

In this paper, we define the ``snow line'' as the radius where the mass fraction of H$_2$O ice exceeds 1\% of total solid mass. 
In the fiducial case, the snow line radius is $r_{\rm snow} \simeq 3.7 \,\rm  au$.
We assume that smaller silicate particles are released after an icy particle passes the snow line.
Even if fluffy silicate aggregates are first formed as icy mantle evaporates \citep{Aumatell2011}, we assume that they are destroyed by collisional fragmentation into grains, for simplicity. 

%In our calculation, although
% In the case of a fixed 
%if we estimate 
% evaporation temperature  
% $\sim 167$ K, the position of the snow line \red{evolution due to decay of $\dot{M}_{\rm g}$} is given by
% \begin{equation}
%     r_{\rm snow}\sim 3.7 \left(\frac{\alpha}{10^{-3}}\right)^{-\frac{2}{9}}\left(\frac{\dot{M}_{\rm g}}{10^{-8}\,M_{\odot}/{\rm yr}}\right)^{\frac{4}{9}} \,{\rm au}.
%     \label{eq:r_snow}
% \end{equation}
% \red{We here calculate
% the evaporation temperature, taking into account of the dependence on 
% %varies corresponding to 
% water vapor pressure. Because the evaporation temperature is lower for lower $\dot{M}_{\rm g}$, the snow line evolution is faster than Eq.~(\ref{eq:r_snow}). } \ida{Correct?}

\subsection{Carbon destruction rate by FUV}\label{subsec:carbon}
%In this section, we consider the carbon destruction by FUV. 
\citet{Alata2014,Alata2015} suggested that hydrocarbon molecular gas is released from the surface of amorphous hydrocarbons by FUV. 
They 
%These amorphous hydrocarbon particles 
are suggested to exist in the ISM and comets as so-called ``amorphous carbons.''

We calculated the carbon fraction change with the carbon depletion timescale:
\begin{equation}
    \delta f_{\rm c}=-2\left(\frac{\delta t}{t_{\rm ph}}+\frac{\delta t}{t_{\rm ox}}\right)f_{\rm c},
\end{equation}
where $t_{\rm ph}$ and $t_{\rm ox}$ are timescale of photolysis and oxidation.
To suppress a too abrupt change of the carbon fraction, when $|\delta f_{\rm c}/f_{\rm c}|\ge 0.01$ with a dynamical timestep, we adopted a shorter chemical timestep $\delta t_{\rm chem}$, such that $|\delta f_{\rm c}/f_{\rm c}|< 0.01.  $  We then recalculated the timescale by updating the particle's position, assuming that the particle moves linearly during the dynamical timestep $\delta t$ \citep{Ishizaki2023}. The detailed calculation model is shown in Appendix \ref{app:tune}.

\subsubsection{Opacity for FUV} \label{subsec:opa}

\begin{figure}
    \centering
    \includegraphics[keepaspectratio,scale=0.23]{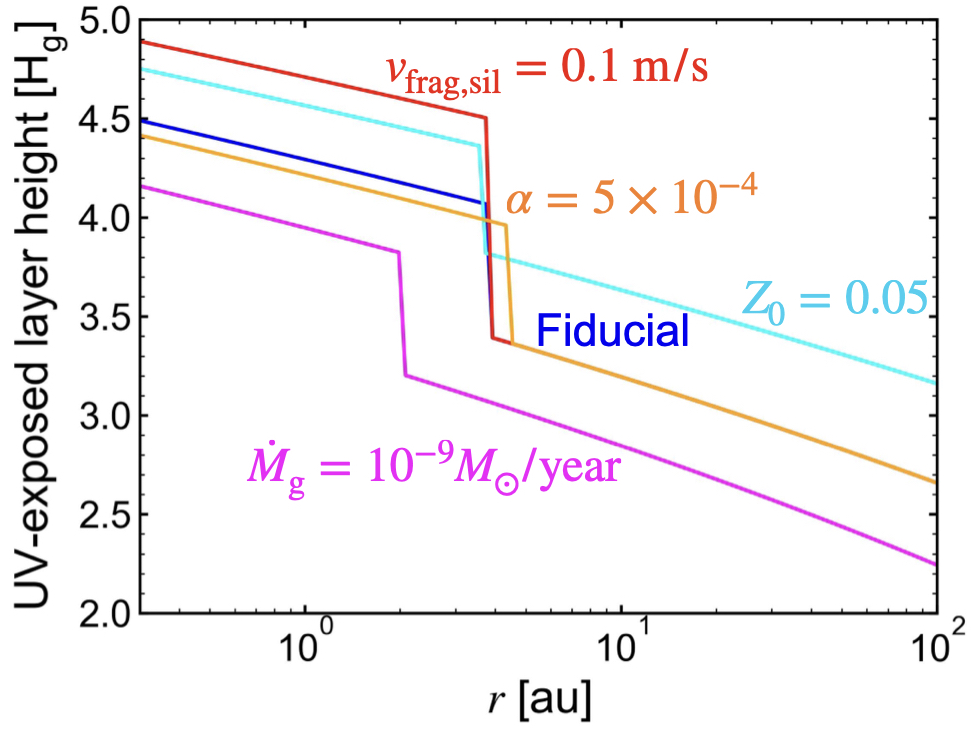}
    \caption{Height of the FUV-exposed layer for the initial condition.}
    \label{fig:height}
\end{figure}

The vertical optical depth ($
\tau_{z \lambda}$) for a given wavelength $\lambda$ is estimated by 
%a vertical integration of the local density of dust particles,
\begin{equation}
    \tau_{z \lambda} =\kappa_0\int^{\infty}_{|z|} \rho_{s_{\lambda}} dz,
\end{equation}
where $\kappa_{0 \lambda}$ is the dust opacity for $\lambda$. 
Because the particles with $s > s_{\lambda}\sim$ do not
contributes to the opacity 
in the Rayleigh regime, 
the effective dust density for the opacity,  $\rho_{s_{\lambda}}$, is 
given by
\begin{equation}
\rho_{s_{\lambda}}\sim\frac{\Sigma_{\rm d}}{\sqrt{2\pi}H_{\rm d}}\exp\left(-\frac{z^2}{2H_{\rm d}^2}\right)\times \left(\frac{s_{\lambda}}{s_{\rm max}}\right)^{0.5},
    \label{eq:tau}
\end{equation}
where $H_{\rm d}$ is
the scale height of dust particles given by
\begin{equation}
    H_{\rm d}=\sqrt{\frac{\alpha}{{\alpha+{\rm St}(s_{\lambda},z)}}}H_{\rm g},
    \label{eq:Hd}
\end{equation}
and we used the relation that the mass fraction ($\zeta$) of particles of $s\leq s_{\lambda}$ is $\sim \sqrt{s_{\lambda}/s_{\rm max}}$
with the assumed size distribution, $dN (s) \propto s^{-3.5}ds$ \footnote{$
\zeta = \int^{s \lambda}_{0.1\mu m} s^3 (dN/ds)\, ds \, \Big/ \int^{s max}_{0.1\mu m} s^3 (dN/s) \,ds \simeq (s_\lambda/s_{\rm max})^{0.5}$}. In summary, the vertical optical depth is rewritten as
\begin{equation}
    \label{eq:tau_z}
     \tau_{z \lambda}=\kappa_0\Sigma_{\rm d}\left(\frac{s_{\lambda}}{s_{\rm max}}\right)^{0.5}{\rm erfc}\left(\frac{|z|}{\sqrt{2}H_{\rm d}}\right).
\end{equation}
We assumed  $\lambda = 0.6 \,\rm \mu$m and $\kappa_0\sim 2.5\times 10^4 \, \rm cm^2g^{-1}$, following \citet{Binkert2023}.
We also considered the FUV radiation from the central star. Because of the flaring of the disk, the optical depth from the central star is given by $\tau_r=\tau_z/\Phi$, where $\Phi$ is the flaring angle set to 0.05 in our calculations.
Figure~\ref{fig:height} shows the bottom of the FUV-exposed layer ($\tau_r\sim1$) for the initial condition of each parameter set. 

\subsubsection{Photolysis rate}\label{subsec:photo}
The photolysis timescale is estimated by the mean collision time of a photon 
% as the timescale of which a photon collides
with a carbon atom of amorphous hydrocarbons \citep{Klarmann2018, Binkert2023}:
\begin{equation}
    t_{\rm ph}=\frac{4s\rho_{\rm c}}{3Y_{\rm ph}F_{\rm FUV}m_{\rm c}},
\end{equation}
where $F_{\rm FUV}$ is the flax of FUV radiation given by
\begin{equation}
    F_{\rm FUV}=\frac{L_{\rm FUV}}{4\pi r^2}\exp(-\tau_r).
\end{equation}
Here, $Y_{\rm ph}$ is the yield of photolysis per incoming photon, and $m_{\rm c}$ and $\rho_{\rm c}$ is the atomic mass and the bulk density of carbons. We set the yield $Y_{\rm ph}=10^{-4}$. As \cite{Binkert2023} suggested photons that penetrate the lower layer could destroy more amorphous hydrocarbons because more particles exist in the lower layer. 
However, this effect could lower the height of photolysis occurring region only by $\lesssim 0.5\,H_{\rm g}$ (see also Fig.3 in \cite{Binkert2023}).

\subsubsection{Oxidation rate} \label{subsec:oxi}
We assume that oxidation occurs only in the exposed layer ($\tau_{\rm r}\le 1$), since \citet{Lee2010} showed that oxygen atoms are abundant in the FUV-exposed layer and the abundance decreased steeply with the decrease of the height. 
We calculate the timescale of oxidation described by
\begin{equation}
    t_{\rm ox}=\frac{4s\rho_{\rm c}}{3n_{\rm ox}\v_{\rm ox}Y_{\rm ox}m_{\rm c}},
\end{equation}
where 
$\v_{\rm ox}$ is the mean thermal velocity of oxygen atoms given by
$\v_{\rm ox} =\sqrt{(8/\pi) (k_{\rm B}T_{\rm FUV}/m_{\rm ox})}$, $m_{\rm ox}$ is the oxygen atom mass, 
and $Y_{\rm ox}$ is the number of carbon atoms removed by an oxygen atom given by \citep{Draine1979}
\begin{equation}
    Y_{\rm ox}=A\exp\left(-\frac{B}{T_{\rm FUV}}\right),
\end{equation}
with $A=2.3$, $B=2580\,{\rm K}$ for $T_{\rm FUV}<440\,{\rm K}$ and $A=170$, $B=4430\,{\rm K}$ for $T_{\rm FUV}>440\,{\rm K}$.
We estimated the number density of oxygen atoms $n_{\rm ox}=\epsilon \,n_{\rm H_2}$, where $n_{\rm H_2}$ is the number density of ${\rm H_2}$ given by
\begin{equation}
    n_{\rm H_2}=\frac{\Sigma_{\rm g}}{\sqrt{2\pi}\mu m_{\rm p}H_{\rm g}}\exp\left(-\frac{z^2}{2H_{\rm g}^2}\right).
\end{equation}
We assumed $\epsilon = 10^{-4}$, following  
%similarly to 
\citet{Klarmann2018}.

\section{Results} \label{sec:resu}

\subsection{Detailed evolution in the fiducial run} \label{sec:detailed_evol}

In this section, we show the result of the fiducial run with the disk parameters and the fragmentation velocities in Table~\ref{tab:fidu},
% We first show the result 
assuming that all the refractory carbons are amorphous hydrocarbons

\begin{figure}
    \begin{tabular}{c}
         \begin{minipage}[t]{\hsize}
             \centering
             \includegraphics[keepaspectratio,scale=0.25]{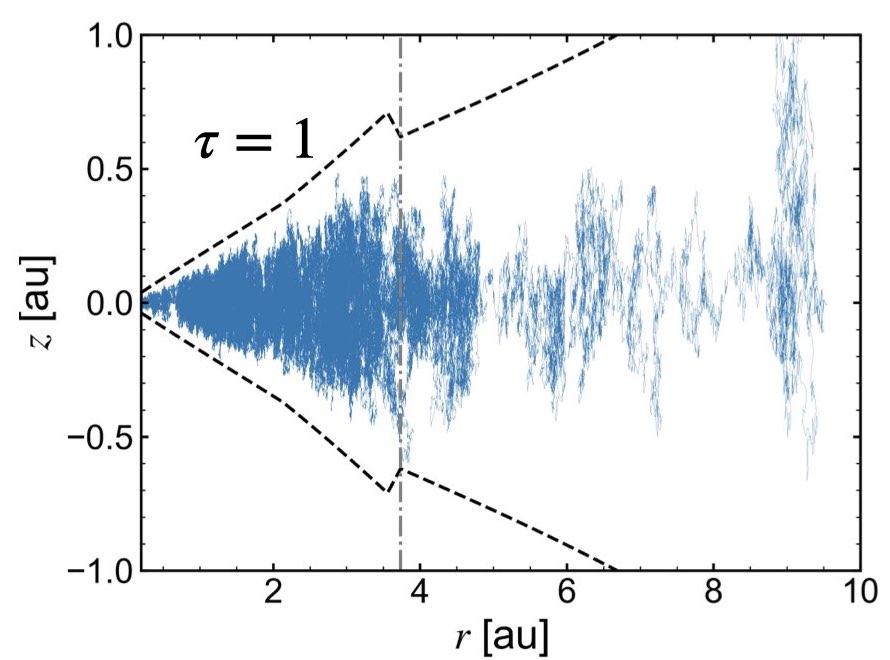}
         \end{minipage}\\
         \begin{minipage}[t]{\hsize}
             \centering
             \includegraphics[keepaspectratio,scale=0.25]{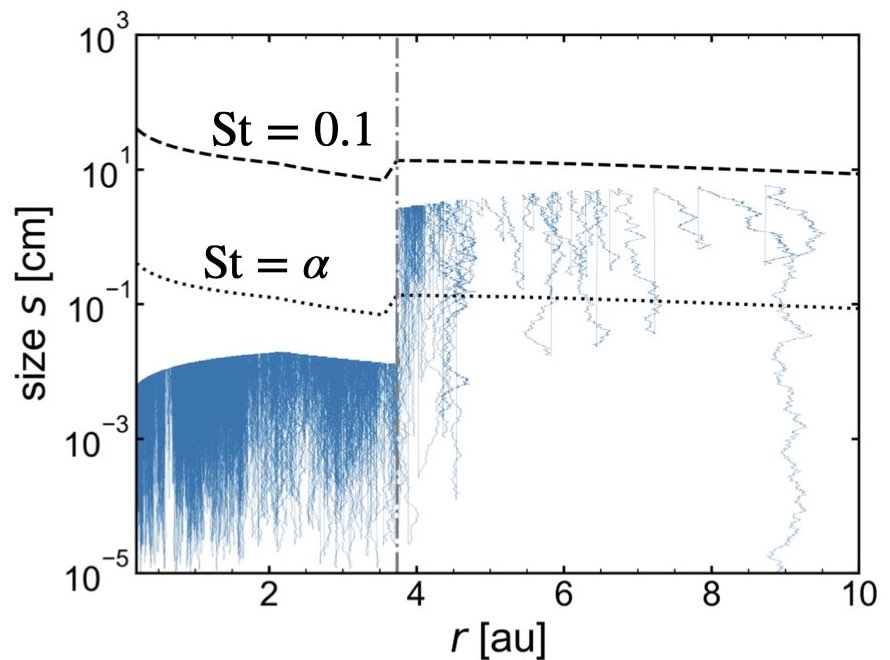}
         \end{minipage}
    \end{tabular}
    \caption{Typical trajectory of a solid particle (upper panel) and the size evolution (lower panel) in a disk with the fiducial parameters. The gray dash-dotted line represents the snow line at $r_{\rm snow} \simeq 3.7 \, \rm au$. The black dotted line of the top panel shows the height of the FUV-exposed layer. The black dotted and dotted lines show the particle sizes corresponding to %the cases of
    ${\rm St}=0.1$ and ${\rm St}=\alpha$, respectively.
    }
    \label{fig:tra}
\end{figure}

Figure~\ref{fig:tra} shows a typical trajectory in the $z$-$r$ and $s$-$r$ planes of a super-particle which starts at $r \simeq 9 \, \rm au$ with $s = 0.1 \,\mu$m. 
In the early phase, growth dominates over drift. In this run, the particle is mostly in a viscously heated region, except for the initial stage (Eq.~(\ref{eq:r_vis-irr}). 
Substituting the disk temperature given by Eq.~(\ref{eq:Tirr}) into Eq.~(\ref{eq:frag}), the fragmentation-limited Stokes number is
${\rm St}_{\rm frag} \simeq 0.059 \,(r/9\,\rm au)^{3/7}$.
When the particle's St value exceeds
${\rm St}_{\rm frag}$,
%size goes down when it reaches the maximum size limited by fragmentation, by 
we randomly re-assign its size ($< s_{\rm frag}$)  to a weight corresponding to the given power law.
The particle grows up to the size corresponding to ${\rm St}\sim 0.06$ around 9 au.
%and drifts inward with settling down to near the mid-plane. 
After the particle passes
%inside 
the snow line at $r_{\rm snow} \simeq 3.7 \, \rm au$, smaller particles are released to be strongly coupled to the gas (${\rm St} < \alpha$) as a result of the evaporation of icy mantle of the pebbles. Finally, the particle is lifted up inside $r_{\rm snow}$.

\begin{figure*}
    \begin{tabular}{cccc}
        \begin{minipage}[t]{0.255\textwidth}
            \centering
            \includegraphics[keepaspectratio,scale=0.36]{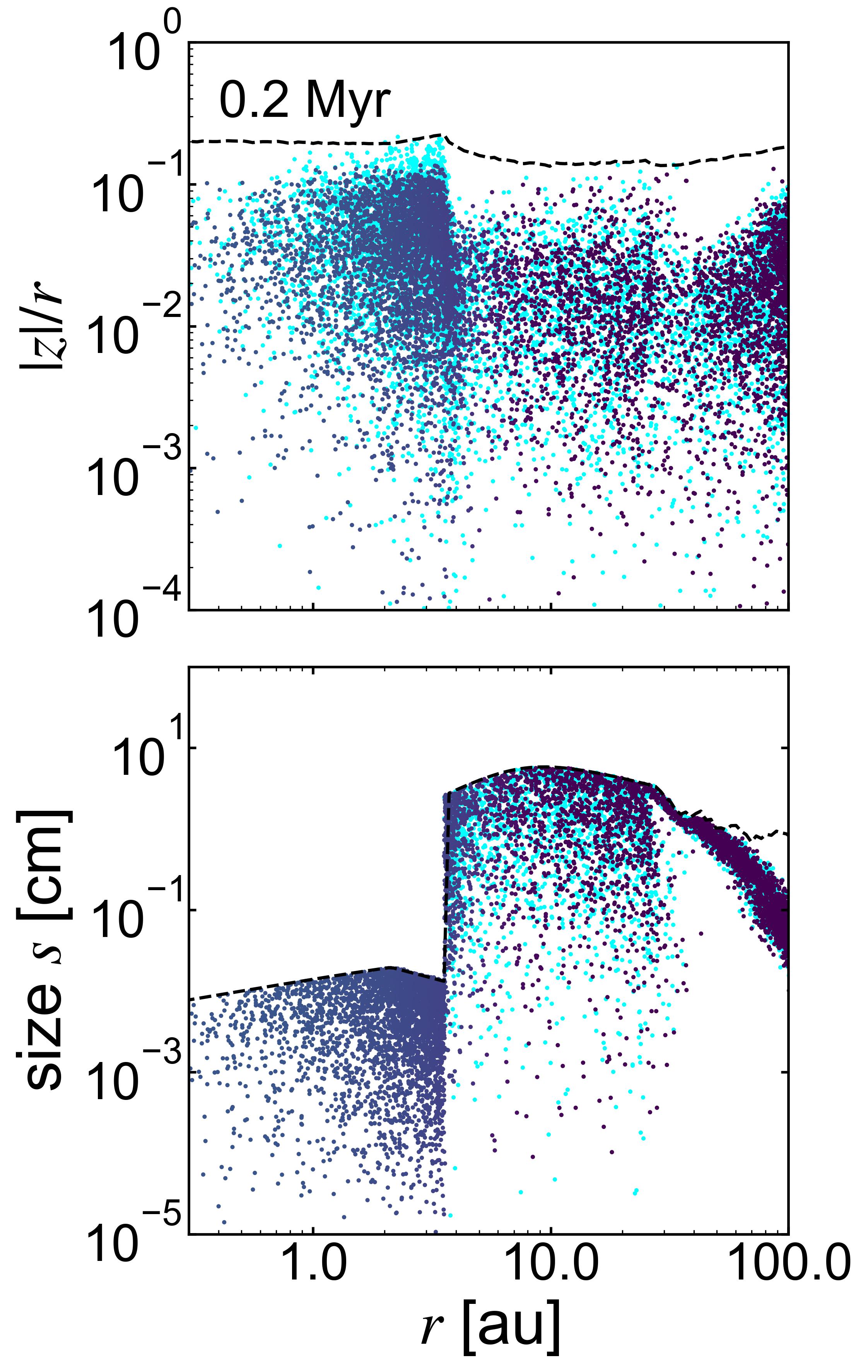}
            \subcaption{}
        \end{minipage}&
        \begin{minipage}[t]{0.2\textwidth}
            \centering
            \includegraphics[keepaspectratio,scale=0.065]{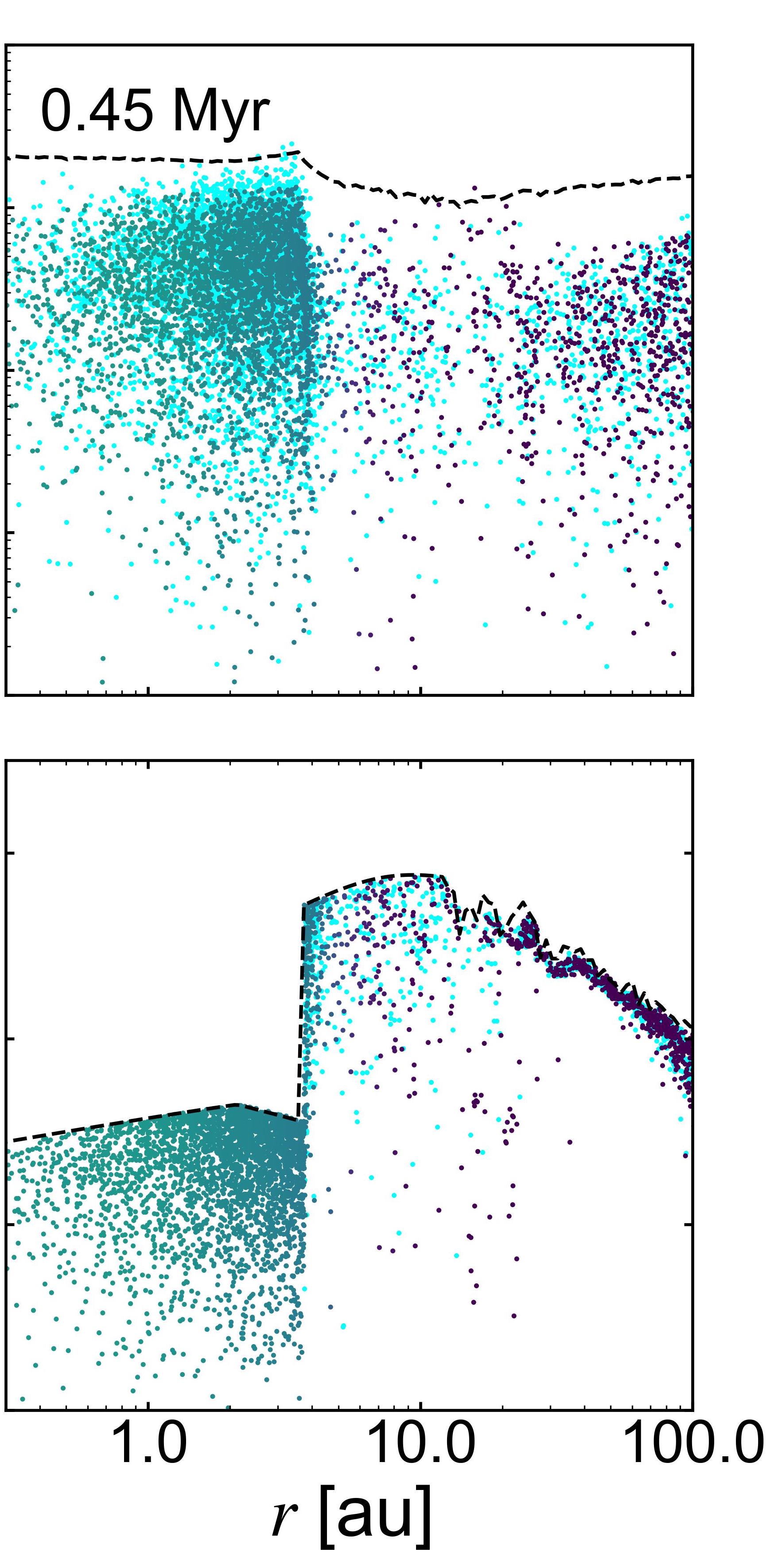}
            \subcaption{}
        \end{minipage}&
        \begin{minipage}[t]{0.2\textwidth}
            \centering
            \includegraphics[keepaspectratio,scale=0.065]{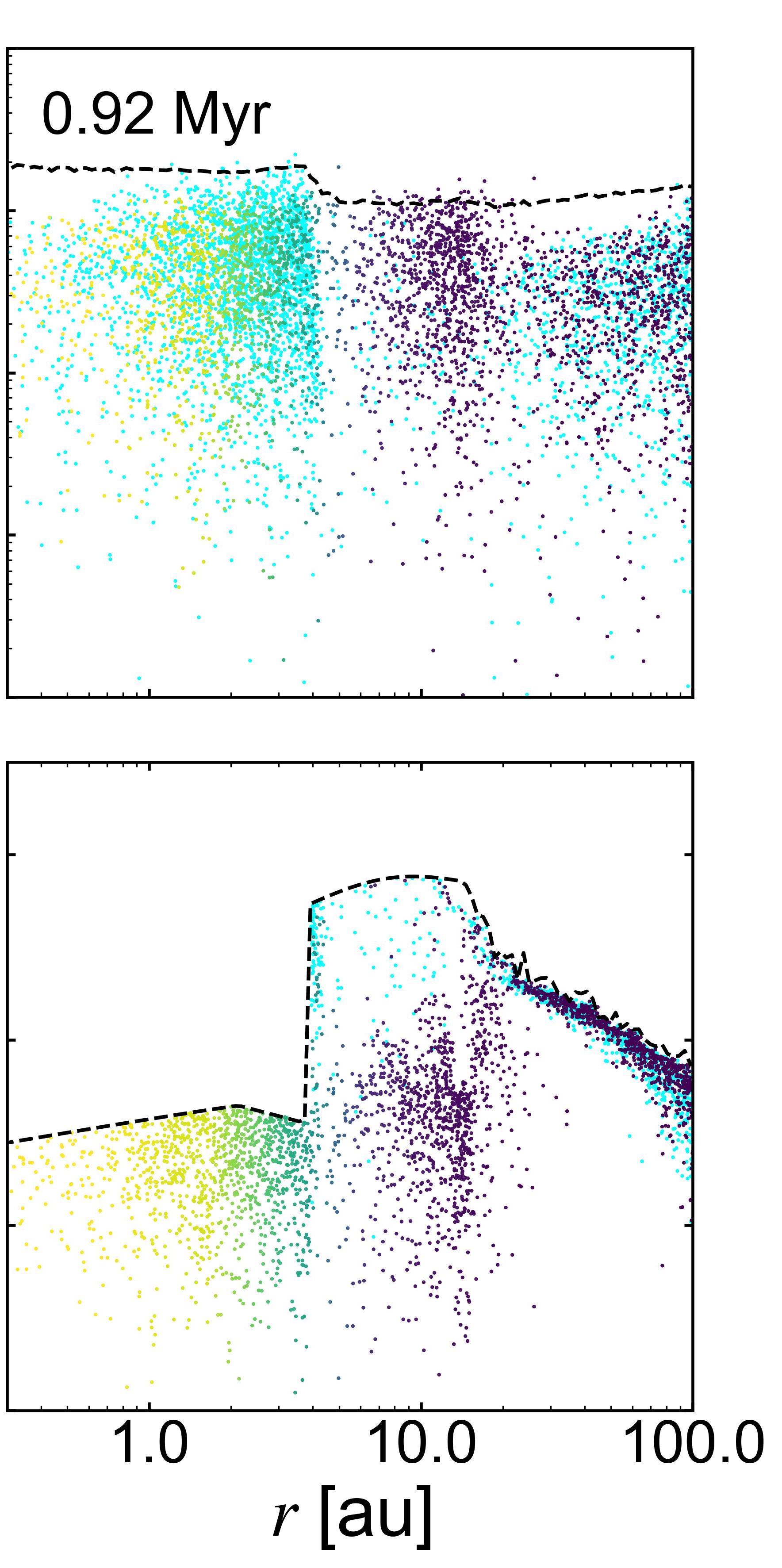}
            \subcaption{}
        \end{minipage}&
        \begin{minipage}[t]{0.238\textwidth}
            \centering
            \includegraphics[keepaspectratio,scale=0.065]{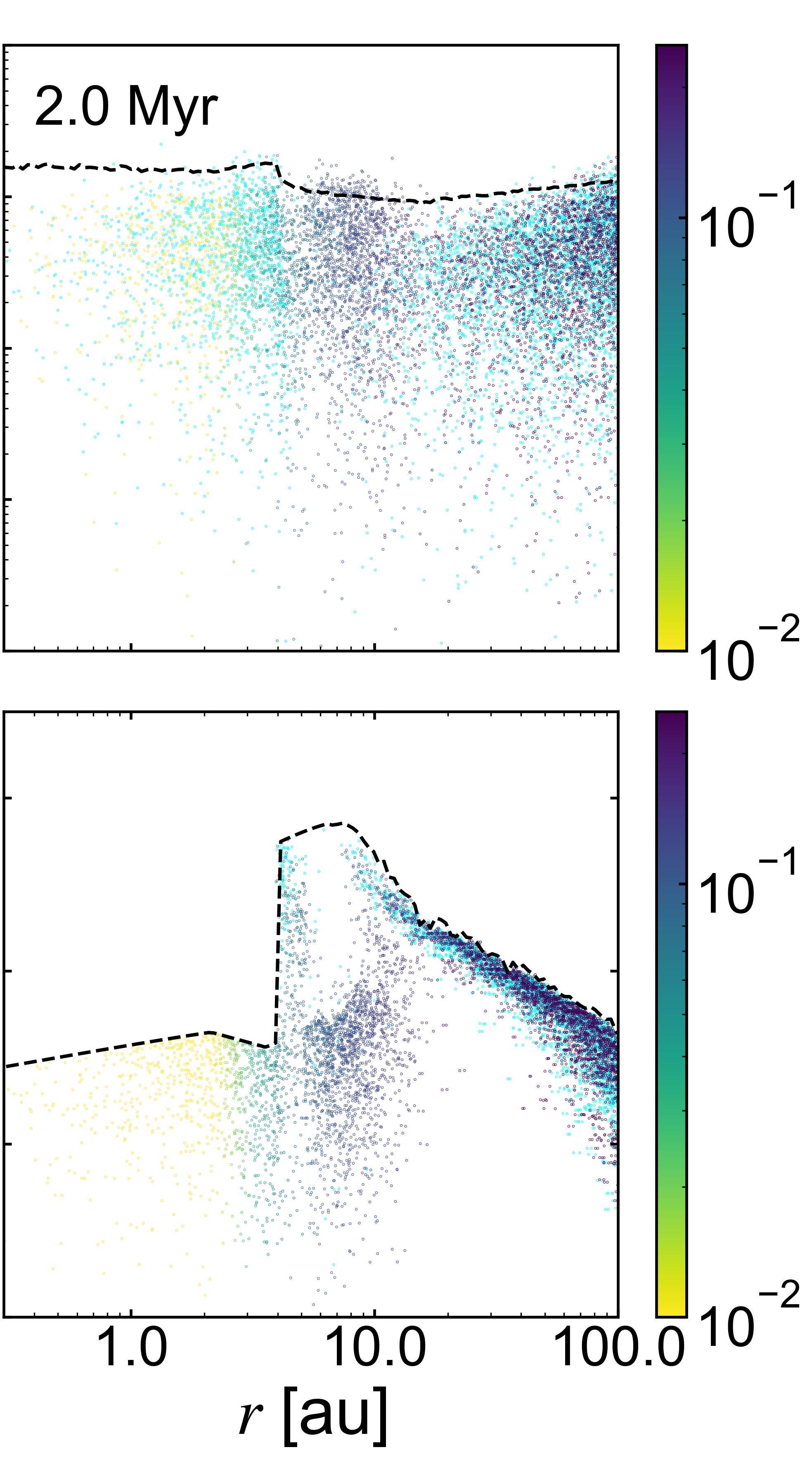}
            \subcaption{}
        \end{minipage}
    \end{tabular}
    \caption{
    Snapshots of solid super-particles in the $r$-$(|z|/r)$ plane and size distribution. The upper panels show the snapshots 
of silicates at (a) 0.2 Myrs ($\sim t_{\rm pf}$), (b) 0.45 Myrs ($\sim t_{\rm pf}+t_{\rm drift,ice}$), (c) 0.92 Myrs($\sim t_{\rm pf}+t_{\rm drift,ice}+t_{\rm diff,snow}$), and (d) 2.0 Myrs. The color bar shows the carbon fraction of each silicate super-particle. The cyan dots show the icy super-particles. The black dashed line shows the height of $\tau_{\rm FUV}=1$.
The lower panels show the size distributions of silicates
 corresponding to the upper panels. The black dashed line shows the local maximum dust size ($s_{\rm max}$) (Eq.~\ref{eq:s_max}).}
\label{fig:snap}
\end{figure*}

Figure~\ref{fig:snap} shows the snapshots of size and vertical distributions of the particles.
The colors of particles show the carbon fraction of each particle. The dark navy dots indicate the silicate super-particles that preserve original carbon fraction similar to the solar value ($f_{\rm c}\sim 0.25$).
The yellow dots inside the snow line represent the silicate super-particles with significantly depleted carbon fraction ($f_{\rm c} \lesssim 0.01$). 
The green ones are in between. 
The size of particles is dramatically changed at the snow line ($\sim 3.7$ au) due to the difference in stickiness between icy ($\v_{\rm frag} = 10 \,\rm m/s$) and silicate ($\v_{\rm frag} = 1 \,\rm m/s$) particles, as shown in Eq.~(\ref{eq:frag}).
The mass averaged size of particles 
is $\sim s_{\rm max}/3$, where $s_{\rm max}$ is the maximum dust size (Eq.~\ref{eq:s_max}).
% If the all particles are in Epstein regime (Eq.~(\ref{eq:St})), the effective (mass averaged) Stokes number in the fragmentation regime is also $\sim {\rm St}_{\rm max}/3$.
% and the settling of icy pebbles to the mid-plane causes high local  $\rho_{\rm d}/\rho_{\rm g}$. 
The particle size is usually limited by fragmentation rather than drift. 
% The maximum Stokes number is given by Eq.~(\ref{eq:frag}).
The upper panel in Fig.~\ref{fig:snap}(a) shows that most of the icy pebbles grow up to the size given by $s_{\rm max}$ (Eq.(\ref{eq:s_max})).
%For the grown icy pebbles and the silicate particles released at the snow line, ${\rm St} > \alpha$ and ${\rm St} < \alpha$, respectively. As a result, 

% Equation~(\ref{eq:frag}) also shows that 
For the icy pebbles, 
${\rm St} > \alpha$, and their motions are not strongly coupled to the gas, while the motions of the released silicate are strongly coupled (${\rm St} < \alpha$), and the vertical distribution is expanded up to $|z|/r \sim 0.1$.
As a result, the aspect ratio of the icy pebble distribution is considerably lower than that of the small silicate particles (the upper panel in Fig.~\ref{fig:snap}(a)), and icy pebbles beyond the snow line are usually shielded from FUV from the host star. 

Because the contrasts in $\v_{\rm frag}$ and ${\rm St}/\alpha$ between inside and beyond $r_{\rm snow}$ lead to the significant depletion of solid carbon inside $r_{\rm snow}$, we explain the growth, drift, and diffusion of particles together with the evolution of $f_{\rm c}$ in details.
The evolution of the radial distribution of the particles is divided into four stages and the carbon fraction is changed according to the stages.

\subsubsection{Quasi-steady dust accretion stage} \label{sec:1st_stage}

\begin{figure}
    \begin{tabular}{c}
         \begin{minipage}[t]{\hsize}
             \centering
             \includegraphics[keepaspectratio,scale=0.23]{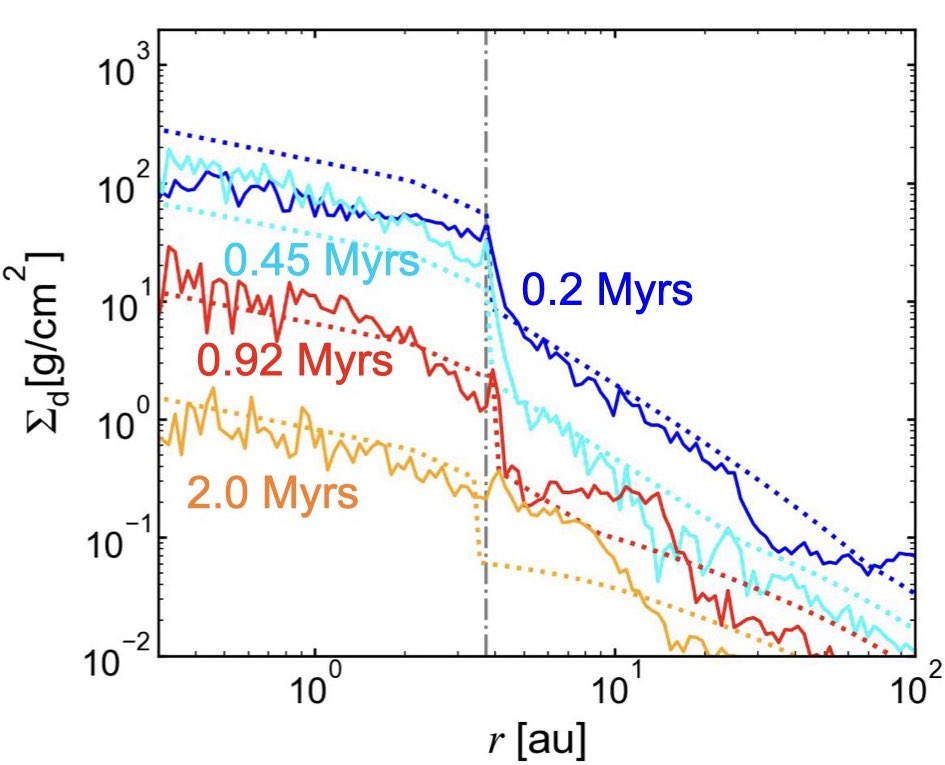}
         \end{minipage}\\
         \begin{minipage}[t]{\hsize}
             \centering
             \includegraphics[keepaspectratio,scale=0.23]{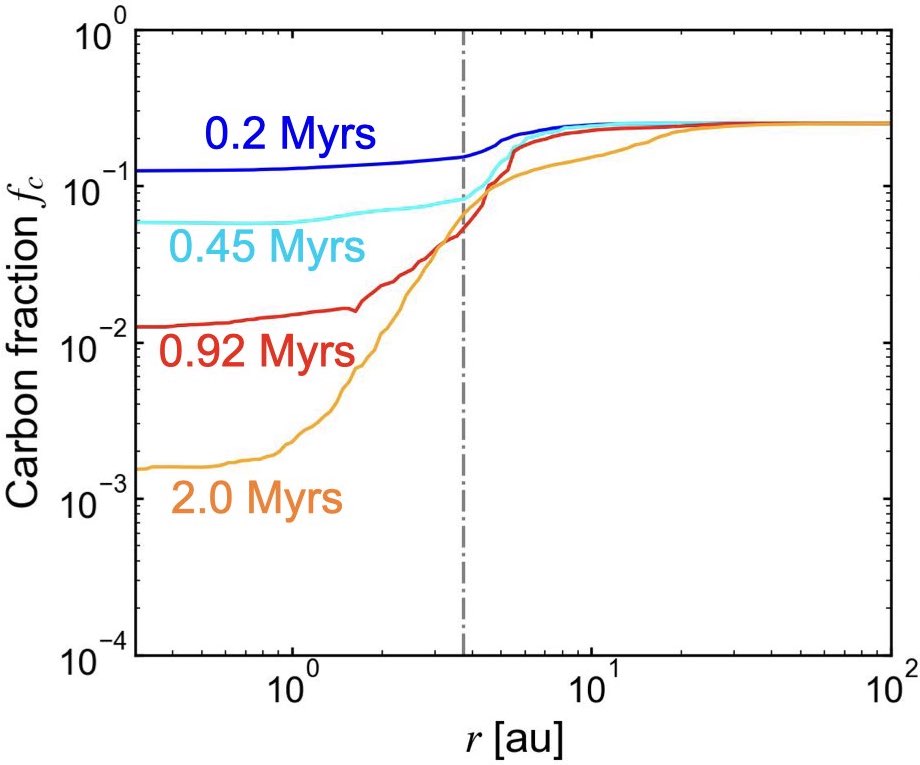}
         \end{minipage}
    \end{tabular}
    \caption{Time evolution of the solid surface density $\Sigma_{\rm d}$ (upper panel) and the carbon fraction, $f_{\rm c}$, (lower panel) of the fiducial run. The gray dash-dotted line shows the snow line. The dotted lines in the upper panel show the analytical estimations given by Eq.~(\ref{eq:Z}). The dash-dotted lines represent the snow line at $r_{\rm snow} \sim 3.7 \, \rm au$.}
    %\ida{Add the snow line}}% The circle and cross markers show the boundaries of limitation of the maximum particle size between fragmentation and drift given by Eq.~(\ref{eq:rdri_fra}) and calculation.}
    \label{fig:fidu}
\end{figure}

% The carbon fraction varies corresponding to 4 stages of

% Icy pebbles grow up fast in the outer disk regions. The pebble formation timescale $t_{\rm pf}$ in Epstein regime at the orbital radius $r$ is estimated by Eq.~(\ref{eq:St}) \citep[]{Sato2016,IdaYmaOku2019}: \begin{equation} t_{\rm pf}\sim 10 \times \frac{4}{\sqrt{3\pi}} Z_0^{-1} \Omega^{-1} \sim  2 \times 10^5\left(\frac{Z_0}{0.01}\right)^{-1}\left(\frac{r}{100\,{\rm au}}\right)^{3/2}\,{\rm years},\label{eq:tpf} \end{equation} where $Z_0$ is the initial solid-to-gas ratio in the disk. This timescale is relatively short at smaller $r$, meaning that pebble formation proceeds in an inside-out manner. 
% After the icy pebbles sufficiently grow up, the drift timescale becomes shorter than the growth timescale and they start drifting inward.

The theoretical estimate for pebble formation timescale due to pairwise coagulation in the Epstein regime is  
\citep[]{Sato2016,IdaYmaOku2019} \begin{equation} 
t_{\rm pf}\sim 
% 10 \times \frac{4}{\sqrt{3\pi}} Z_0^{-1} \Omega^{-1} \sim  
2 \times 10^5\left(\frac{Z_0}{0.01}\right)^{-1}\left(\frac{r}{100\,{\rm au}}\right)^{3/2}\,{\rm years},\label{eq:tpf} 
\end{equation} 
where $Z_0$ is the initial solid-to-gas ratio in the disk.
Thus, pebble formation proceeds in an inside-out manner.
At $t \sim 0.2$ Myrs, the pebbles at $r \la 100 \, \rm au$ have already started drifting and the quasi-steady accretion of solids with the silicate particle pile-up inside the snow line has been established.

Because the mean radial velocity of the silicate particles is lower than that of the icy pebbles, the silicate particles released by sublimation of the icy mantle of the pebbles pile up inside the snow line (see the line at $t=0.2$ Myr in the top panel of Fig.~\ref{fig:fidu}).
Since the solid mass flux should be continuous at the snow line, the ratio of solid surface densities from just inside the snow line $\Sigma_{\rm d,in}$ to just outside $\Sigma_{\rm d,out}$ is given by
\begin{equation}
\label{eq:in_out}
    \frac{\Sigma_{\rm d,in}}{\Sigma_{\rm d,out}}=\frac{0.5 \dot{M}_{\rm d}}{2\pi r \v_{r,\rm in}} \frac{2\pi r \v_{r,\rm out}}{\dot{M}_{\rm d}}= 0.5 \, \frac{\v_{r,\rm out}}{\v_{r,\rm in}},
\end{equation}
where we assume half of the solid mass is converted to the water vapor inside the snow line. 
Equation~(\ref{v_r0}) is rewritten as
\begin{equation}
\label{v_r1}    
\v_r \simeq -\frac{\Lambda}{1+\Lambda^2\,{\rm St}^2} 
\left(2\Lambda\,{\rm St}\,C_\eta + \frac{3}{2} \alpha \right) \, h_{\rm g}^2 r \Omega.
\end{equation}
Inside the snow line, since the released silicate particles are small enough to be coupled with the gas ($\rm St \ll \alpha$), their radial velocity is 
\begin{equation}
\label{v_r1}    
\v_{r,\rm in} \simeq - \frac{3\Lambda}{2} \alpha \, h_{\rm g}^2 \, r \Omega.
\end{equation}
Outside the snow line, ${\rm St} \gg \alpha$ for icy pebbles and   
the radial drift velocity is given by 
\begin{align}
\label{eq:v_drift2}
\v_{r,{\rm out}} & \simeq 
 - 2\Lambda^2 \,{\rm St} \, C_{\eta}\,h_{\rm g}^2r\Omega. 
 % \nonumber\\ & \simeq - \frac{2\Lambda^2 \, C_\eta}{3} \times 2.3 \times 10^{-2} \left( \frac{\alpha}{10^{-3}}\right)^{-1} \left( \frac{\v_{\rm frag}}{10 \, \rm m/s} \right)^2 \left( \frac{T}{150 \, \rm K} \right)^{-1} \times \,h_{\rm g}^2r\Omega \nonumber\\  &  \simeq - 7.8\times 10^{-5}\Lambda^2 \, \left( \frac{\alpha}{10^{-3}}\right)^{-1} \left( \frac{\v_{\rm frag}}{10 \, \rm m/s} \right)^2 \left( \frac{r}{1 \, \rm au} \right)^{\frac{1}{2}}\, {\rm au/y},
\end{align}
With $\Lambda \simeq 1$, we obtain
\begin{align}
     \frac{\Sigma_{\rm d,in}}{\Sigma_{\rm d,out}} & \simeq 0.5\times \frac{4C_{\eta}{\rm St_{frag}}}{3\times 3 \, \alpha}  \simeq 6 \left( \frac{\alpha}{10^{-3}}\right)^{-2} 
 \left( \frac{\v_{\rm frag,ice}}{10 \, \rm m/s} \right)^2
 \left( \frac{T(r_{\rm snow})}{170 \, \rm K} \right)^{-1}, \label{eq:jump_in_Sigma}
\end{align}
where we substitute Eq.~(\ref{eq:frag}) into $\rm St \sim St_{frag}/3$ (taking the size distribution into account) and use $C_{\eta}=11/8$. The numerical result at $t = 0.2$ Myr in the top panel of Fig.~\ref{fig:fidu} shows that the jump in $\Sigma_{\rm d}$ is consistent with the Eq.~(\ref{eq:jump_in_Sigma}).

The released silicate particles are occasionally stirred up highly by the disk gas turbulence because ${\rm St} \ll \alpha$, while they radially go back and forth. 
As a result, some fraction of amorphous hydrocarbons in these particles are destructed by photolysis and oxidation.

However, $f_{\rm c}$  decreases only slightly at $t = 0.2$ Myrs
% only by a factor of 4 during this quasi-steady accretion phase
(Fig.~\ref{fig:fidu}(b)) because fresh silicate particles full of amorphous hydrocarbons are steadily supplied and well mixed with particles where the amorphous hydrocarbons become destructed. This is consistent with the result found by \citet{Klarmann2018}.

\begin{figure}
    \centering
    \includegraphics[keepaspectratio,scale=0.25]{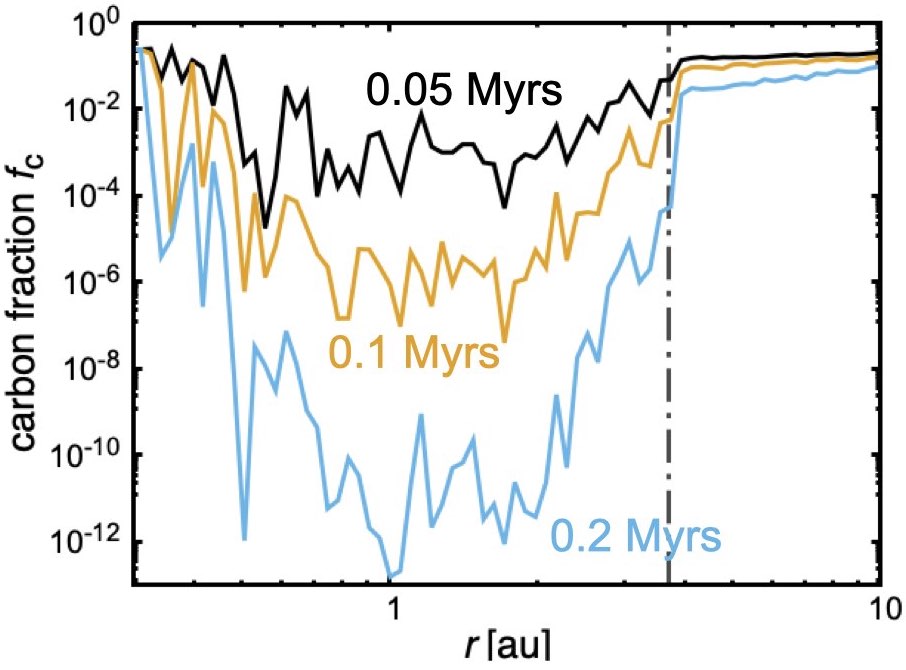}
    \caption{Time evolution of carbon fraction, $f_{\rm c}$, when the particles diffuse only vertically and neither drift nor diffuse radially.}
    \label{fig:nodrift}
\end{figure}

To highlight the effect of the continuous supply, we carried out a hypothetical case where all the radial motion is set to zero.
% The time evolution of carbon fraction is shown in Fig.~\ref{fig:nodrift}.}
Because the pebble flux is also stopped, the fresh refractory carbon supply is also stopped. The carbon fraction changes only through vertical diffusion, photolysis, and oxidation.
As shown by the black line in Fig.~\ref{fig:nodrift}, this process efficiently depletes $f_{\rm c}$ on a timescale of 50 kyrs.

% However, if the fresh refractory carbon supply is taken into account, the reduction of the carbon fraction is deleted.

\subsubsection{Icy pebble flux decaying stage} \label{sec:2nd_stage}

However, the quasi-steady accretion ends, and the carbon depletion pattern changes, when the pebble formation front reaches the disk's outer edge at the characteristic radius $r_{\rm disk}$. Then, because most of the reservoir of solid materials is consumed, the pebble flux rapidly decays \citep{Sato2016,IdaYmaOku2019}. 

The duration of the quasi-steady supply of icy pebbles is given by $t_{\rm pf}(r_{\rm disk}) + t_{\rm drift,ice}$, where $t_{\rm drift,ice}$ is the drift timescale from the disk outer edge to the snow line.
The drift velocity is given by
\begin{align}
\label{eq:v_drift2}
\v_{r} & \simeq 
 2\Lambda^2 \,{\rm St} \, C_{\eta}\,h_{\rm g}^2r\Omega \nonumber\\
% & \sim \frac{2\Lambda^2 \, C_\eta}{3} \times 2.3 \times 10^{-2} \left( \frac{\alpha}{10^{-3}}\right)^{-1} \left( \frac{\v_{\rm frag}}{10 \, \rm m/s} \right)^2 \left( \frac{T}{150 \, \rm K} \right)^{-1} \times \,h_{\rm g}^2r\Omega \nonumber\\
  &  \simeq 7.8\times 10^{-5}\Lambda^2 \, \left( \frac{\alpha}{10^{-3}}\right)^{-1}
  \left( \frac{\v_{\rm frag}}{10 \, \rm m/s} \right)^2
  \left( \frac{r}{1 \, \rm au} \right)^{1/2}\, {\rm au/y},
\end{align}
%\ida{Write the $T$-independent expression.}
%where we substituted Eq.~(\ref{eq:frag}) into $\rm St \sim St_{max}/3$ and used $C_{\eta}=11/8$. 
Because $h_{\rm g}^2 \propto T$, the $T$-dependence is canceled out with $T^{-1}$ in $\rm St \simeq St_{frag}/3$. The $\alpha$-dependence appears from ${\rm St} \propto 1/\alpha$ in the fragmentation limit.
The factor $\Lambda=(1+\rho_{\rm d}/\rho_{\rm g})^{-1}$ is estimated by
\begin{align}
    \frac{\rho_{\rm d}}{\rho_{\rm g}}&=\frac{\Sigma_{\rm d}}{\Sigma_{\rm g}}\frac{H_{\rm g}}{H_{\rm d}}
    \sim \frac{\Sigma_{\rm d}}{\Sigma_{\rm g}} \sqrt{\frac{\rm St}{\alpha}}\\
%    &\sim 0.048\left(\frac{\Sigma_{\rm d}/\Sigma_{\rm g}}{0.01}\right) \left( \frac{\alpha}{10^{-3}}\right)^{-1} \left( \frac{\v_{\rm frag}}{10 \, \rm m/s} \right) \left( \frac{T}{150 \, \rm K} \right)^{-\frac{1}{2}},
    & \simeq 0.13\left(\frac{\Sigma_{\rm d}/\Sigma_{\rm g}}{0.01}\right) \left( \frac{\alpha}{10^{-3}}\right)^{-1} 
 \left( \frac{\v_{\rm frag}}{10 \, \rm m/s} \right)
 \left( \frac{r}{100 \, \rm au} \right)^{3/14}.\end{align}
Here, we use Eq.~(\ref{eq:Hd}) for $H_{\rm d}$ assuming ${\rm St}\gg\alpha$
and the irradiation-dominated $T$ distribution is applied in the regions outside the snow line (Eq.~(\ref{eq:Tirr})).
Then, $t_{\rm drift,ice}$ is estimated as
\begin{equation}
\begin{split}
    t_{\rm drift,ice} & =-\int^{r_{\rm disk}}_{r_{\rm snow}}\frac{dr}{\v_{r}}\sim 3.1\times 10^5\left( \frac{\alpha}{10^{-3}} \right) \left( \frac{10\,{\rm m/s}}{\v_{\rm frag,ice}}\right)^{-2}\\ &\times\left(\frac{1+\rho_{\rm d}/\rho_{\rm g}}{1.1}\right)^2\left(\sqrt{\frac{r_{\rm disk}}{100\,{\rm au}}}-0.1 \sqrt{\frac{r_{\rm snow}}{1\,{\rm au}}}\right) {\rm years}.  
\end{split}
    \label{eq:tdrift}
\end{equation}
% This timescale shows when the new icy pebble formation finishes. Thus, icy pebble flux is expected to start decaying after $t_{\rm pf}$. Figure~\ref{fig:snap} (a) shows the vertical and size distributions at $t\sim t_{\rm pf}$. Most of the icy pebbles grew up to the size given by Eq.(\ref{eq:size}) and settled down \red{to} a lower layer than silicate dust outside the snow line. 
In the fiducial run, $t_{\rm drift,ice}$ is estimated to be about $2.5\times 10^5$ years. %\ida{How did you derive this from the above equation? You assume $\rho_{\rm d}/\rho_{\rm g}\sim 1.5$?  Why the ratio is so high beyond the snow line?}
% This timescale shows how long it takes from the disk edge to the snow line for the icy pebbles. 

The second stage starts at 
%the supply of icy pebbles into the snow line is expected to last \red{for} %until 
$t \sim t_{\rm pf}+t_{\rm drift,ice}$
($\sim 0.45$ Myr in the fiducial run).
In this stage, while $\Sigma_{\rm d}$ 
%the silicate surface density 
inside the snow line has not been reduced, 
the carbon fraction $f_{\rm c}$ starts decreasing.
% is significantly depleted. 
Figure~\ref{fig:snap}(b) shows 
the vertical and size distributions at $t = 0.45$ Myr. 
% The average of Stokes number in the fragmentation regime is $\sim {\rm St}_{\rm max}/3$ \ida{How did you derive this?}
% and the settling of icy pebbles to the mid-plane causes high local  $\rho_{\rm d}/\rho_{\rm g}$. \Oka{If the particle size is limited by fragmentation, the maximum Stokes number is given by Eq.~(\ref{eq:frag}).} 
The total mass of dust particles inside $r_{\rm snow}$ is larger than that outside it, 
%the snow line 
since most of the icy pebbles carrying silicate particles had drifted to the regions inside $r_{\rm snow}$. 
The upper panel of Fig.~\ref{fig:fidu} shows the evolution of $\Sigma_{\rm d}$.
The $\Sigma_{\rm d}$ beyond $r_{\rm snow}$ significantly drops from the first phase ($t = 0.2 \, \rm Myr$) to the second phase ($t = 0.45 \, \rm Myr$), while the $\Sigma_{\rm d}$ decay is not significant inside $r_{\rm snow}$. As a result, $\Sigma_{\rm d}$ jumps at $r_{\rm snow}$ by as much as a factor of 30.
%In the second stage, the carbon fraction decreases due to decaying of the supply of amorphous hydrocarbons from outside the snow line. 
However, carbon depletion proceeds only by a factor of a few
from the first to the second stage, because
%, although the icy pebble flux has already started decaying in this stage, the 
supply of icy pebbles still remains at a reduced rate.

\subsubsection{Silicate dust surface density decaying stage} \label{sec:3nd_stage}

In the third stage, $\Sigma_{\rm d}$ 
%the solid surface density 
decreases also inside $r_{\rm snow}$. 
%the snow line. 
Because the influx of silicate particles from the outer region is already diminished,
the decrease occurs on
%After that, the abundance of silicate dust inside the snow line decreased with 
a diffusion timescale at $r_{\rm snow}$
%silicate particles at the snow line 
\citep{Okamoto2022}:
\begin{equation}
\begin{split}
    t_{\rm diff,snow} &\sim \left. \frac{r^2}{\alpha H_{\rm g}^2\Omega_{\rm K}} \right|_{r_\textrm{snow}} = 
   \left. \frac{1}{\alpha h_{\rm g}^2 \Omega_{\rm K}} \right|_{r_\textrm{snow}}\\  &\sim 4.7 \times 10^5 \left( \frac{\alpha}{10^{-3}} \right)^{-10/9} \left( \frac{\dot{M}_{\textrm{g}}}{10^{-8}\,M_{\odot}\,\textrm{yr}^{-1}}\right)^{2/9} \textrm{years}.
\end{split}
   \label{eq:tdiff}
\end{equation}
In the fiducial run, $t\sim t_{\rm pf}+t_{\rm drift,ice}+t_{\rm diff,snow}\sim 0.92$ Myrs. 
The upper panel of Fig.~\ref{fig:fidu} shows that at $t = 0.92$ Myrs, 
 $\Sigma_{\rm d}$ 
 %the silicate surface density 
 inside $r_{\rm snow}$ 
 %the snow line 
 is decreased by a factor of several from that at $t = 0.45$ Myr.

In this stage, $f_{\rm c}$ 
shows a pronounced hyperbolic-tangent-function-like ($\tanh$-like) depletion pattern near the snow line by more than one order of magnitude (the line of $t = 0.92 \, \rm Myr$). 
The greater depletion is due to the diminished influx of silicate particles from the outer region.
% After the second stage, less fresh amorphous hydrocarbons are coming from outside the snow line while the silicate particles are piled up inside the snow line. That leads the less mass fraction of amorphous hydrocarbons inside the snow line. Therefore, the carbon fraction starts decreasing after $t=0.45$ Myr (see also a blue line in Fig.~\ref{fig:para}).
% \Oka{In the third stage, the carbon fraction distribution is changed from the first steady accretion stage: it shows a pronounced hyperbolic tangent function like dropping near the snow line by as much as one order of magnitude (the line of $t = 0.92 \, \rm Myr$ in Figure~\ref{fig:fidu}(b)).
% In this stage, the supplied fresh amorphous hydrocarbons is negligible to the total mass of the silicate particles with amorphous hydrocarbons destructed, which is responsible for the preservation of the significantly low carbon fraction inside the snow line.} 
The flat $f_{\rm c}$ pattern at $\lesssim 1\,\rm au$ 
%inside the snow line until this stage. That 
is caused by higher $\Sigma_{\rm d}$ there. 
Because  
% solid surface density inside the snow line. In this region, as 
photolysis is inefficient there, $f_{\rm c}$ conserves 
that in the outer region.
%is determined by the carbon fraction of solids drifted from the outer disk region.

\begin{figure}
    \centering
    \includegraphics[keepaspectratio,scale=0.23]{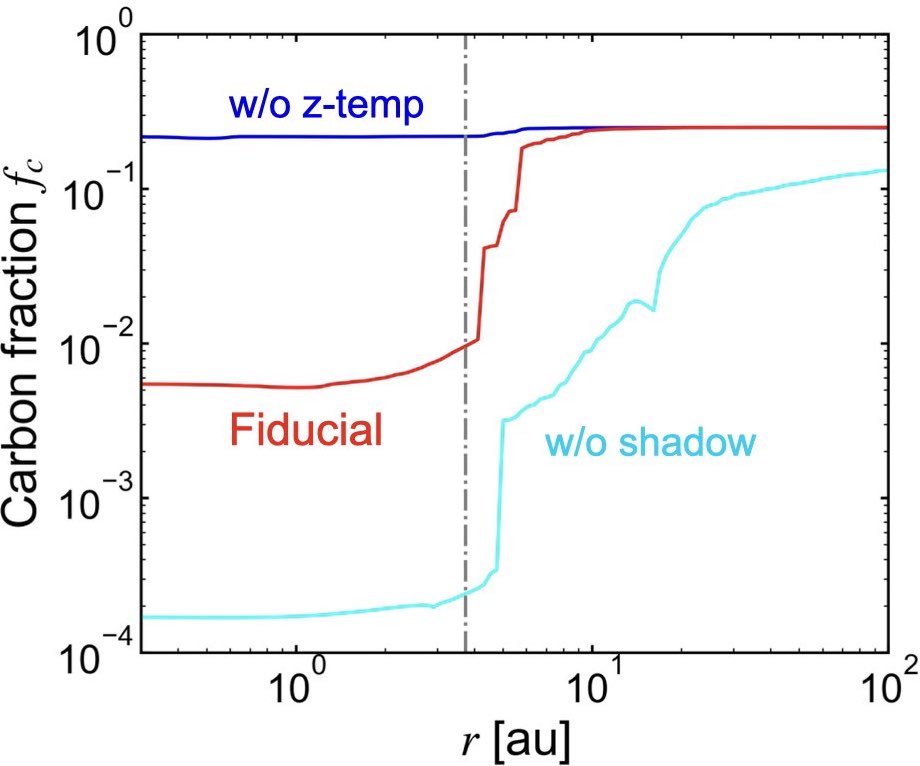}
    \caption{Carbon fraction, $f_{\rm c}$, at 0.68 Myr in the fiducial run, without the vertical temperature profile, and without the shadow area outside the snow line.}
    \label{fig:noshadow}
\end{figure}

% \red{
% The shadow effect that amorphous hydrocarbons are not destroyed outside $r_{\rm snow}$ due to the FUV shielding that we adopt (Sect.~\ref{subsubsec:photo_oxy}), is not a main origin for the pronounced decrease in $f_{\rm c}$ around $r_{\rm snow}$.}
The pronounced decrease in $f_{\rm c}$ around $r_{\rm snow}$ is not originated mainly by the shadow effect, that amorphous hydrocarbons are not destroyed outside $r_{\rm snow}$ due to the FUV shielding that we adopt (Sect.~\ref{subsubsec:photo_oxy}).
% The dropping off of the carbon fraction at the snow line is independent of the existence of the shadow area. 
In Fig.~\ref{fig:noshadow}, the fiducial case result is compared with the result without the shadow effect
at 0.68 Myrs($\sim t_{\rm pf}+t_{\rm drift,ice}+0.5t_{\rm diff,snow}$),
%the carbon fraction 
% with our fiducial parameters and without the shadow area behind the snow line. 
where the similar drop-off in $f_{\rm c}$ around $r_{\rm snow}$ is shown.
While $f_{\rm c}$ is decreased at $r \gtrsim r_{\rm snow}$, 
the non-shadow effect just lowers the baseline of $f_{\rm c}$ in the inner region. 
%While the value is lower than the case with the shadow area in the entire disk region, the dichotomy between the inner and outer disk region is not changed. That suggests that the structure is produced not by the shadow area behind the snow line.

We have argued that the $\tanh$-like $f_{\rm c}$ pattern is formed by a jump in $\Sigma_{\rm d}$ across $r_{\rm snow}$.
After 0.92 Myr, since $\Sigma_{\rm d}$ is depleted also inside $r_{\rm snow}$, it could be predicted that $f_{\rm c}$ goes up again. However,
Fig.~\ref{fig:fidu} shows that $f_{\rm c}$ further decreases while the enhanced jump in $\Sigma_{\rm d}$ at $\sim 
r_{\rm snow}$ disappears. 
% It is thought to disappear when the jump also disappears\Oka{, and $f_{\rm c}$ is thought to start increasing. At $2.0$ Myrs, although the jump of solid surface density disappear at the snow line, $f_{\rm c}$ decrease rather than increase. 
This is caused by 
more efficient photolysis, associated with the opacity decrease due to the $\Sigma_{\rm d}$ decay,
which overwhelms the effect of the reduced jump of $\Sigma_{\rm d}$ at $\sim 
r_{\rm snow}$.
As this effect also narrows the area where $f_{\rm c}$ is flat, the $\tanh$-like $f_{\rm c}$ pattern is diminished.
%cancels the effect of the shrinking the jump of $\Sigma_{\rm d}$.}
%(the pink line in the top panel of Fig.~\ref{fig:fidu}). At this time, the carbon fraction outside the snow line decreases due to the diffusion of the small particles. That teases dependence on $r$ of the carbon fraction around the snow line.

%\subsubsection{Carbon fraction regulated by a jump in solid surface density across the snow line}
\subsection{Analytical formula for $\Sigma_{\rm d}$ across the snow line}
\label{subsec:Sig_jump}

We found that the carbon fraction inside the snow line is regulated by the ratio of $\Sigma_{\rm d}$
%the solid surface density between 
inside and outside the snow line, except for the final phase with highly depleted $\Sigma_{\rm d}$.
To explain the numerical results, we analytically estimate the $\Sigma_{\rm d}$
%solid surface density
distribution.

The solid-to-gas ratio with $\Lambda \ll 1$ is given by (Eq.~(\ref{v_r1})):
\begin{equation}
\label{eq:sigdg}
%    Z = \frac{\Sigma_{\rm d}}{\Sigma_{\rm g}}=\frac{2\pi r|u_{\nu}|}{2\pi r|\v_{r}|}\frac{\dot{M}_{\rm d}}{\dot{M}_{\rm g}}=\frac{(1+\Lambda^2{\rm St}^2)\cdot 3\alpha h_{\rm g}^2/2}{\Lambda h_{\rm g}^2(2\Lambda C_{\eta}{\rm St}+3\alpha/2)}\frac{\dot{M}_{\rm d}}{\dot{M}_{\rm g}},
    Z = \frac{\Sigma_{\rm d}}{\Sigma_{\rm g}}=\frac{2\pi r|u_{\nu}|}{2\pi r|\v_{r}|}\frac{\dot{M}_{\rm d}}{\dot{M}_{\rm g}} \simeq  \frac{3\alpha h_{\rm g}^2/2}{h_{\rm g}^2(2 C_{\eta}{\rm St}+3\alpha/2)}\frac{\dot{M}_{\rm d}}{\dot{M}_{\rm g}},
\end{equation}
In a steady accretion disk region, $\dot{M}_{\rm d}/\dot{M}_{\rm g}$ is constant with $r$. When ${\rm St} \ll \alpha$, corresponding to the region at $ r < r_{\rm snow}$,
%%the solid and gas accretion rates should be steady or quasi-steady, value. Thus, in this case, the solid surface density should be proportional to gas surface density.
\begin{equation}
Z\simeq \dot{M}_{\rm d}/\dot{M}_{\rm g}.   
\label{eq:Zinside_r_snow}
\end{equation} 

% since $\dot{M}_{\rm d}/\dot{M}_{\rm g}$ should be constant, the solid-to-gas ratio is estimated as
% \begin{equation}    \frac{\Sigma_{\rm d}}{\Sigma_{\rm g}}\propto {\rm St}^{-1},\end{equation}
%\Oka{where we assume the $\Lambda$ and $C_{\eta}$ are almost constant in whole disk region and ${\rm St}^2\ll1$. 

On the other hand, for ${\rm St} \gg \alpha$ at $r > r_{\rm snow}$, %the snow line, 
$Z \propto {\rm St}^{-1}$. 
As Eqs.~(\ref{eq:frag}) and (\ref{eq:drift}) show, the drift limit is more stringent at relatively large $r$, where the dust density is lower and collisions are less frequent.
For $r > r_{\rm dri-fra}$
%where $r_{\rm dri-fra}$ is
(the transition radius between the drift and fragmentation limits),
${\rm St} \sim {\rm St_{drift}}/3 \propto Z \, T^{-1} r^{-1}$ (Eq.~(\ref{eq:drift}).
Deleting St from this relation and $Z \propto {\rm St}^{-1}$, we obtain

\begin{align}
Z(r) & \simeq Z(r_{\rm disk})
\left(\frac{T(r)}{T(r_{\rm disk})} \right)^{1/2}
\left(\frac{r}{r_{\rm disk}}\right)^{1/2}, 
% \\ Z(r) & \simeq Z(r_{\rm disk}) \left(\frac{T(r)}{T(r_{\rm disk})}\right) \hspace{1.5cm} [r < r_{\rm dri-fra}].
\end{align}
where $r_{\rm disk}$ is the disk radius.
In the outer disk region ($r>r_{\rm vis-irr}$), irradiation is dominant and $T\propto r^{-3/7}$. Thus, 
\begin{equation}
     Z = \frac{\Sigma_{\rm d}}{\Sigma_{\rm g}}\simeq Z(r_{\rm disk}) \cdot\left(\frac{r}{r_{\rm disk}}\right)^{2/7}.
\end{equation}
Substituting this equation into Eq.~(\ref{eq:drift}), we obtain
\begin{equation}
    {\rm St}_{\rm drift}= 0.13\left( \frac{Z(r_{\rm disk})}{10^{-3}} \right)
    \left( \frac{r}{1 \, \rm au} \right)^{-5/7}\left( \frac{r_{\rm disk}}{100 \, \rm au} \right)^{-2/7}
    \left( \frac{T}{150 \, \rm K} \right)^{-1}.
    \label{eq:drift2}
\end{equation}
Comparing this with Eq.~(\ref{eq:frag}), the transition radius $r_{\rm dri-fra}$ is given by
\begin{equation}
    r_{\rm dri-fra}=11\left( \frac{Z(r_{\rm disk})}{10^{-3}} \right)^{7/5}\left( \frac{r_{\rm disk}}{100 \, \rm au} \right)^{-2/5}\left( \frac{\v_{\rm frag}}{10 \, \rm m/s} \right)^{-14/5} \left( \frac{\alpha}{10^{-3}}\right)^{7/5}\,{\rm au}. 
    \label{eq:rdri_fra}
\end{equation}
%In summary, the solid-to-gas ratio beyond the snow line is given by $\Sigma_{\rm d}/\Sigma_{\rm g} \simeq Z(r_{\rm disk})\cdot (r_{\rm dri-fra}/r_{\rm disk})^{2/7}$ for $r >r_{\rm dri-fra}$, and $\simeq Z(r_{\rm disk}) \cdot (r_{\rm dri-fra}/r_{\rm disk})^{2/7} \cdot (T(r)/T(r_{\rm dri-fra}))$ for $ r_{\rm snow} < r < r_{\rm dri-fra}.$  
For $r < r_{\rm dri-fra}$, $Z \propto \rm St^{-1}$ with Eq.~(\ref{eq:frag}) shows
\begin{align}
Z(r) & \simeq Z(r_{\rm dri-fra})
\left(\frac{T(r)}{T(r_{\rm dri-fra})} \right).
\end{align}

Inside the snow line, $Z$
%the solid-to-gas ratio 
is constant with $r$ (Eq.~(\ref{eq:Zinside_r_snow})). Because $\Sigma_{\rm g} \propto r^{-1/2}$ in this region (Eqs.~(\ref{eq:Sigma_T}) and (\ref{eq:Tvissil})), $\Sigma_{\rm d} \propto r^{-1/2}$. 
The ratio of $\Sigma_{\rm d}$ just inside the snow line to just outside it is given by Eq.~(\ref{eq:jump_in_Sigma}).
Thus, in the steady accretion disk regions, %established,the solid surface density at $r$ is estimated as
\begin{equation}
    % Z(r) = \frac{\Sigma_{\rm d}(r)}{\Sigma_{\rm g}(r)} =\left\{
    Z(r) = \left\{
    \begin{array}{ll}
    \displaystyle{ Z(r_{\rm disk})\left(\frac{r}{r_{\rm disk}}\right)^{2/7} }
    & [r>r_{\rm dri-fra}] \\
    \displaystyle{ Z(r_{\rm dri-fra}) 
    \left(\frac{T(r)}{T(r_{\rm dri-fra})}\right)} & [r_{\rm snow} < r<r_{\rm dri-fra}]\\ 
    \displaystyle{ 
    \begin{split}
        &Z(r_{\rm snow}) \times 
        6 \left( \frac{\alpha}{10^{-3}}\right)^{-2} 
        \left( \frac{\v_{\rm frag,ice}}{10 \, \rm m/s} \right)^2\\
        &\times \left( \frac{T(r_{\rm snow})}{170 \, \rm K} \right)^{-1}
        \left(\frac{r}{r_{\rm snow}}\right)^{-1/2}  
    \end{split}
    } & [r < r_{\rm snow}]. \\ 
    \end{array}
    \right.
    \label{eq:Z}
\end{equation}
Here, $Z(r_{\rm dri-fra}) = Z(r_{\rm disk})(r_{\rm dri-fra}/r_{\rm disk})^{2/7}$
and 
$Z(r_{\rm snow}) = Z(r_{\rm dri-fra})(T(r_{\rm snow})/T(r_{\rm dri-fra}))$.
The steady accretion across the snow line would continue until $\Sigma_{\rm d}$ inside the snow line starts to decay ($t\sim t_{\rm pf}+t_{\rm drift,snow}\sim 0.45$ Myrs).

Equation~(\ref{eq:Z}) shows that once $Z(r_{\rm disk})$ is given, the $\Sigma_{\rm d}$ distribution is estimated ($
\Sigma_{\rm d} = Z \, \Sigma_{\rm g}; \: \Sigma_{\rm g}$ is given by the simulation parameter $\dot{M}_{\rm g}$).
Before the pebble formation front arrives at $r_{\rm disk}$, $Z(r_{\rm disk})$ is given by its initial value $Z_0 \sim 0.01$.
After that, $Z(r_{\rm disk})$ is rapidly reduced.
\citet{IdaYmaOku2019} derived the time evolution of $Z(r_{\rm disk})$ as
\begin{equation}
    \label{eq:Z_disk}
    Z(r_{\rm disk}) \sim Z_0 (1+t/t_{\rm pf})^{-\gamma},
\end{equation}
where
\begin{equation}
    \gamma\sim 1+0.15\,(300\,{\rm au}/r_{\rm disk}).
\end{equation}
The dashed lines in the upper panel of Fig.~\ref{fig:fidu} show the analytical estimate of $\Sigma_{\rm d}$ given by substituting Eq.~(\ref{eq:Z_disk}) into Eq.~(\ref{eq:Z}).
The $\Sigma_{\rm d}$ distribution at $\gtrsim r_{\rm snow}$ is well fitted by the analytical estimate for all of the four different times (0.2--1.5 Myr).
At $t = 0.45$ and 0.92 Myrs, 
it is fitted also in the region inside the snow line except for the innermost region.
For this duration, $\Sigma_{\rm d}$ is also decaying and the steady accretion is established.

\subsection{Important parameters for carbon depletion}

\subsubsection{The $\v_{\rm frag}$ change across the snow line}\label{subsec:v_frag}

Detailed descriptions of the evolution of particles in Sect.~\ref{sec:detailed_evol} strongly suggest that the conditions for the tanh-like pattern of $f_{\rm c}$ are an enhanced jump in $\Sigma_{\rm d}$ at $\sim r_{\rm snow}$ and early %significant 
decay of icy pebble flux 
% cay of 
that reduces the supply of fresh amorphous hydrocarbons to the region inside $r_{\rm snow}$.
The both features are originated from $\v_{\rm frag,sil} < \v_{\rm frag,ice}$ such that ${\rm St} \sim \alpha$ at $r < r_{\rm snow}$ and ${\rm St} \gg \alpha$ at $r > r_{\rm snow}$.

The decay of icy pebble flux starts at $t \sim t_{\rm pf} + t_{\rm drift,ice}$
where $t_{\rm pf}$ and $t_{\rm drift,ice}$ are given by Eqs.~(\ref{eq:tpf}) and~(\ref{eq:tdrift}).
For $Z_0 = 0.01$ and $r_{\rm snow}=3.7\,\rm au$,
we have\begin{align} 
& t_{\rm pf} \sim 
2 \times 10^5 
\left(\frac{r_{\rm disk}}{100\, \rm au}\right)^{3/2} \:{\rm years},\label{eq:tpf2} \\
& t_{\rm drift,ice} \sim 2.5 \times 10^5\left( \frac{\alpha}{10^{-3}} \right) \left( \frac{10\,{\rm m/s}}{\v_{\rm frag,ice}}\right)^{-2}
\left(\frac{r_{\rm disk}}{100\, \rm au}\right)^{1/2} \:{\rm years}.  
    \label{eq:tdrift2}
\end{align}
The drift timescale is shorter for larger $\v_{\rm frag,ice}$.
For the early decay of the supply of fresh amorphous hydrocarbons to occur,
$t_{\rm drift,ice}$ must be short enough compared with global disk gas depletion timescale $t_{\rm disk} \sim 3 \times 10^6(\alpha/10^{-3})^{-1}(r_{\rm disk}/100\, \rm au)$ yrs, which is equivalent to 
% necessarily occurs during the presence of the disk gas. As shown in Eq.~(\ref{eq:tdrift}), the decay of the supply of fresh amorphous hydrocarbons delays from the pebble formation decay by the timescale of $t_{\rm drift,ice}$. The timing is delayed more for smaller $\v_{\rm frag,ice}$.
\begin{equation}
\label{eq:decay_limit}
 \v_{\rm frag,ice} \ga 3.2 
 \left( \frac{\alpha}{10^{-3}} \right) 
 \left( \frac{r_{\rm disk}}{100\,{\rm au}} \right)^{-1/4} \rm m/s.
\end{equation}

\begin{figure}
    \begin{tabular}{c}
         \begin{minipage}[t]{\hsize}
             \centering
             \includegraphics[keepaspectratio,scale=0.23]{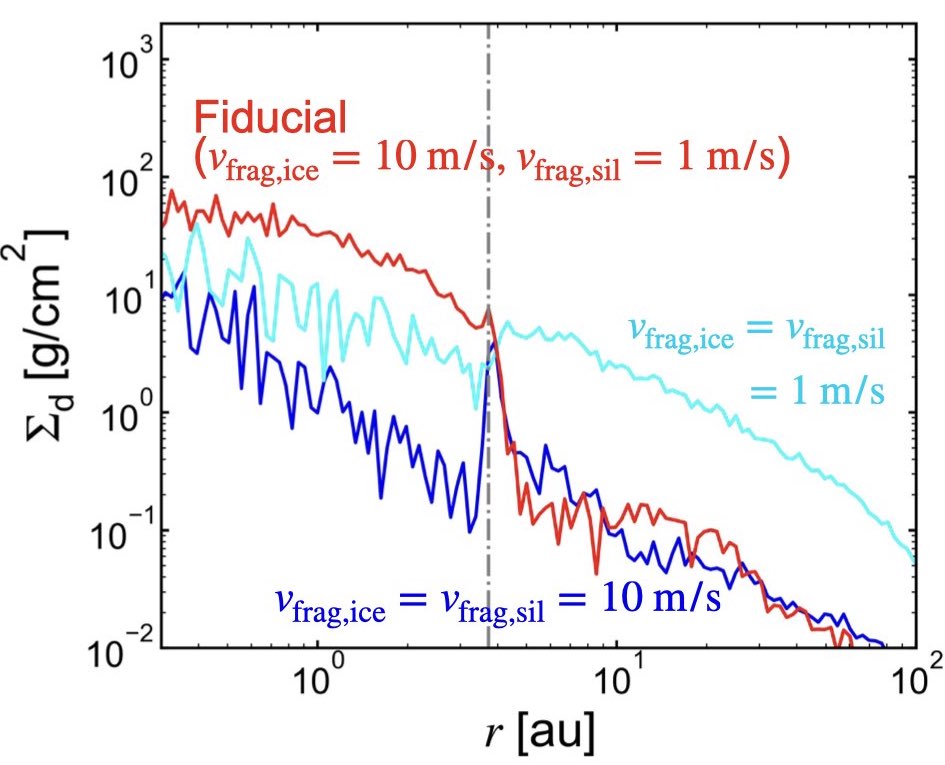}
         \end{minipage}\\
         \begin{minipage}[t]{\hsize}
             \centering
             \includegraphics[keepaspectratio,scale=0.23]{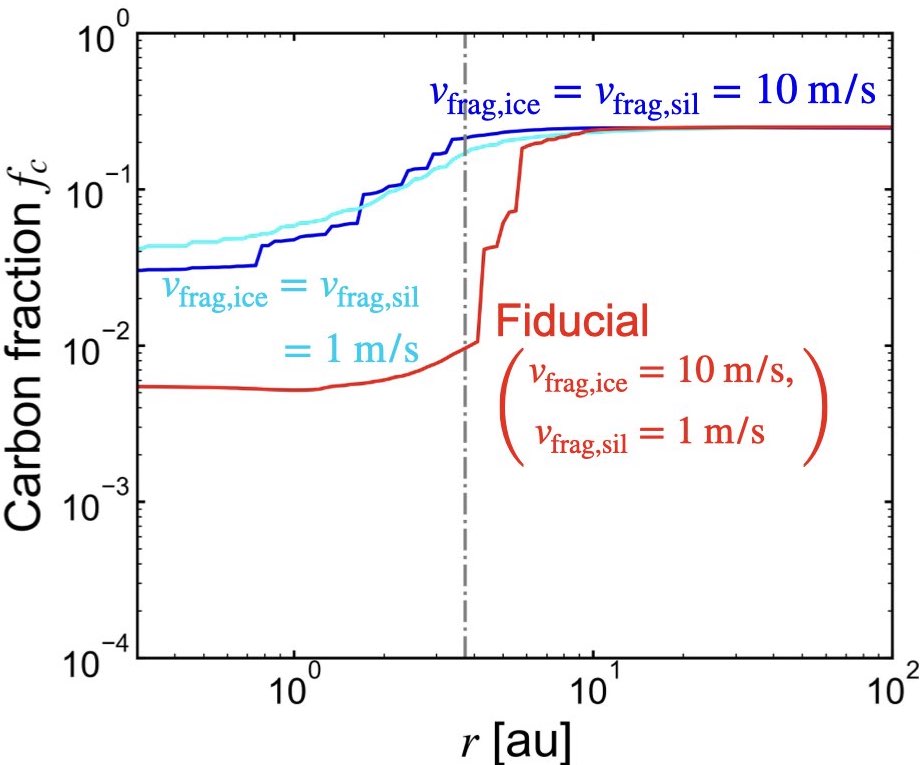}
         \end{minipage}
    \end{tabular}
    \caption{%\ida{g/cm$^2$ must NOT be Italic.}
    Distributions of $\Sigma_{\rm d}$ and $f_{\rm c}$ at 0.68 Myrs for $\v_{\rm frag,ice} = \v_{\rm frag,sil} =1 \, \rm m/s$ and $10\,\rm m/s$ and the fiducial case ($\v_{\rm frag,ice} = 10\,\rm m/s$ and $\v_{\rm frag,sil} =1 \, \rm m/s$).}
    % \ida{Is this run different from the fiducial run? If not, you should make clear.}}}
    \label{fig:nopile}
\end{figure}

Here, we confirm the above conditions by performing the runs with different combinations of $\v_{\rm frag,ice}$ and $\v_{\rm frag,sil}$.
Figure~\ref{fig:nopile} shows $\Sigma_{\rm d}$ and $f_{\rm c}$ at 0.68 Myr
%the solid surface densities and the carbon fractions 
for $\v_{\rm frag,ice} = \v_{\rm frag,sil} =10\,\rm m/s$ (the blue line), $\v_{\rm frag,ice} = \v_{\rm frag,sil} =1 \,\rm m/s$ (the cyan line), and the fiducial case
(the red line) at 0.68 Myr, which is the intermediate time of the third stage ($\sim t_{\rm pf}+t_{\rm drift,ice}+0.5t_{\rm diff,snow}$). 
To highlight the $\v_{\rm frag}$ effect,
the shadow effect is applied even in the case of $\v_{\rm frag,ice} = \v_{\rm frag,sil}$.
%When it is same the fragmentation velocities for both silicate and ice, the shadow area behind the snow line should not appear as the silicate particles are not piled up as shown in the top panel of Fig.~\ref{fig:nopile}. However, in this section, in order to show the effect of the different fragmentation velocities clearly, we show the result in the case that we assumed the existence of the shadow area despite the value of fragmentation velocities. We show 
The results without the shadow effect are shown in Appendix~\ref{app:same}.

In the run with $\v_{\rm frag,ice} = \v_{\rm frag,sil} =1 \,\rm m/s$, for both silicate particles and ice pebbles, ${\rm St} = 2.1\times 10^{-4}$ at the snow line. Because ${\rm St} < \alpha$, silicate particles do not accumulate inside the snow line and $f_{\rm c}$ 
%carbon fraction 
decreases only slightly 
%with the decrease of $r$ gently 
inside $r_{\rm snow}$.
%the snow line.}

When $\v_{\rm frag,ice} = \v_{\rm frag,sil} =10$ m/s, for %Stokes numbers of 
both silicate particles and ice pebbles, ${\rm St} = 2.1\times 10^{-2}$ at the snow line, and
%, and St is grater than $\alpha$ in the entire disk region. In this situation, 
particles are also not piled up except around the snow line.
%as shown by the blue line in the top panel of Fig.~\ref{fig:nopile}. 
Since the pile-up around the snow line is caused by recondensation of water vapor diffusing out from inside the snow line, it does not significantly influence the motions of the silicate particles, and $f_{\rm c}$ decreases only slightly as well.
%As shown by the blue line in Fig.~\ref{fig:nopile}(b), carbon fraction also decreased with the decrease of $r$ gently.}

On the other hand, 
when $\v_{\rm frag,ice} = 10 \,\rm m/s$ and $\v_{\rm frag,sil} =1 \,\rm m/s$, ${\rm St} = 2.1\times 10^{-4} < \alpha$ for silicate particles and ${\rm St} = 2.1\times 10^{-2} > \alpha$. Accordingly, the pile-up occurs and $f_{\rm c}$ significantly decreases with the tanh-like shape. 

\subsubsection{The vertical temperature profile}\label{subsec:z_temp}

\begin{figure}
    \begin{tabular}{c}
        \begin{minipage}[t]{\hsize}
             \centering
             \includegraphics[keepaspectratio,scale=0.35]{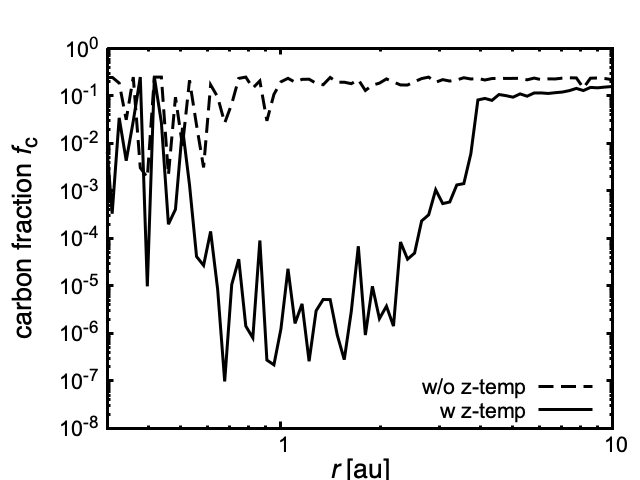}
         \end{minipage}\\
        \begin{minipage}[t]{\hsize}
             \centering
             \includegraphics[keepaspectratio,scale=0.35]{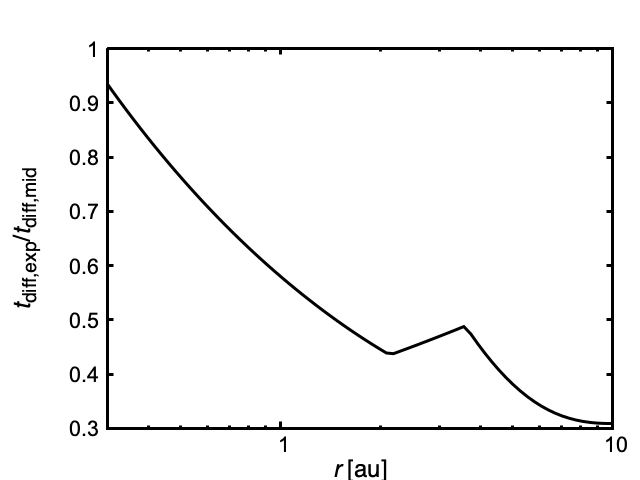}
         \end{minipage}
    \end{tabular}
    \caption{Carbon fraction at 0.1 Myr with and without vertical temperature profile (upper panel) and ratio of vertical diffusion timescale at FUV-exposed layer to mid-plane (lower panel).
    In these runs, the radial drift is neglected to highlight the effect of the vertical temperature profile.} 
    %\ida{The $r$ range should be the same for the upper and lower panels. Otherwise, readers are confused.}}
    \label{fig:z_diff}
\end{figure}

%In this section, we discuss the influence of the vertical temperature profile. 
The upper panel of Fig.~\ref{fig:z_diff} shows $f_{\rm c}$
%carbon fraction 
at 0.1 Myr with and without vertical temperature profile (see Fig.~\ref{fig:T_z}).
To highlight this effect, 
we stop the radial motion of particles. The carbon fraction $f_{\rm c}$ is the minimum at $r \sim 1-2 \, \rm au$. 
%than the inner disk region. This is
% \ida{I don't understand what you mean$\rightarrow$} because the particle at the higher layer can diffuse the upper layer. 
The lower panel
%Figure~\ref{fig:z_diff}(b) 
shows the ratio of vertical diffusion timescale in the FUV-exposed layer to that at the mid-plane. The ratio
%at the FUV-exposed layer 
is the smallest at $r \sim 2 \,\rm au$ inside $r_{\rm snow}$.
%around 2 au than the inner disk region. 
Because the temperature is significantly higher in the upper layer than around the mid-plane at $\sim 1$ au, the diffusion there leads to   
%Therefore, 
more photodegradation of amorphous hydrocarbons.
%are at the higher layer. 
The carbon fraction outside the snow line ($\sim 3.7$ au) does not decrease
%. This is 
because the icy pebbles are not particularly stirred up. While the radial drift is neglected
in Fig.~\ref{fig:z_diff}, a significant impact of the vertical temperature profile on the carbon depletion is also shown
in Fig.~\ref{fig:noshadow}, where the radial drift is included.

\subsubsection{Other parameter dependencies}

% \ida{Because the main points are Sects.~3.3.1 and 3.3.2, this section should be simpler (ASSARI).}

%In this section, 
We also carried out runs with a different value for each parameter from the fiducial run, $\alpha=5\times 10^{-4}, \v_{\rm frag,sil}=0.1 \,\rm m/s$, and $ Z_0=0.05$.
In these cases, 
% will show the results of other disk parameter sets in case 
${\rm St}\la \alpha$ inside $r_{\rm snow}$ and 
%the snow line and 
${\rm St}\gg \alpha$ outside it, which is equivalent to the two conditions of the pile-up of silicate particles and earlier decay of icy pebble flux are satisfied.
% the snow line. As discussed in Section~\ref{sec:detailed_evol}, the carbon fraction dropping-off requires the conditions of: 1) the lower fragmentation velocity in the inner disk region and 2) rapidly decay of the icy pebble flux. The former results in the high solid surface density inside the snow line, and the latter results in the low solid surface density outside the snow line.
%\ida{Why at 0.7 Myrsbut nit 0.68 Myr?}
%shows the carbon fraction for different parameter sets at 0.7 Myr, when all results are in the third stage (Section~\ref{sec:3nd_stage}). We chose the parameter sets which satisfy ${\rm St}<\alpha$ inside the snow line and ${\rm St}>\alpha$ outside it. The detail results for each parameter are shown in Appendix.~\ref{app:para}. The carbon fractions show the step-function like structure 
In these cases, the tanh-like shaped $f_{\rm c}$ pattern is produced with small variations in the bottom values (Fig.~\ref{fig:para_car}), as expected. The detail results are shown in Appendix~\ref{app:para}.

The only exception is 
% except for 
the case with $\dot{M}_{\rm g} = 10^{-9}\,M_{\odot}/{\rm yr}$, where the disk opacity is significantly lower.
%Even when
As the silicate particles drift to inner regions, the carbon destruction proceeds to result in qualitatively different $f_{\rm c}$ shape. 
% as shown in Fig.~\ref{fig:height}. As discussed in Section~\ref{sec:3nd_stage}, the flat distribution inside the snow line requires the higher opacity in this region. \Oka{Figure~\ref{fig:para} shows time evolution of the lowest carbon fraction for the case that the carbon distribution shows the hyperbolic tangent structure. For all cases, the carbon fraction starts decreasing after the icy pebble flux decaying stage (Section~\ref{sec:2nd_stage}). For the lower turbulence case ($\alpha=5\times 10^{-4}$), the carbon fraction is higher than the other cases because less particles can be lifted up to the FUV-exposed layer by turbulence. The minimum values do not change after the second stage except for the case with the lower fragmentation velocity for silicates. In this case, as there are more small particles inside the snow line, more small amorphous hydrocarbons can experience photolysis. Therefore, the carbon fraction decreases more with the decrease of the solid surface density inside the snow line. In the other cases, although photolysis should be also efficient with the decrease of the solid surface density, the mass fraction of amorphous hydrocarbon coming from outside the snow line increases at the same time. As these effects cancel each other, the minimum values become almost steady.}

%\Oka{As shown in Figs.~\ref{fig:para_car} and ~\ref{fig:para}, the carbon fraction depends on the fragmentation velocity of silicates and the disk parameters, especially the gas accretion rate strongly.}

\begin{figure}
    \centering
    \includegraphics[keepaspectratio,scale=0.23]{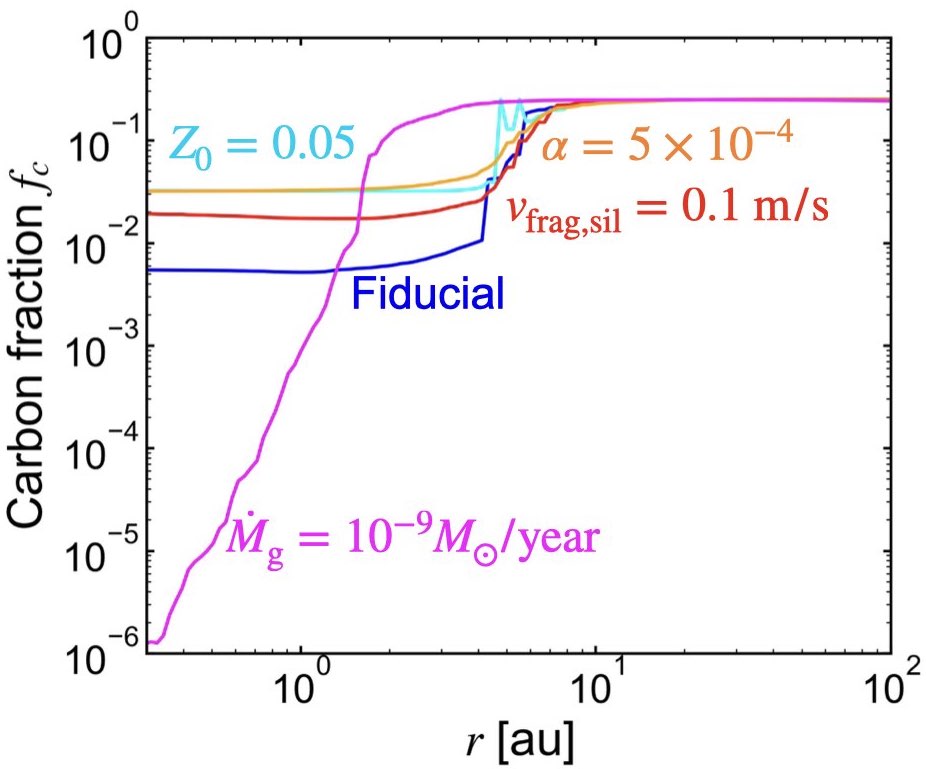}
    \caption{Carbon fraction at 0.68 Myrs for different parameter sets.}
    \label{fig:para_car}
\end{figure}
 
% \begin{table*}
% \caption{Minimum carbon fraction in 1.5 Myr.}
% \begin{center}
% \begin{tabular}{llc}\hline \hline
% Variation & Value & minimum $f_{\rm c}$ \\ \hline
% Fiducial &  & $4.9\times 10^{-3}$ \\
% lower turbulence &  $\alpha=5\times 10^{-4}$ & $0.03$ \\
% % Low gas accretion rate & $\dot{M}_{\rm g}=10^{-9}\,M_{\odot}/{\rm yr}$ & $3.1\times 10^{-4}$ \\
% High initial solid-to-gas ratio &  $Z_0=0.05$ & $0.01$ \\
% Low fragmentation velocity for silicate & $\v_{\rm frag,sil}=0.1\,{\rm m/s}$ & $5.8\times 10^{-4}$ \\
% \hline
% \end{tabular}
% \end{center} 
% \label{tab:para}
% \end{table*}
% \vspace{1em}

% \begin{figure}
%     \centering
%     \includegraphics[keepaspectratio,scale=0.2]{photolysis_param_2024_07_23.jpeg}
%     \caption{Time evolution of the minimum carbon fraction.
%     \ida{This figure is not referred to in the main text.}}
%     \label{fig:para}
% \end{figure}

\section{Discussion}\label{sec:dis}
% \ida{I moved only 5 lines in Sect. of ``Re-condensation of water vapor on the particles' surface" to Sect. 2.3 and the other part was commented out.} 
%\subsection{Re-condensation of water vapor on the particles' surface} We assumed water vapor re-condenses on the surfaces of both ice and silicate dust particles. However, \citet{Ros2019} suggested water vapor tends to re-condense on the icy pebbles' surface rather than the silicate dust surface. While this fact could change our results, it would not change the fast decaying of the icy pebble flux. Despite that \citet{Okamoto2022} did not consider the re-condensation of water vapor on the dust surface, we showed the ratio of dust coming from outside the snow line to dust existing inside the snow line decrease in case of the fast decaying of icy pebbles flux. Thus, the carbon fraction inside the snow line might decay if water vapor does not re-condense on the silicate dust surface as long as most icy pebbles drift inward rapidly.
\subsection{Pyrolysis of complex organics}
\label{subsec:pyrolysis}
\begin{figure*}
    \centering
    \includegraphics[keepaspectratio,scale=0.25]{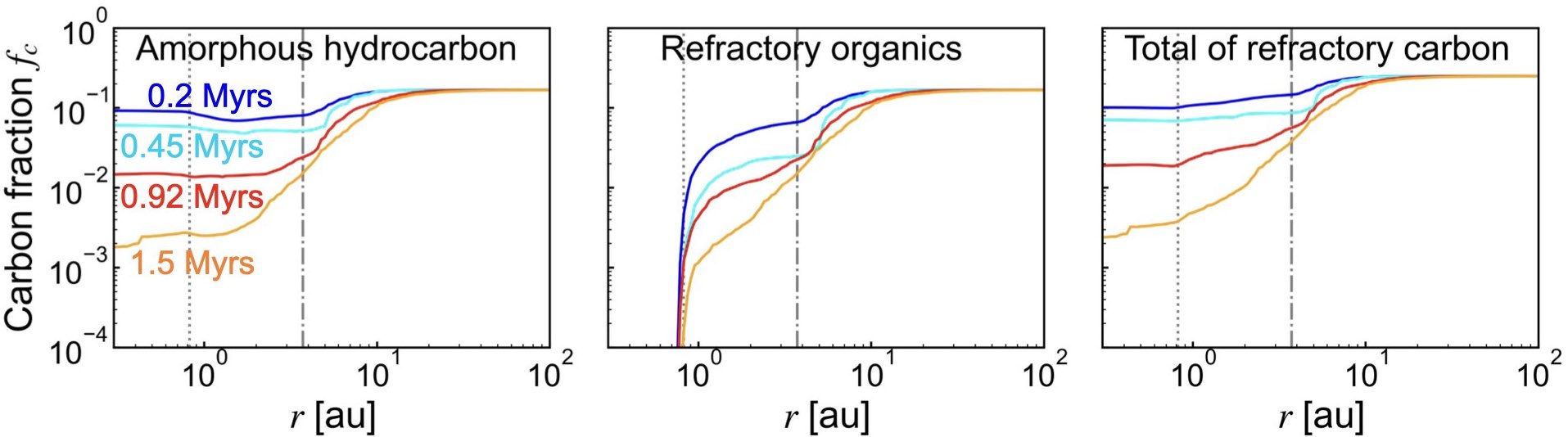}
    \caption{Carbon fraction, $f_{\rm c}$, with a mixture of complex organics with $T_{\rm des} = 540\:\rm K$ (50 wt.\%) and amorphous hydrocarbons (50 wt.\%). The left figure shows the mass fraction of amorphous hydrocarbons, the middle panel shows that of complex organics, and the right panel shows the total carbon fraction. The gray dash-dotted and dotted lines snow the snow and pyrolysis lines.}
    \label{fig:pyl}
\end{figure*}

% \ida{I deleted the arguments here that duplicated with those in Introduction.}
% \citet{Li2021} suggested the pyrolysis of complex organics could explain the earth's carbon depletion. \citet[]{Binkert2023} considered pyrolysis of complex organics such as Kerogen and suggested the FU-Ori type outburst made most of the organics are converted to gas at 1 au. However, as argued in Section~\ref{sec:intro}, some products like amorphous hydrocarbons may still remain after pyrolysis \citep{Chyba1990}. Therefore, photolysis of amorphous hydrocarbon should be as important as pyrolysis for carbon abundance in the solar system.
So far we have not included complex organics, because it is likely that some fraction of the products of their pyrolysis are amorphous hydrocarbons (Sect.~\ref{sec:intro}). 
In Fig.~\ref{fig:pyl}, we show the result with initial solid carbon of a mixture of complex organics with $T_{\rm des} = 540\:\rm K$ (50 wt.\%) and amorphous hydrocarbons (50 wt.\%). 
The simulation parameters are the same as the fiducial case.
%Pyrolysis can change our result inside the snow line. Pyrolysis occurs in the hotter disk region ($T>500\:\rm K$). This reaction can 
The pyrolysis of complex organics occurs even near the disk midplane. 
In the runs in Fig.~\ref{fig:pyl}, however, we still assume that half the mass of the complex organics is converted into amorphous hydrocarbons through this reaction.
Even though at 0.92 Myr, the total carbon fraction \(f_{\rm c}\) inside the pyrolysis line shows a slightly lower value than outside of it, the difference is not significant enough to explain the discrepancy between chondrites and the bulk Earth. 

The initial ratio of complex organics and amorphous hydrocarbons and the 
conversion rate of complex organics to amorphous hydrocarbons through pyrolysis remain unclear. Adjusting these values could allow the \(f_{\rm c}\) pattern to better reproduce the lower \(f_{\rm c}\) observed in the bulk Earth compared to chondrites. 
% \ida{Add discussion on the $f_{\rm c}$ pattern such as ``It can explain the lower carbon fraction of bulk earth than chondrites ...." }

\subsection{FUV radiation outside the snow line}\label{subsec:out}

We assumed carbon destruction does not occur outside the snow line because FUV radiation from the host star is hindered by piled-up dust inside the snow line. However, the ``shadow area'' may only extend up to $\sim 10$ au \citep{Ohno2021}. 
% The disk is also exposed to interstellar FUV radiation. From the results of our calculation, \Oka{the opacity for FUV is lower outside the snow line than inside the snow line}. Thus, more amorphous hydrocarbons outside the snow line could be destroyed by the radiation. 

However, this could not be effective enough to erase the dichotomy in $f_{\rm c}$ 
%of the carbon fraction 
between the inner and outer disk regions. As shown in Fig.~\ref{fig:noshadow}, the $f_{\rm c}$ distribution shows the tanh-like pattern also in the case 
%hydrophobic tangent function-like structure 
without the shadow area. 
This result suggests that the limited shadow region and the interstellar FUV radiation do not break the higher carbon fraction at the outer disk region than the inner disk region. However, as discussed in Sect~\ref{subsec:obs}, the result without the shadow area may be inconsistent with the observation of the comets. 
% \Oka{when the interstellar FUV radiation is weaker than the solar FUV radiation.}%
% That is because interstellar radiation is much weaker and not more effective than stellar radiation. Even if we include the carbon destruction outside the shadow area, the carbon would not be destroyed efficiently since the stellar FUV radiation becomes weaker in power-law $\propto r^{-2}$.

\subsection{Oxidation}
We assumed amorphous hydrocarbon is oxidized only if the dust exists in the FUV-exposed layer ($\tau\lesssim 1)$. However, a few oxygen atoms might exist in the lower layer \citep[e.g.,][]{Lee2010,Anderson2017}. Therefore, oxidation in the lower layer might reduce the carbon abundance.

Moreover, the carbon particle might be oxidized by OH radicals \citep[e.g.,][]{Finocchi1997,Gail2017}. OH radicals are thought to be formed by thermal dissociation of ${\rm H_2O}$ at the hotter region ($T\gtrsim 1100\,{\rm K}$) \citep[e.g.,][]{Gail2017}. The abundance of OH radicals in the inner disk region has some uncertainty as \cite{Finocchi1997} estimated $10^{-10}-10^{-7}$ relative to ${\rm H_2}$ and \cite{Gail2017} calculated $10^{-12}-10^{-9}$ relative to ${\rm H_2}$ at $T\gtrsim 1100\,{\rm K}$. Although the OH abundance is not expected to be high, oxidation by OH might also reduce more carbon particles \citep{Gail2017}. The calculation with a more accurate estimation of OH radical and oxygen atom abundances is left to a future study.

\subsection{Other processes: Gap or pressure bump in the gas disk}
\cite{Klarmann2018} concluded that carbon depletion needs some mechanisms that stop the inward solid particles' flux, such as a gap or a pressure bump in the gas disk. 
However, the significant reduction in  
%as stopping 
the inward solid particles' flux 
by the gap or pressure bump should lower
%can make lower 
$\Sigma_{\rm d}$ and the disk optical depth, the flat $f_{\rm c}$ pattern would not appear (Fig.~\ref{fig:para_car}).

\cite{Klarmann2018} calculated the evolution of $\Sigma_{\rm d}$ with single-sized particles.
The different sizes between icy pebbles and silicate particles released at the snow line cause pile-up of the silicate particles without the gap or pressure bump \citep{Saito2011, Ida&Guillot2016,2017Schoonenberg,Ida2021}.
In this paper, we have shown that this effect can result in significant carbon depletion. 

%in the disk with the gap or the pressure bump.
%out considering the different stickiness between silicate and icy particles. However, we suggest this difference can contribute to carbon depletion of the inner solar system without a gas gap or pressure bump.

% \Oka{The dropping off of the carbon fraction is caused by the higher solid surface density inside the snow line. If the fragmentation velocity for icy particles is the same as that for silicate particles, and there is a gas gap, the solid surface density inside the gap should be lower than outside the gap. In this case, photolysis occurs efficiently inside the gap. The carbon fraction should be dependent on $r$ strongly similarly to the case with low gas accretion rate shown in Fig.~\ref{fig:para_car}.}

% The important point is the carbon abundance of CC is similar to that of a matrix in NC (Figure~\ref{fig:obs}) while there is an isotope dichotomy between these chondrites \citep[e.g.,][]{Budde2016}. Thus, there must have been a different mechanism that caused the carbon depletion than that caused the isotope dichotomy.

\subsection{Comparison to the observational data}
\label{subsec:obs}

\begin{figure}
    \includegraphics[keepaspectratio,scale=0.4]{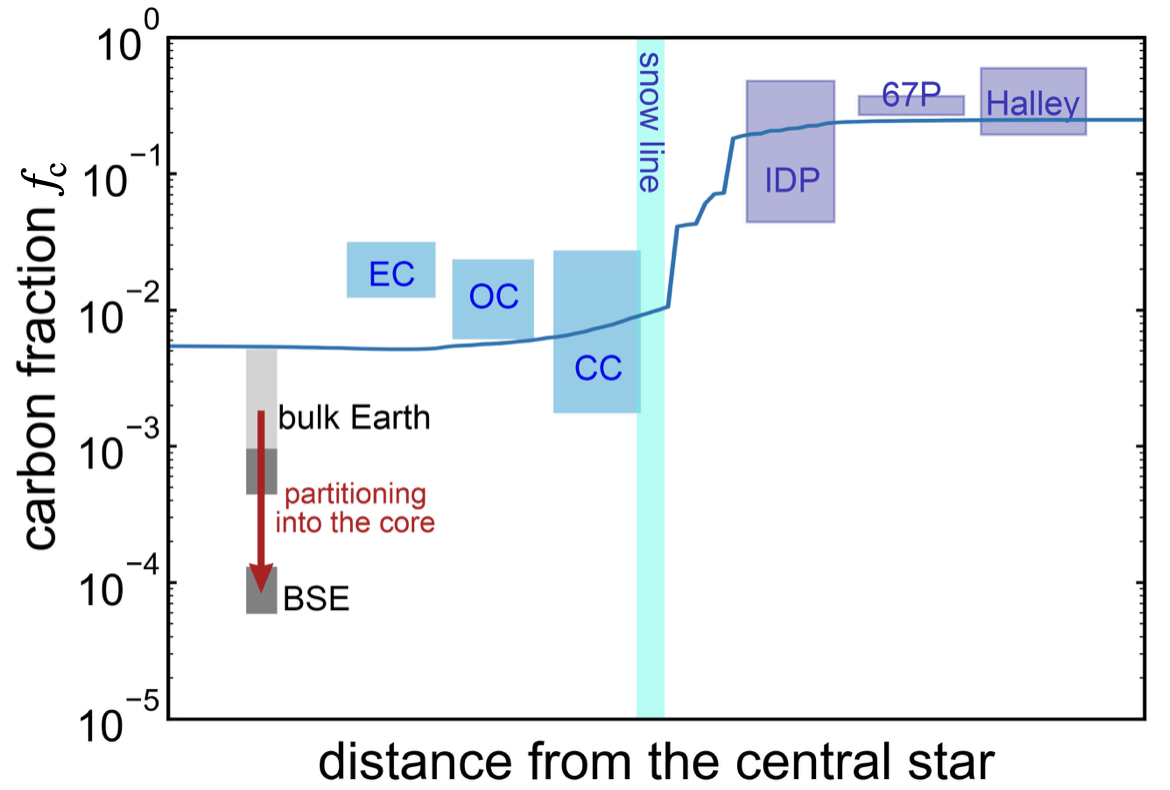}
    \caption{Expect the formation region for each rocky body based on our calculation results.}
    \label{fig:obs_ex}
\end{figure}

It is shown that chondrite meteorites in the matrix scaling universally show $f_{\rm c} \sim 0.01$
(Fig.~\ref{fig:obs} and \citet{Alexander2007}).
Our simulations show a flat bottom value of the tanh-like shape as
$f_{\rm c} \sim 0.02$--0.03 inside the snow line, which is consistent with the chondrite data, if their parent bodies were formed inside the snow line (Fig. ~\ref{fig:obs_ex}). 
The further smaller $f_{\rm c}$ of the bulk Earth could be potentially accounted for by including complex organics as well as amorphous hydrocarbons and some adjustments of their initial abundance ratio and conversion rate from complex organics to amorphous hydrocarbons through pyrolysis (Sect.~\ref{subsec:pyrolysis}).
Comets have $f_{\rm c}$ as high as ISM, 
which is also consistent with the tanh-like pattern that our calculations produce (see also Sect.~\ref{subsec:out}).

As discussed in Sect.~\ref{sec:intro}, 
the observed $f_{\rm c}$ on the WD photo-spheres, which should reflect the bulk compositions of asteroids or rocky planets around the WDs, is distributed from the solar value far down to
the bulk Earth value.
We found that the conditions to produce the tanh-like depletion pattern of solid carbon are not too severe, but may be sensitive to details of dust particle properties (and possibly to the disk parameters).  
This may produce a diversity of $f_{\rm c}$ values in a range of a few orders of magnitude.
It could be responsible for the observed large variety of C/Si ratios on the WD photo-spheres.
More detailed comparisons are left for future works.

%photolysis should be suppressed at the comet formation region. In this study, we consider the shadow area behind the snow line. However, \cite{Ohno2021} showed that the shadow area spreads only by 10 au. The higher carbon fraction of the comets could be explain by considering the volatile carbonaceous molecules, such as carbon monoxide and methane gases. We did not calculate the motion of these gases released by photolysis and/or oxidation. The comet formation region may be cold enough for these gases to condense on the solid surface. Therefore, the higher carbon abundance of the comets could be reproduced by condensation of these gases.

\section{Conclusion}\label{sec:con}
Rocky bodies in the inner Solar System are depleted in carbon compared to the Sun, by one to several orders of magnitude, although most solid carbonaceous components should be refractory in the ISM and the comets. Possible mechanisms to destruct refractory carbonaceous components are photolysis \citep{Alata2014,Alata2015} 
and oxidation for highly refractory carbons such as amorphous hydrocarbons \citep[e.g.,][]{Draine1979} 
in the upper FUV-exposed layer and pyrolysis for modestly complex organics \citep[e.g.,][]{Chyba1990}. 
% \cite{Lee2010} suggested that there were enough oxygen atoms to oxidize the carbon particle in the FUV-exposed layer of the disk and \cite{Anderson2017} suggested refractory carbon particles could be destructed enough in the upper layer of the disk. 
However, previous theoretical models \citep{Klarmann2018,Binkert2023} did not succeed in reproducing the observed significant carbon depletion in the case of highly refractory carbons.

% suggested that the carbon abundance of the disk could not decline due to vertical and radial dust transport. Moreover, \cite{Binkert2023} suggested carbon abundance of the inner solar system could be explained only if there was a FU-Ori type outburst event and complex organics completely sublimated into the gas phase. These previous studies did not consider the difference in stickiness between icy and silicate particles. 

We  performed a 3D Monte Carlo simulation to track superparticles trajectories in a steady accretion gas disk, 
taking account of the effects that were not considered in the previous models.
We found that the carbon fraction, $f_{\rm c}$, is lower by two orders of magnitude inside the snow line when the following conditions are present. 
\begin{enumerate}
    \item The fragmentation limit velocity of silicate particles is lower than that of icy pebbles so that ${\rm St} \lesssim \alpha$ for silicate particles and ${\rm St} \gg \alpha$ for icy pebbles.
    \item The gas temperature is considerably higher in the upper optically thin layer than near the mid-plane.
\end{enumerate}
The first condition leads to a pile-up of silicate particles inside the snow line and earlier decay of icy pebble flux incoming to the snow line than the depletion of the silicate particles.
The latter is shown by globally calculating the growth and radial drift of pebbles in the disk.
The second condition leads to more effective vertical diffusion in the upper optically thin layer. Because the FUV-exposed layer is as high as $\sim 4 \, H_{\rm g}$, the effective vertical diffusion can significantly enhance photolysis.

The matrix-scaled $f_{\rm c}$ is similar (one to two orders of magnitude depletion) among carbonaceous, ordinary, and enstatite chondrites, while the orbital radii of their parent bodies may be diverse. This feature is well reproduced by our simulations with highly refractory carbon (amorphous hydrocarbons).
The bulk Earth could have further lower $f_{\rm c}$
(three to four orders of magnitude depletion).
 This feature might also be reproduced if the refractory carbon source is a mixture of amorphous hydrocarbons and (modestly) complex organics. %\ida{$\leftarrow$ OK?}}  

The first and second conditions are not too severe, but still sensitive and their effects could be responsible for the observed large variety of C/Si on white dwarf photo-spheres, which would be polluted by rocky bodies that existed around the white dwarfs. Theoretical investigations of the carbon deletion problem in the inner Solar System can offer insights into the origin of the large variety of C/Si in rocky bodies in exoplanetary systems.    
If the  $f_{\rm c}$ values of Earth-size planets in exoplanetary habitable zones have a variation by a few (or more) orders of magnitude, the surface environments (and even interiors) of these planets should exhibit a broad scope of diversity. More studies are needed to explore these issues in depth.

\begin{acknowledgement}
We thank the referee for his insightful and encouraging comments. We also thank  
Shogo Tachibana, Hideko Nomura, and Chris Ormel for helpful discussions.
This work was supported by JST Tokyo-Tech SPRING, Grant Number JPMJSP2106, and JSPS Kakenhi 23KJ0885 and 21H04512.
\end{acknowledgement}

\bibliography{MC_carbon}
\bibliographystyle{aa}

%\newpage
\begin{appendix}

\section{Tuning for a higher resolution of smaller particles}\label{app:tune}

Small particles tend to be lifted up to the higher layer more easily.
As discussed in Sect.~\ref{subsec:opa}, the mass fraction of particles smaller than 0.1 $\rm \mu m$ is given by $({\rm 0.1\:\mu m}/s_{\rm max})^{0.5}$. If $s_{\rm max}>$ 1 mm, their mass fraction is less than 1\%. We assume the super-particles as aggregations of solid particles with a single size and that all super-particles have equal mass. When the mass fraction of the particles with $s\le 0.1\rm \:\mu m$ is less than 1\%, the number of super-particles should be higher than 100 to ensure the presence of the small particles. However, as shown in Eq.~(\ref{eq:N_par}), the typical number of super-particles inside 1 au is less than 100. Thus, the probability of the super-particles with $s\le 0.1\:\mu m$ existing inside 1 au is very low.

To adjust the influence of small particles, we add `small' super-particles with a fixed size of 0.1 $\rm \mu m$. These are used only for the calculation of $f_{\rm c}$, and the $\Sigma_{\rm d}$ is calculated by the number of the `large' super-particles alone.

In each timestep, we calculate the mass-averaged carbon fraction between the small and large super-particles given by 
\begin{equation}
    f_{\rm c}=f_{\rm c,s}\left(\frac{0.1\:\rm \mu m}{s_{\rm max}}\right)^{0.5}+f_{\rm c,l}\left(1-\left(\frac{0.1\:\rm \mu m}{s_{\rm max}}\right)^{0.5}\right),
\end{equation}
where $f_{\rm c,s}$ and $f_{\rm c,l}$ are the mass-averaged carbon fractions for small and large super-particles, respectively. We consider the redistribution of carbon between the small and large particles through coagulation and fragmentation of solids and assume that all super-particles in each bin have the average carbon fraction in the next timestep.

\section{The cases of $v_{\rm frag,sil} = v_{\rm frag,ice}$ without the shadow area}\label{app:same}

\begin{figure}[htbp]
    \centering
    \includegraphics[keepaspectratio,scale=0.25]{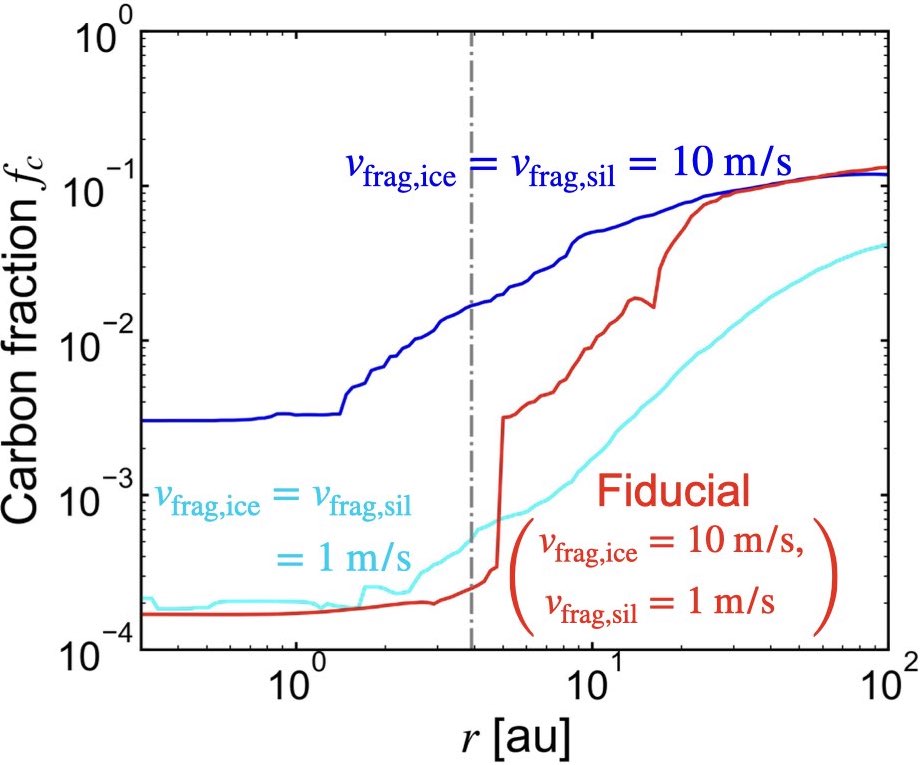}
    \caption{Same as Fig.~\ref{fig:nopile} but without the shadow area behind the snow line.}
    \label{fig:app_vfrag}
\end{figure}

As argued in Sect.~\ref{subsec:v_frag}, for $v_{\rm frag,sil} = v_{\rm frag,ice}$, the shadow area should not appear behind the snow line.
%when the fragmentation velocity for silicate is the same as that of ice. 
Figure~\ref{fig:app_vfrag} shows $f_{\rm c}$ at 0.68 Myr without the shadow area.
%for the different fragmentation velocities. 
The different fragmentation velocities between icy and silicate particles make a larger dichotomy in $f_{\rm c}$ between the inner and outer disk region than the other cases even if we do not consider the shadow area. The flat region in the same fragmentation velocity cases might be produced by the inefficiency of photolysis there as shown in Fig.~\ref{fig:z_diff}.

\section{Results with different parameter sets}\label{app:para}

\begin{figure*}[htbp]
    \begin{tabular}{cccc}
        \begin{minipage}[t]{0.22\textwidth}
            \centering
            \includegraphics[keepaspectratio,scale=0.13]{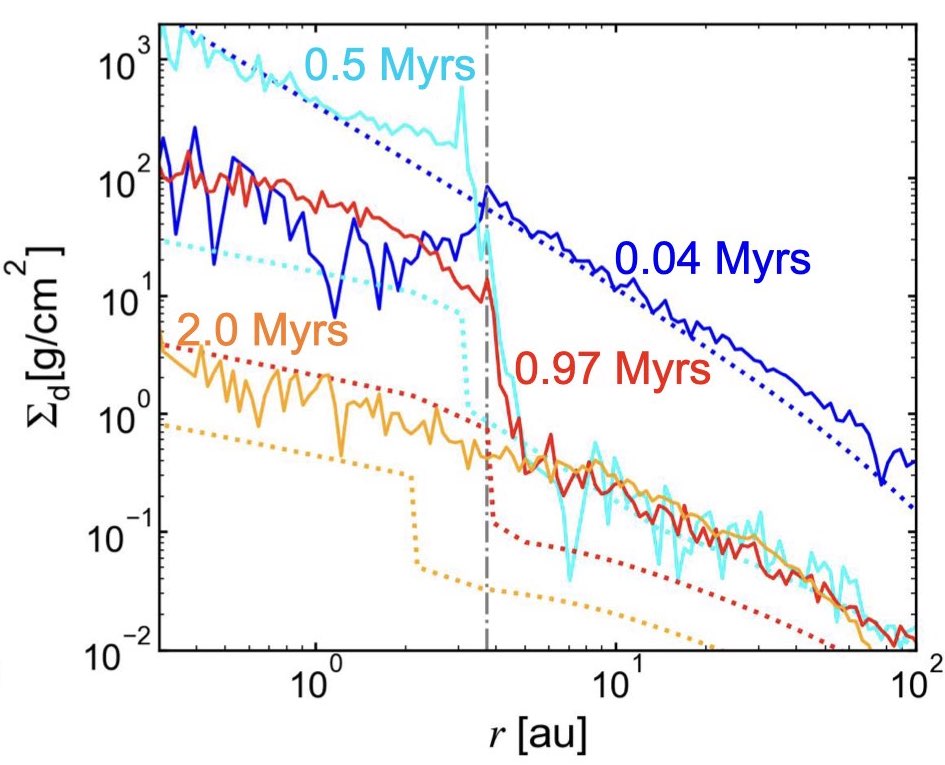}
        \end{minipage}&
        \begin{minipage}[t]{0.22\textwidth}
            \centering
            \includegraphics[keepaspectratio,scale=0.13]{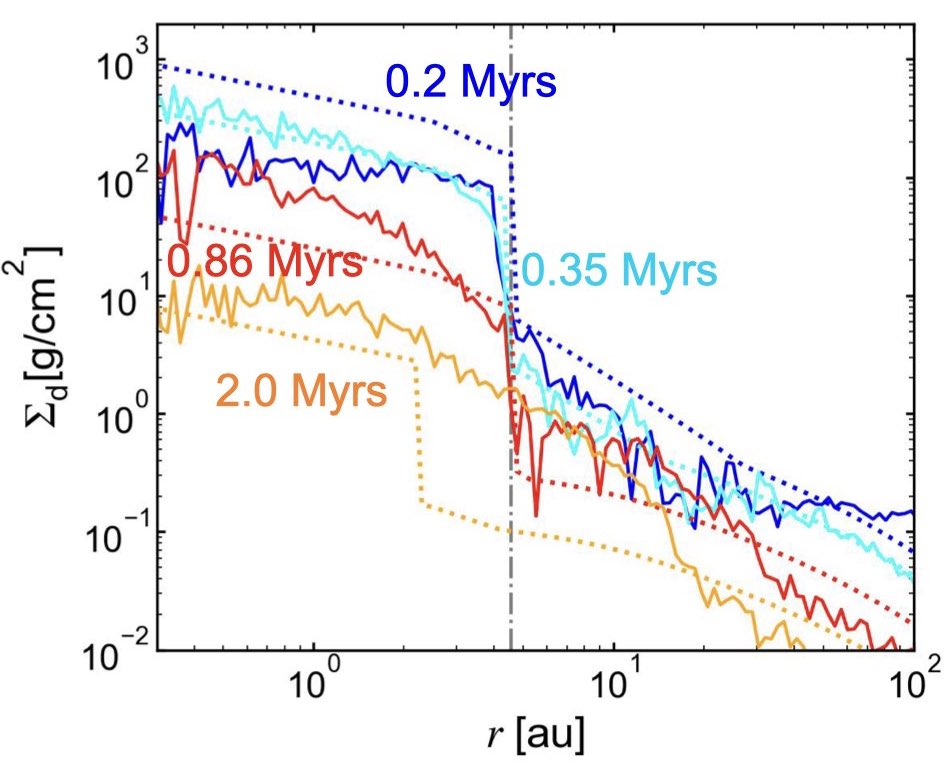}
        \end{minipage}&
        \begin{minipage}[t]{0.22\textwidth}
            \centering
            \includegraphics[keepaspectratio,scale=0.13]{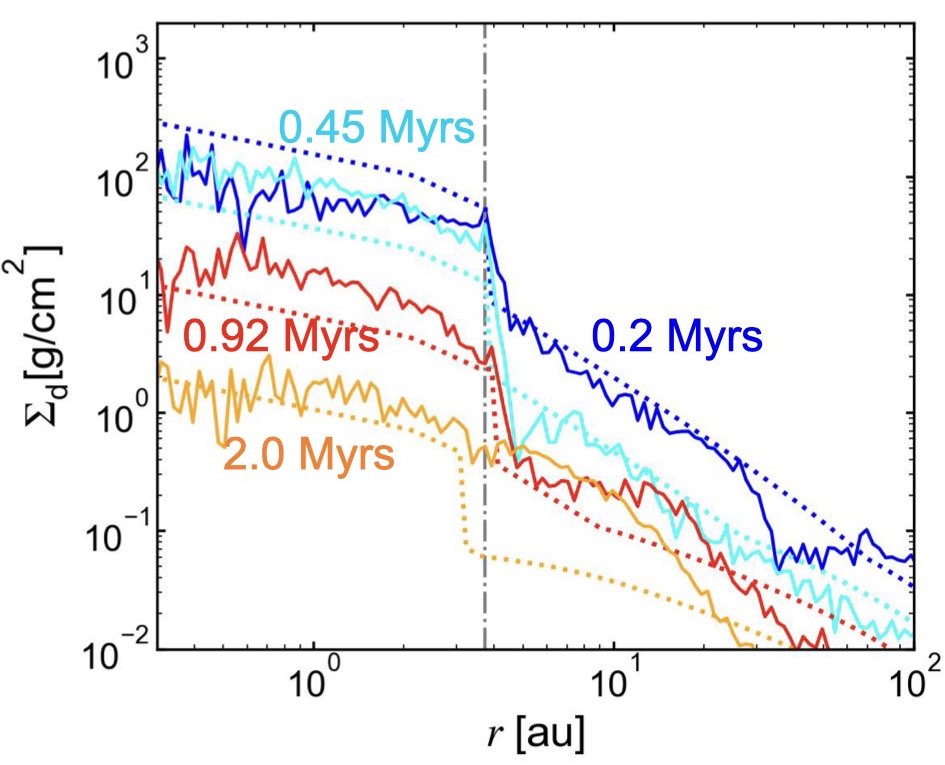}
        \end{minipage}&
        \begin{minipage}[t]{0.22\textwidth}
            \centering
            \includegraphics[keepaspectratio,scale=0.13]{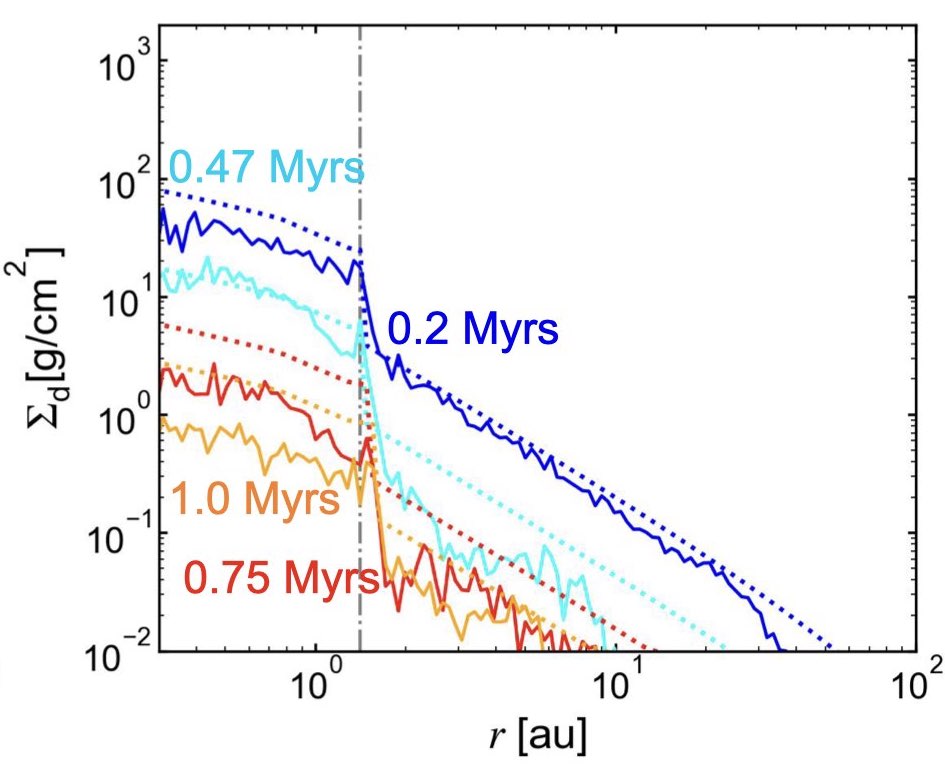}
        \end{minipage}\\
        \begin{minipage}[t]{0.22\textwidth}
            \centering
            \includegraphics[keepaspectratio,scale=0.133]{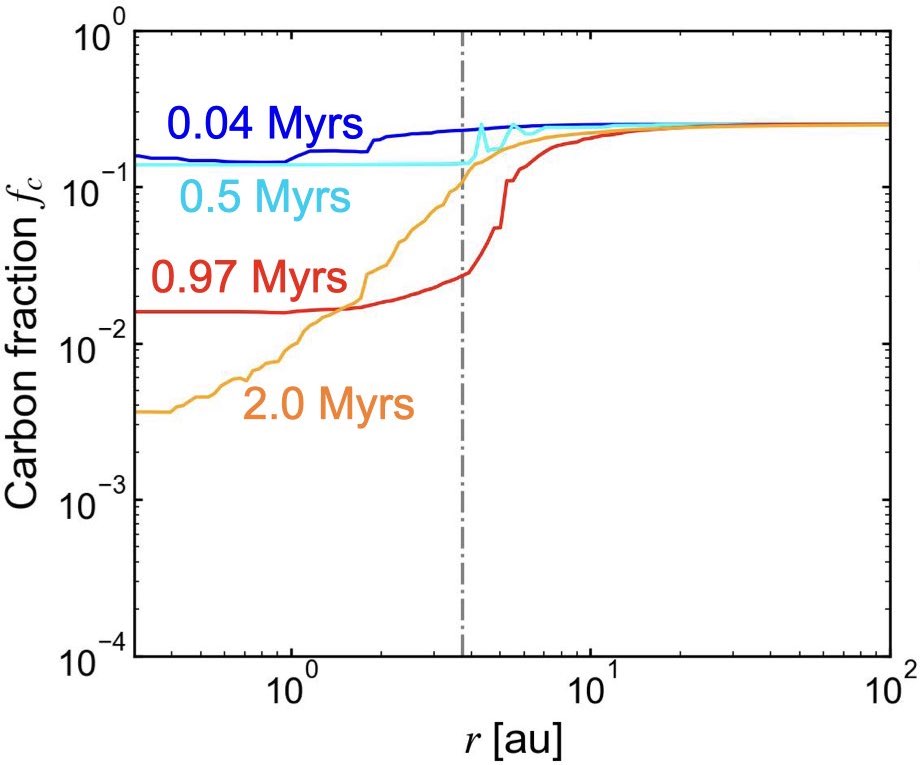}
            \subcaption{$Z=0.05$}
        \end{minipage}&
        \begin{minipage}[t]{0.22\textwidth}
            \centering
            \includegraphics[keepaspectratio,scale=0.133]{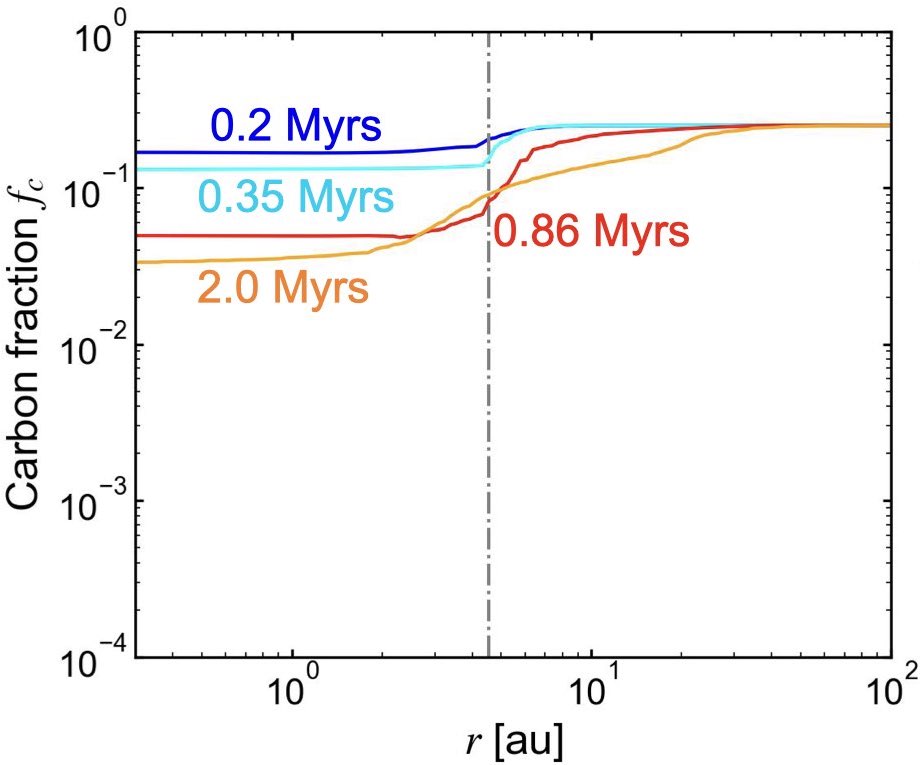}
            \subcaption{$\alpha=5\times 10^{-4}$}
        \end{minipage}&
        \begin{minipage}[t]{0.22\textwidth}
            \centering
            \includegraphics[keepaspectratio,scale=0.133]{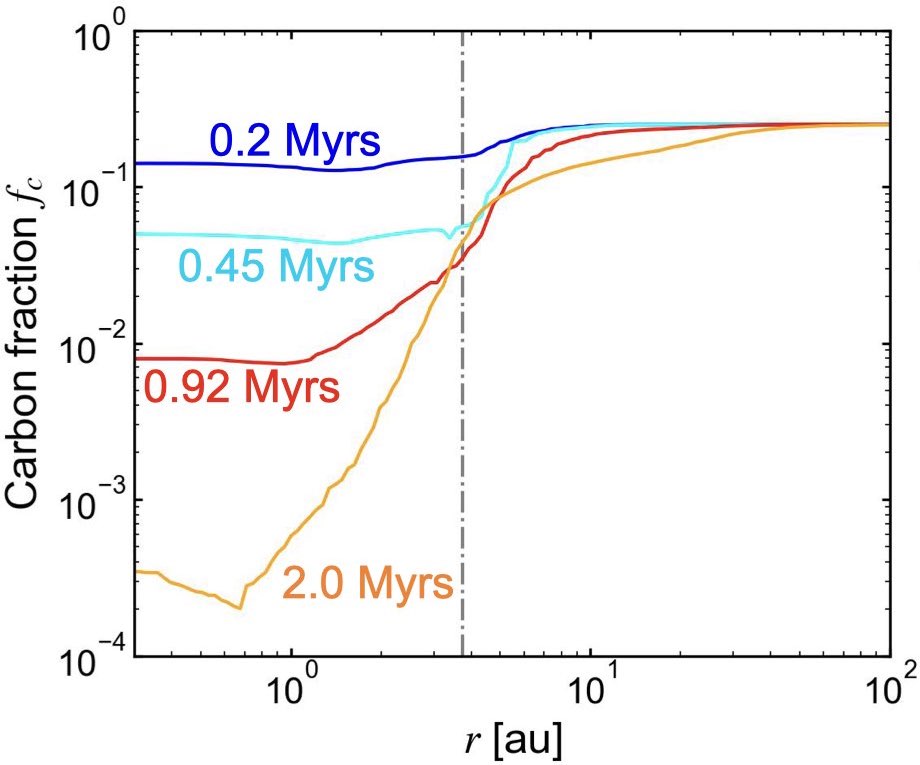}
            \subcaption{$\v_{\rm frag,sil}=0.1\:\rm m/s$}
        \end{minipage}&
        \begin{minipage}[t]{0.22\textwidth}
            \centering
            \includegraphics[keepaspectratio,scale=0.133]{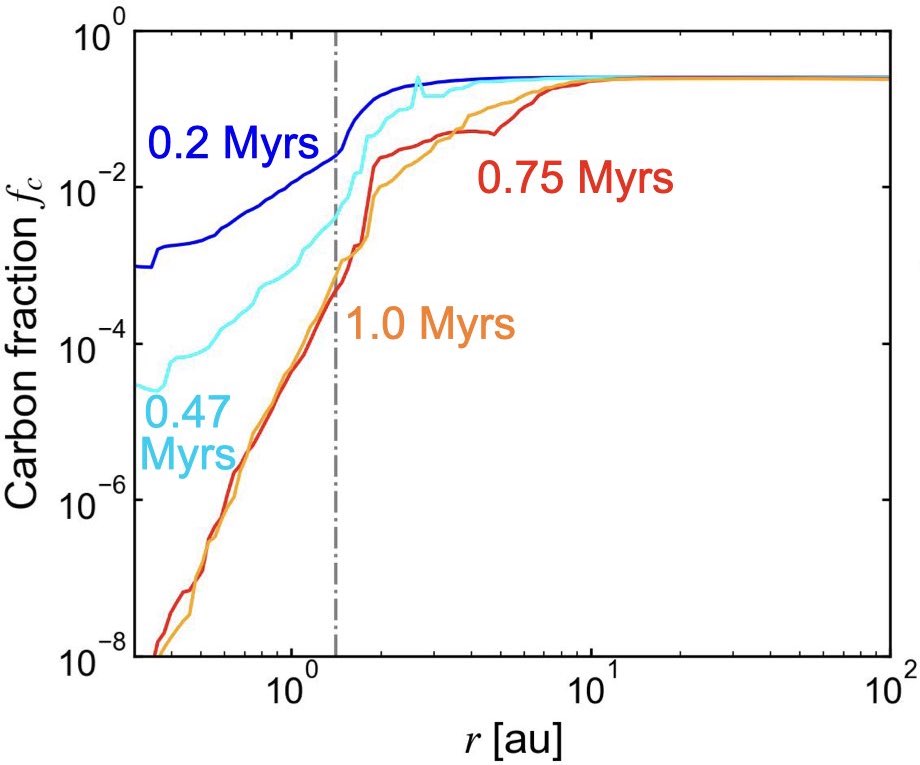}
            \subcaption{$\dot{M}_{\rm g}=10^{-9}M_{\odot}/\rm year$}
        \end{minipage}
    \end{tabular}
    \caption{Time evolution of the solid surface densities (top row) and carbon fractions (bottom row) for each parameter set. The sub-captions show the parameter which we changed from the fiducial parameter set. The dashed lines in the top row show the analytic estimation given by Eq.~\ref{eq:Z}.}
\label{fig:app_para}
\end{figure*}

Figure~\ref{fig:app_para} shows the time evolution of the solid surface densities ($\Sigma_{\rm d}$) and $f_{\rm c}$ for different parameter sets. The times of the four lines are corresponding to the four stages discussed in Sect.~\ref{sec:detailed_evol}. In the lower gas accretion rate stage (final stage), $\dot{M}_{\rm g}=10^{-9}\,M_{\odot}/\rm year$, we stopped the calculation at 1 Myr, when the total solid mass inside the snow line is lower than the Earth's mass.

\end{appendix}
\end{document}